\documentclass[a4paper,hidelinks, 12pt]{article}
\usepackage[utf8]{inputenc}
\usepackage{a4wide}
\usepackage{graphics}
\usepackage{amsmath}
\usepackage{amsfonts}
\usepackage{amssymb}
\usepackage{subfigure}
\usepackage{cite}
\usepackage{xcolor}
\usepackage{soul}
\usepackage[toc,page]{appendix}
\usepackage{mathtools}
\usepackage{hyperref}
\usepackage{enumitem}
\usepackage{threeparttable}
\usepackage{etoolbox} 
\usepackage{longtable}
\usepackage{multirow}
\usepackage{threeparttablex}
\usepackage{booktabs}
\usepackage{pdflscape}
\usepackage{float}

\allowdisplaybreaks
\NewCommandCopy{\originalsubsection}{\subsection}
\RenewDocumentCommand{\subsection}{sd()O{#4}m}{%
  \IfBooleanTF{#1}
    {\originalsubsection*{#4}}%
    {%
     \IfNoValueTF{#2}
       {
        \renewcommand{\thesubsection}{\thesection.\arabic{subsection}}%
       }
       {
        \addtocounter{subsection}{-1}%
        \renewcommand{\thesubsection}{\thesection#2}%
       }
     \originalsubsection[#3]{#4}%
    }%
}

\numberwithin{equation}{section} 

\makeatletter
\@addtoreset{equation}{section}
\let\TPT@hookin\@gobble
\let\TPT@hookarg\@gobble
\newcommand{\labeltext}[2]{%
  \@bsphack
  \csname phantomsection\endcsname 
  \def\@currentlabel{#1}{\label{#2}}%
  \@esphack
}
\makeatother

\begin{document}

\begin{titlepage}
\begin{center}

{\large \bf {Systematic analysis of 3HDM symmetries}}

\vskip 1cm

A. Kun\v cinas,$^{a,b,}$\footnote{E-mail: Anton.Kuncinas@tecnico.ulisboa.pt}
P. Osland$^{c,}$\footnote{E-mail: Per.Osland@uib.no} and 
M. N. Rebelo$^{a,}$\footnote{E-mail: rebelo@tecnico.ulisboa.pt}

\vspace{1.0cm}

$^{a}$Centro de F\'isica Te\'orica de Part\'iculas, CFTP, Departamento de F\'\i sica,\\ Instituto Superior T\'ecnico, Universidade de Lisboa,\\
Avenida Rovisco Pais nr. 1, 1049-001 Lisboa, Portugal,\\
$^{b}$Instituto de Física Corpuscular (IFIC), CSIC‐Universitat de València,\\ Parc Científic UV, c/ Catedrático José Beltrán, 2, E-46980 Paterna (València), Spain\\
$^{c}$Department of Physics and Technology, University of Bergen, \\
Postboks 7803, N-5020  Bergen, Norway\\
\end{center}

\vskip 2cm

\begin{abstract}

\noindent

Symmetries play a crucial role in shaping the structure and predictions of multi-Higgs-doublet models. In three-Higgs-doublet models considerable effort has been put into classifying possible symmetry groups and the conditions for their realisation, yet the completeness of existing classifications remains an open question. In this work, we revisit the problem of identifying realisable symmetries by re-examining conventional Higgs family and general CP transformations from an alternative perspective. Our analysis identifies certain limitations in previous approaches and introduces a clearer, more systematic framework for model builders. We expand our classification by investigating more generalised symmetry structures---the recently identified GOOFy transformations, which act non-trivially on the Higgs doublets and their conjugates. Our analysis consolidates known results, uncovers previously overlooked structures, and expands the set of symmetries in three-Higgs-doublet models, offering both a clearer theoretical foundation and a practical reference for symmetry-based model building.

\end{abstract}

\end{titlepage}

 \tableofcontents
 
 \newpage

\section{Introduction}

Symmetries are central to our understanding of the physical laws---shaping everything from fundamental interactions to conservation rules. Guidelines to choose a particular symmetry can only come from experiment. At this stage a lot of freedom remains due to the lack of definite experimental constraints.

This limitation is particularly noticeable because we depend heavily on the symmetry groups to predict physical properties. It raises an open question: are symmetries truly fundamental or might they be a reflection of deeper structures?

In quantum field theory, a symmetry plays a dual role: it constrains the form of the Lagrangian and governs the possible patterns of spontaneous breaking. Multi-Higgs-doublet models (NHDMs) provide a fertile ground for exploring this interplay, as their scalar sectors admit a rich variety of symmetry groups. In practice, however, such symmetries are typically imposed by hand, rather than derived from underlying principles.

In recent years, substantial progress has been made in identifying realisable symmetries within the framework of 3HDMs~\cite{Ferreira:2008zy,Ivanov:2011ae,Ivanov:2012ry,Ivanov:2012fp,Keus:2013hya,Ivanov:2014doa,Ivanov:2015mwl,Pilaftsis:2016erj,deMedeirosVarzielas:2019rrp,KunMT,Darvishi:2019dbh,Kuncinas:2020wrn,Bree:2024edl,Kuncinas:2024zjq,Doring:2024kdg,Bryan:2026xvs}. Nevertheless, the set of possible symmetries remains largely unexplored, \textit{e.g.}, the recent identification of the so-called GOOFy (abbreviated from the initials of the authors) transformations~\cite{Ferreira:2023dke}; yet, another approach, entirely different, inspired by the information-theoretic properties of scalar models, can be found in Refs.~\cite{Carena:2023vjc,Carena:2025wyh,Busoni:2025dns}, but it is not tackled here. In this work, we revisit the question of realisable symmetries in 3HDMs from a slightly different perspective, as some previously adopted approaches may not always capture the full picture~\cite{KunMT,Kuncinas:2020wrn,Kuncinas:2024zjq}. We note that the compact/unitary symplectic group, $USp(n)$, will not be addressed in this study.

The paper is organised as follows. Our analysis begins in Section~\ref{Sec:Paradigm} by establishing a consistent set of definitions and transformation rules, providing a clear foundation for discussing relevant symmetry transformations in the NHDM context and setting the stage for the classification work presented in subsequent sections. In Sections~\ref{Sec:HF_tr} and~\ref{Sec:GCP_tr} we discuss two different transformations: Higgs family (sometimes referred to as Higgs flavour transformations) and general CP transformations, which are then summarised in Section~\ref{Sec:GT_GCP_summary}. We next turn to the discussion of the GOOFy transformations in Sections~\ref{Sec:GOOFY_3HDMs} and~\ref{Sec:TGOOFY_3HDMs}, examining their implications. We present a summary in Section~\ref{Sec:Summary}. In Appendix~\ref{App:U1_D4} we consider one of the previously missed models, $U(1) \times D_4$, rigorously, followed by the discussion of the irreducible $S_3$ representations in Appendix~\ref{App:S3_sign}. In Appendix~\ref{App:generators} we list generators used to identify ``conventional" GOOFy transformations. Finally, in Appendix~\ref{App:Gen_TGOOFy} we present GOOFy-stabilised potentials.

\section{Terminology and framework}\label{Sec:Paradigm}

In this section we outline the framework and conventions used throughout our analysis. Our aim is to provide a consistent set of definitions, transformation rules, and conventions that will be employed throughout the rest of the paper. We begin by introducing an extended representation for the scalar fields, followed by a discussion of the relevant symmetry transformations in the NHDM context, with a focus on $N=3$, 3HDM. This sets the stage for the explicit examples and classification work presented in later sections.

\paragraph*{Scalar space \\}

We define an extended reducible space for a set of $N$ scalar $SU(2)$ doublets and a corresponding set of their conjugates via
\begin{equation}\label{Eq:H_def_h_h*}
H \equiv h_i \oplus h_i^\ast = (h_1,\, \dots,\,h_N,\, h_1^\ast,\, \dots ,\,h_N^\ast)^\mathrm{T} \in \mathbb{C}^{2N},
\end{equation}
where transposition does not act on the doublets inside the parenthesis. This definition is useful for describing transformations that mix fields with their complex conjugates. The direct sum refers to the structure of the algebraic representation space $H$ (unified multiplet), not the physical degrees of freedom.

\paragraph*{Symmetry transformations\\}

The reparametrisation space of an NHDM is given by an arbitrary basis change
\begin{equation}\label{Eq:h_to_Uh}
h_i \;\longrightarrow\; U_{ij} h_j, \qquad U \in U(N).
\end{equation}
Any field redefinition acting on the doublets induces a transformation on the set of monomials $h_i^\dagger h_j$, and thus on the coefficients of the scalar potential. These transformations are in general \textit{not} symmetry transformations: they relate equivalent basis choices for the same physical model. However, if a scalar potential is left invariant under a particular transformation, such transformation is promoted to a symmetry transformation. 

The standard global symmetry transformations of the scalar multiplet are restricted to those that treat the fields $h_i$ and their complex conjugates in a related way. Imposing this condition allows for two possible  fundamental classes:
\begin{itemize}
\item Higgs family (HF) transformations amount to basis rotations among the doublets,
\begin{equation}\label{Eq:H_to_HF}
H  \mapsto U^\mathrm{HF} H \equiv \begin{pmatrix}
\mathcal{U} & 0 \\
0 & \mathcal{U}^\ast
\end{pmatrix} H.
\end{equation}
\item General CP (GCP) transformations~\cite{Lee:1966ik,Ecker:1981wv,Ecker:1987qp,Neufeld:1987wa} mix doublets with conjugate doublets, 
\begin{equation}\label{Eq:H_to_GCP}
H \mapsto U^\mathrm{GCP} H \equiv \begin{pmatrix}
0 & \mathcal{U} \\
\mathcal{U}^\ast & 0
\end{pmatrix} H.
\end{equation}
\end{itemize}

Here, $\mathcal{U}$ is an $N\times N$ unitary matrix acting in the space of scalar doublets, with $N$ the number of doublets in the model (we focus on $N=3$). In practice we specify generators in terms of $\mathcal{U}$, while the \emph{true} group in the $H$ space is generated by $2N\times 2N$ matrices.

The block structures in eqs.~\eqref{Eq:H_to_HF} and \eqref{Eq:H_to_GCP} define the actual action on $H$. In the HF case, a given $\mathcal{U}$ extends trivially to the $H$ space. In the GCP case, the extension inherently incorporates complex conjugation. If $\mathcal U$ is complex, applying it twice yields an HF transformation proportional to $\mathcal U  \mathcal U^\ast$, and not $\mathcal U  \mathcal U$. 

To avoid clutter with redundant $2N\times 2N$ matrices, we will henceforth specify only the $N\times N$ blocks $\mathcal{U}$. To unambiguously define their action on the scalar multiplet, we shall use the notation:
\[
\langle \mathcal{U} \rangle_\mathrm{HF}
\quad\text{and}\quad
\langle \mathcal{U}\rangle_\ast \quad .
\]
These denote the subgroups generated by the specific embedding of $\mathcal{U}$ as either an HF transformation or a GCP transformation, respectively. We use the ``$\ast$" subscript to indicate the presence of GCP elements within a symmetry group, serving as a reminder that these transformations inherently incorporate complex conjugation.

\paragraph*{The \boldmath$SU(2)$ bilinear singlets\\}

For convenience we define an $SU(2)$ bilinear singlet built out of the scalar doublets,
\begin{equation}
h_{ij} \equiv h_i^\dagger h_j.
\end{equation}
The scalar potential is then given by
\begin{equation}\label{Eq:V_NHDM_gen}
V= \mu_{ij}^2 h_{ij} + \lambda_{ijkl} h_{ij}h_{kl}, \quad \mu_{ji}^2 = (\mu_{ij}^2)^\ast,~ \lambda_{jilk}=\lambda_{ijkl}^\ast,~\lambda_{klij}=\lambda_{ijkl}.
\end{equation}
For the studied case of 3HDM there are 9 bilinear, $\mu_{ij}^2$, and 45 quartic, $\lambda_{ijkl}$, terms.

The $SU(2)$ bilinear singlet transforms under the HF and GCP transformations as
\begin{subequations}
\begin{align}
\mathrm{HF:} \quad h_{ij} \mapsto \mathcal{U}_{ik}^\ast \mathcal{U}_{jl} h_{kl},\label{Eq:HF_tr} \\
\mathrm{GCP:} \quad h_{ij} \mapsto \mathcal{U}_{ik}^\ast \mathcal{U}_{jl} h_{lk}. \label{Eq:GCP_tr}
\end{align}
\end{subequations}
The trivial GCP transformation can be understood as a $\mathbb{Z}_2$ (or $S_2$) symmetry that acts by swapping the indices of the $SU(2)$ bilinear singlets, $\text{GCP}_\mathrm{trivial}:~h_{ij} \mapsto h_{ji}$.

Having identified how $h_{ij}$ transform, the next step would be to argue which form the transformation matrices $\mathcal{U}$ could take. Although there are various methods for addressing symmetries in NHDMs, we will take a more straightforward, less elegant route---opting for a brute-force approach to identify realisable symmetries; this amounts to systematically reducing the general scalar potential by successively imposing symmetry transformations. 

\paragraph*{Realisable and accidental symmetries\\}

In NHDMs, a \textit{realisable symmetry} is defined as a symmetry group $\mathcal{G}_\mathrm{r}$ that can be imposed on the scalar potential without automatically leading to a larger group. In other words, there exists a valid region of parameter space where $\mathcal{G}_\mathrm{r}$ represents the maximal symmetry of the potential.

However, if the parameter constraints required to impose invariance under a group are insufficient to distinguish it from a larger one, the true invariance of the potential becomes an \textit{accidentally enhanced} group. This enhancement occurs because the renormalisable potential lacks the complexity to support the smaller group without automatically satisfying the symmetries of a larger one.

Consequently, a group $\mathcal{G}_\mathrm{nr}$ is \textit{non-realisable} if the constraints required to satisfy $\mathcal{G}_\mathrm{nr}$-invariance accidentally render the potential invariant under a larger group $\mathcal{G}_\mathrm{r}$, such that $\mathcal{G}_\mathrm{nr} < \mathcal{G}_\mathrm{r}$. In this scenario, $\mathcal{G}_\mathrm{nr}$ acts merely as a proper subgroup of the true symmetry group. Therefore, non-realisable groups cannot characterise the exact symmetries of an NHDM potential.

In Ref.~\cite{Ivanov:2011ae}, it is proved that any finite Abelian group $\mathcal{G}$ realisable in NHDM satisfies $|\mathcal{G}| \le 2^{N-1}$. In particular, any cyclic group $\mathbb{Z}_p$ that can be realised must satisfy $p \le 2^{N-1}$. Groups violating this bound are necessarily non-realisable: imposing them forces the potential to acquire a larger symmetry.

For example, in 2HDMs~\cite{Ivanov:2005hg,Ivanov:2006yq,Gerard:2007kn,Ivanov:2007de,Ferreira:2009wh,Ferreira:2010yh,Battye:2011jj,Pilaftsis:2011ed,Haber:2018iwr,Bento:2020jei,Ferreira:2020ana,Ferreira:2022gjh,Bednyakov:2025sri,Solberg:2025ybf}, imposing a $\mathbb{Z}_n$ symmetry with $n \geq 3$ automatically leads to invariance under the $U(1)$ Peccei–Quinn group~\cite{Peccei:1977hh}. Hence, $\mathbb{Z}_n$ for $n \geq 3$ are non-realisable in the scalar sector. It is important to note, however, that when extending the framework to the fermionic sector, a symmetry that is non-realisable for the scalar potential alone may become realisable in the full Lagrangian (see, \textit{e.g.}, the $\mathbb{Z}_3$-symmetric 2HDM in Refs.~\cite{Kuncinas:2025mcn,Ferreira:2025ymc}).

\paragraph*{Identification of symmetry groups\\}
The classification of finite subgroups of $SU(3)$ began several decades ago~\cite{Miller:1916,Fairbairn:1964sga,Bovier:1980ga,Bovier:1980gc,Fairbairn:1982jx} and has more recently continued in Refs.~\cite{Luhn:2007uq,Escobar:2008vc,Ishimori:2010au,Grimus:2010ak,Ludl:2010bj,Ludl:2011gn,Holthausen:2012wt}. A comprehensive list of non-isomorphic finite subgroups of $U(3)$ with orders up to 2000 (with some special exceptions) is given in Ref.~\cite{Jurciukonis:2017mjp}, where the $\mathsf{SmallGrp}$ library~\cite{SmallGrp} of $\mathsf{GAP}$~\cite{GAP4} was used to classify groups based on generators and irreducible representations. The classification is given in terms of explicit generator matrices. These generators often depend on discrete parameters (for instance, integers $\{m,n\}$ that appear in phases such as $e^{2i\pi m/n}$). Such a parameter-dependent generator should be viewed as a family: fixing the parameters produces a generator of a particular discrete subgroup, while promoting those parameters to continuous variables yields a continuous family of transformations and hence a continuous subgroup of $U(3)$. Because of this, the generator-based description in Ref.~\cite{Jurciukonis:2017mjp} is useful both for discrete classifications and for identifying candidate continuous subgroups of $U(3)$. We will predominantly rely on these generator families throughout the paper.

A few clarifications are in order. Real orthogonal matrices $O(n,\mathbb{R})$ are embedded naturally into $U(n)$, $O(n, \mathbb{R}) \hookrightarrow U(n, \mathbb{C})$, since in this case $O^\mathrm{T}O=\mathcal{I}$ implies $O^\dagger O=\mathcal{I}$. In contrast, the complex orthogonal group $O(n,\mathbb{C})$ is not a subgroup of $U(n)$.

The scalar potential depends only on the $SU(2)$ singlets $h_{ij}$, which form a positive-definite Hermitian matrix. The Hermiticity condition is preserved by requiring that field transformations must be unitary. Therefore, the non-unitary groups like $O(n, \mathbb{C})$ would not be physically allowed.

The maximal admissible symmetry of the 3HDM is $PSU(3)$ rather than $U(3)$. Since the scalar potential is constructed entirely from the $SU(2)$-invariant bilinears $h_{ij}$, it is trivially invariant under the global hypercharge transformations $U(1)_Y: h_i \mapsto e^{i\alpha} h_i$. For a scalar potential invariant under $U(3)$, we should bear in mind that it is possible to identify elements of $U(3)$ that differ by an overall $U(1)$ phase, leading to the projective unitary group $PU(3) \equiv  U(3) / U(1)$. Equivalently, restricting to unit-determinant transformations gives $SU(3)$. The center of this group, $Z(SU(N)) = \mathbb{Z}_N = \left\langle e^{2\pi i/N}\mathcal{I} \right\rangle$ acts as a discrete overall phase rotation, and its action therefore lies entirely within the $U(1)_Y$ gauge orbit. The resulting physical symmetry group acting on the scalar-potential parameters is thus the projective group $PSU(3) \cong SU(3)/\mathbb{Z}_3$. Using the group isomorphism $U(3) \cong \left( U(1) \times SU(3) \right) /\mathbb{Z}_3$ and quotienting by the overall phase yields $PU(3) \equiv PSU(3)$.

\paragraph*{Invariant potentials\\}

We construct specific scalar potentials by iteratively reducing the parameter space of the most general 3HDM potential through the imposition of symmetry requirements. We start with the most general potential $V_0(H)$ and sequentially impose invariance under groups $\{ \mathcal{H}_1, \dots, \mathcal{H}_j\}$, resulting in the following sequence of constrained potentials:
\begin{equation}\label{Eq:Inv_Cond}
V_{0}(H) \xrightarrow[\text{impose } \mathcal{H}_1]{} V_{\mathcal{H}_1}(H) \xrightarrow[\text{impose } \mathcal{H}_2]{} 
 V_{\{ \mathcal{H}_1,\mathcal{H}_2\}}(H) \xrightarrow[\text{impose } \mathcal{H}_3]{}
\, \dots \, \xrightarrow[\text{impose } \mathcal{H}_j]{} V_{\mathcal{G}}(H).
\end{equation}
The constraints imposed by the generators of $\mathcal{H}_i$ reduce the number of independent parameters step-by-step, leaving the potential unchanged if the symmetry was already present.

The resulting potential cannot generate new monomials (beyond $h_{ij}$ and $h_{ij}h_{kl}$), it is just the couplings that can be constrained. By comparing the coefficients of the monomials on both sides of each step in eq.~\eqref{Eq:Inv_Cond}, we obtain a set of linear constraints on the couplings.

For a chosen symmetry subgroup $\mathcal{H}_i \leq \mathcal{G}$, generated by $\mathcal{H}_i = \left \langle g_i \right\rangle$ (a group generated by a single generator), we thus demand that the corresponding potential remains invariant under the action of $\mathcal{H}_i$:
\begin{equation}
V_{\mathcal{H}_i}(H) = V_{\mathcal{H}_i}(g_i H).
\end{equation}

In general, the full target symmetry group $\mathcal{G}$ may consist of several components,
\begin{equation}
\mathcal{G} \cong \mathcal{H}_1 \,\ast\, \mathcal{H}_2 \,\ast\, \cdots  \ast \mathcal{H}_j = \langle g_{1},\, g_{2},\, \dots,\, g_{j} \rangle.
\end{equation}
Here, we deliberately employ the symbol ``$\ast$" as a convenient shorthand to denote an unspecified group product or extension. While this typically represents a standard direct ``$\times$" or semidirect ``$\rtimes$" product, it also accommodates cases where the constituent subgroups share a non-trivial intersection. In such instances, the algebraic structure encompasses central products, necessitating a quotient by the overlapping elements to define the full group. The final product is strictly determined by the commutation and conjugation relations amongst the constituent generating sets.

Finally, it is important to note that the imposed constraints may lead to an accidental enhancement of the symmetry; thus, the final symmetry group of the potential can strictly exceed the initially specified target group $\mathcal{G}$.

\paragraph*{Different representations\\}
By examining the finite subgroups of $U(3)$, we not only construct various symmetry-invariant scalar potentials, but also obtain different representations. For example, let us consider the dihedral group, with the group presentation:
\begin{equation}
D_n = \left\langle g_1,\, g_2~ \big| ~ g_1^n = g_2^2 = (g_2 g_1)^2 = e\right\rangle.
\end{equation}
Consider the following representations of $D_4$, implemented as an HF transformation:
\begin{itemize}
\item For 
\begin{equation}
D_4 = \left\langle  \begin{pmatrix}
0 & -1 & 0\\
1 & 0 & 0\\
0 & 0 & 1
\end{pmatrix},\, \begin{pmatrix}
-1 & 0 & 0\\
0 & 1 & 0\\
0 & 0 & 1
\end{pmatrix}  \right\rangle_\mathrm{HF},
\end{equation}
the invariant scalar potential is given by:
\begin{equation}\label{Eq:D4_g1}
\begin{aligned}
V_{D_4}={}& \mu_{11}^2 (h_{11} + h_{22}) + \mu_{33}^2 h_{33}\\
& + \lambda_{1111} (h_{11}^2 + h_{22}^2) + \lambda_{3333} h_{33}^2 + \lambda_{1221} h_{12}h_{21} + \lambda_{1331} (h_{13}h_{31} + h_{23}h_{32})\\
& + \lambda_{1122} h_{11} h_{22} + \lambda_{1133} (h_{11} + h_{22})h_{33}  + \lambda_{1212} (h_{12}^2 + h_{21}^2) \\
&+ \left\lbrace \lambda_{1313} (h_{13}^2 + h_{23}^2) + \mathrm{h.c.} \right\rbrace.
\end{aligned}
\end{equation}

\item For 
\begin{equation}
D_4^\prime = \left\langle  \begin{pmatrix}
-i & 0 & 0\\
0 & i & 0\\
0 & 0 & 1
\end{pmatrix},\, \begin{pmatrix}
0 & 1 & 0\\
1 & 0 & 0\\
0 & 0 & 1
\end{pmatrix}  \right\rangle_\mathrm{HF},
\end{equation}
the invariant scalar potential is given by:
\begin{equation}\label{Eq:D4_g2}
\begin{aligned}
V_{D_4^\prime}={}& \mu_{11}^2 (h_{11}^\prime + h_{22}^\prime) + \mu_{33}^2 h_{33}^\prime\\
& + \lambda_{1111} (h_{11}^{\prime \,2} + h_{22}^{\prime \,2}) + \lambda_{3333} h_{33}^{\prime \,2} + \lambda_{1221} h_{12}^\prime h_{21}^\prime + \lambda_{1331} (h_{13}^\prime h_{31}^\prime + h_{23}^\prime h_{32}^\prime)\\
& + \lambda_{1122} h_{11}^\prime h_{22}^\prime + \lambda_{1133} (h_{11}^\prime + h_{22}^\prime)h_{33}^\prime  + \lambda_{1212} (h_{12}^{\prime \,2} + h_{21}^{\prime \,2}) \\
& + \left\lbrace \lambda_{1323} h_{13}^\prime h_{23}^\prime + \mathrm{h.c.} \right\rbrace.
\end{aligned}
\end{equation}
\end{itemize}

These two are connected via a unitary transformation,
\begin{equation}\label{Eq:D1_to_D4pr}
\begin{pmatrix}
h_1 \\
h_2 \\
h_3
\end{pmatrix} = \frac{1}{\sqrt{2}}\begin{pmatrix}
e^{i \pi/4} & e^{i \pi/4} & 0 \\
-e^{-i \pi/4} & e^{-i \pi/4} & 0 \\
0 & 0 & \sqrt{2}
\end{pmatrix} \begin{pmatrix}
h_1^\prime \\
h_2^\prime \\
h_3^\prime
\end{pmatrix}.
\end{equation}
For simplicity we shall define an $SU(2)$ transformation matrix,
\begin{equation}
\mathcal{R}_{(\alpha,\,\beta,\,\theta,\,\phi)} \equiv \begin{pmatrix}
e^{i (\alpha + \beta + \phi/2)} \cos \theta & e^{i(\alpha - \beta + \phi/2)} \sin \theta & 0\\
-e^{-i(\alpha - \beta - \phi/2)} \sin \theta & e^{-i (\alpha + \beta - \phi/2)} \cos \theta & 0\\
0 & 0 & 1
\end{pmatrix}.
\end{equation}
In this notation the basis transformation matrix is given by \mbox{$h_i = (\mathcal{R}_{(\frac{\pi}{4},\,0,\,\frac{\pi}{4},\, 0)})_{ij} \,h^\prime_j.$}

\vspace{6pt}
\paragraph*{Basis transformations and redundancies\\}
Although different representations of the same symmetry group can yield distinct forms of the scalar potential, care must be taken to avoid redundant degrees of freedom in the couplings. Eliminating such redundancies, often possible via basis transformations, can reveal an enhanced underlying symmetry, in the sense that the scalar potential becomes explicitly invariant (in a particular basis) under additional symmetry generators.

This raises an interesting question: how can one identify the number of independent couplings? We address  this question in Appendix~\ref{App:Redundancies}.

\vspace{6pt}
\paragraph*{Bilinear formalism\\}
The final ingredient we wish to discuss, before proceeding to our analysis, is the bilinear formalism in the context of NHDMs. Initially developed to study vacuum structures~\cite{Velhinho:1994np}, it was subsequently extended in Refs.~\cite{Nagel:2004sw,Ivanov:2005hg,Maniatis:2006fs,Ivanov:2006yq,Maniatis:2006jd,Maniatis:2007vn,Ivanov:2007de,Nishi:2007dv}. While we do not present a formal derivation here, the bilinear formalism is a convenient framework that offers significant insight into the underlying symmetries of the scalar potential and its properties. For those unfamiliar with the method, we refer to Ref.~\cite{deMedeirosVarzielas:2019rrp} for a detailed discussion, including its application in identifying symmetries in a basis-invariant manner for 3HDMs. We found it useful to classify scalar potentials based on the multiplicities of the eigenvalues of the $\Lambda_{ij}$ matrix, as discussed in Ref.~\cite{deMedeirosVarzielas:2019rrp}.

To establish the basic notation, the nine complex $SU(2)$-invariant bilinears $h_{ij}$, $\{i,j\} \in \{1, 2, 3\}$ (strictly speaking not independent), can be decomposed into a single real singlet $r_0$ and a real vector $r_a$, with $a$ running from 1 to 8, using the Gell-Mann matrices $\lambda_a$:
\begin{equation}
h_{ij} = \frac{1}{\sqrt{3}} r_0 \delta_{ij} +  \sum_{a=1}^{8} r_a (\lambda_a)_{ij},
\end{equation}
where
\begin{subequations}
\begin{align}
r_0 ={}& \frac{1}{\sqrt{3}} \sum_{i=1}^{3} h_{ii},\\
r_a ={}& \frac{1}{2} \sum_{\{i,\,j\}=1}^{3} (\lambda_a)_{ij}\,  h_{ij}.
\end{align}
\end{subequations}

In this basis, the most general gauge-invariant scalar potential is expressed as a quadratic form:
\begin{equation}
V = M_0 r_0 + M_i r_i + \Lambda_{0} r_0^2 + L_{i} r_0 r_i + \Lambda_{ij} r_i r_j,
\end{equation}
where all parameters are strictly real by Hermiticity of the scalar potential. For the purpose of symmetry classification, we restrict our attention to the quartic couplings and, specifically, the $8 \times 8$ symmetric sub-matrix $\Lambda_{ij}$. Because $r_0$ represents the total norm of the scalar multiplets, it is a singlet under $SU(3)$; it remains trivially invariant under any HF or GCP transformation.

Degeneracy in the eigenvalues of $\Lambda_{ij}$ corresponds to rotational freedom in the bilinear space, and hence to symmetry enhancement. We represent the eigenvalue multiplicity patterns compactly: the notation $v_m$ denotes that there are $v$ distinct eigenvalues, each appearing with multiplicity $m$. For example, the pattern $1_8$ indicates that there is a single eigenvalue with multiplicity 8, while $8_1$ indicates eight eigenvalues with multiplicity one. Similarly, $3_2 2_1$ corresponds to three eigenvalues with multiplicity two and two eigenvalues with multiplicity one. The ordering is irrelevant, $3_2 2_1$ has the same meaning as $2_1 3_2$.

\vspace{6pt}
\paragraph*{Workflow summary\\}

For completeness, we briefly summarise our procedure for obtaining all realisable symmetry patterns. Starting from the family classes of generators given in Ref.~\cite{Jurciukonis:2017mjp}, we impose them iteratively on the most general 3HDM potential, discarding cases that reproduce previously obtained potentials. We do not restrict the search to finite groups but treat the continuous generators on par with the discrete ones. Surviving potentials are then split into sets based on the eigenvalue structure in the bilinear space, the number of couplings (some of which may be redundant), and the basis-invariant criteria such as those derived in Ref.~\cite{deMedeirosVarzielas:2019rrp}. Distinct representations of the same group are identified using explicit $SU(3)$ basis transformations. The underlying symmetry of each remaining potential is determined from the full set of generators leaving it invariant, with accidental enhancements identified straightforwardly using the classification of Ref.~\cite{Jurciukonis:2017mjp}. This systematic approach will yield the complete catalogue of realisable symmetry groups and their associated scalar potentials.

\section{HF transformations}\label{Sec:HF_tr}

While several claims have been made about having identified \textit{all} possible realisable symmetries of 3HDMs, the reality appears more nuanced. For example, the $O(2) \times U(1)$ symmetry was identified in the context of the minimisation conditions for different $S_3$ implementations~\cite{KunMT,Kuncinas:2020wrn}. The generators leading to $[U(1) \times U(1)] \rtimes S_3$ were partially identified in Refs.~\cite{Ivanov:2012ry,Ivanov:2012fp} and only later in Ref.~\cite{Ivanov:2020jra} the underlying symmetry was properly named. Recently, all continuous groups were systematically checked in Ref.~\cite{Kuncinas:2024zjq}, and in addition to including these additional symmetries, a completely new group, denoted $U(1)\times D_4$, was identified.

In Ref.~\cite{Kuncinas:2024zjq}, we proposed the $U(1) \times D_4$ group as a previously missed realisable symmetry group. The starting point in Ref.~\cite{Kuncinas:2024zjq} was the semi-direct product structure $[U(1) \times \mathbb{Z}_2] \rtimes S_2$. However, through a private communication with I. Ivanov, we were informed that due to a non-trivial intersection between the $U(1)$ and $D_4$ subgroups, the resulting group structure might more appropriately be expressed as a quotient group. According to the discussion in Appendix~\ref{App:U1_D4}, the symmetry group relevant for our construction $\widetilde{\mathcal{G}}$ is the central product (denoted by ``$\circ$") obtained by identifying a finite central subgroup of $U(1)$ with the center $Z(\mathcal{G})$ of a finite group $\mathcal{G}$ such that
\begin{equation*}
\widetilde{\mathcal{G}} \cong U(1) \circ_{Z(\mathcal{G})} \mathcal{G} \quad \text{with} \quad Z(\mathcal{G}) \cong \mathbb{Z}_n, ~ \mathcal{G}/Z(\mathcal{G}) \cong V_4, ~ n \in 2\mathbb{Z}.
\tag{\ref{Eq:Def_tilde_G_true}}
\end{equation*}

The identification of $Z(\mathcal{G})$ with the corresponding subgroup of $U(1)$ ensures that the common $\mathbb{Z}_n$ group is not double-counted. Explicit realisations mentioned in Ref.~\cite{Kuncinas:2024zjq} include $\mathcal{G}=D_4$ or $Q_8$ for $n=2$, and the Pauli group $\mathcal{P}_1$ for $n=4$. For $n=2m$ or $4m$ with odd $m$ one obtains a trivial splitting into direct product extensions $\mathbb{Z}_m \times \mathcal{G}_{2^k}$, where $n = 2^k m$. Crucially, all such central products induce the same 3HDM scalar potential; hence in the phenomenological analysis we can treat them as a single potential-equivalence class, using the compact label $U(1) \circ V_4$ to denote this family of central products.

We started our scan for possible realisations of HF transformations, which are subgroups of $U(3)$, from the families of generators given in Ref.~\cite{Jurciukonis:2017mjp}. First, we applied continuous generators and then particularised to discrete cases. Although mathematical completeness cannot be claimed in a strict sense, our scan for discrete subgroups of $U(3)$ up to order 2000 exhibits a clear saturation behaviour: once all distinct scalar potentials invariant under finite realisable symmetries were obtained, further enlargement in the order of the symmetry groups no longer produced new potentials.

Some of the potentials obtained following this procedure lead to continuous enhancements of the symmetry, including all those obtained previously, by applying continuous generators. Both continuous and discrete realisations thus obtained converge to the $SU(3)$-invariant limit of the 3HDM. This suggests that nothing new would come after order 2000, taking into account that we are considering a renormalisable, truncated potential. In fact, no new potentials were obtained beyond order 36. Our procedure confirms that the maximal Abelian case is $\mathbb{Z}_4$, see Ref.~\cite{Ivanov:2011ae}, and that the largest realisable discrete symmetry group is $\Sigma(36) \cong (\mathbb{Z}_3\times\mathbb{Z}_3)\rtimes\mathbb{Z}_4$ (see eq.~\eqref{Eq:Sigma_36_isom}) as identified in Ref.~\cite{Ivanov:2012ry}. This consistency together with the null emergence of new potentials and agreement with existing classifications strongly indicates that all relevant 3HDM symmetries have been identified.
 
This conclusion refers strictly to realisable symmetries of the scalar potential. Note that, when the Yukawa Lagrangian is included, the initial HF symmetry may admit distinct fermion embeddings, \textit{e.g.}, $\mathbb{Z}_{n \geq 5} \leq U(1)$, yielding different textures and phenomenology.

\section{GCP transformations}\label{Sec:GCP_tr}

For simplicity, let us introduce two abbreviations for the most general scalar potentials:
\begin{itemize}
\item $V_\mathbb{C}$---the most general potential with complex parameters, as long as Hermiticity is preserved;
\item $V_\mathbb{R}$---the $V_\mathbb{C}$ potential restricted to real parameters, which is obtained by imposing the trivial CP transformation, corresponding to $h_{ij} \mapsto h_{ji}$.
\end{itemize}

In the literature, this $V_\mathbb{R}$ potential is referred to as CP1, CP2, or CPa in the 3HDM context, despite describing identical physics. On the other hand, after imposing an HF symmetry directly on the $V_\mathbb{C}$ potential, one can arrive at a real scalar potential, which would be a restricted version of $V_\mathbb{R}$. In this case the HF symmetry automatically enforces the trivial CP symmetry.

Applying the same generators to $V_\mathbb{C}$ and $V_\mathbb{R}$ may result in distinct forms  scalar potentials. Certain symmetries cannot be realised if one starts from $V_\mathbb{R}$: starting from $V_\mathbb{R}$ can artificially restrict the realisation of symmetries. A pedagogical illustration is provided by the $A_4$ and $S_4$ groups. These potentials share the same form, but the distinction is that the $A_4$-symmetric potential allows complex couplings. When these couplings are restricted to be real, the symmetry is automatically enhanced to $S_4$, since $S_n\cong A_n\rtimes \mathbb{Z}_2$, for $n \geq 2$. Explicitly,
\begin{subequations}
\begin{align}
\begin{split}
V_{A_4} ={}& \mu_{11}^2 h_{ii} + \lambda_{1111} h_{ii}^2 + \lambda_{1122} h_{ii}h_{jj} + \lambda_{1221} h_{ij}h_{ji} + \left\lbrace \lambda_{1212} h_{ij}^2 + \mathrm{h.c.} \right\rbrace,
\end{split}\\
\begin{split}
V_{S_4} ={}& \mu_{11}^2 h_{ii} + \lambda_{1111} h_{ii}^2 + \lambda_{1122} h_{ii}h_{jj} + \lambda_{1221} h_{ij}h_{ji} + \lambda_{1212} ( h_{ij}^2 + h_{ji}^2),
\end{split}
\end{align}
\end{subequations}
where implicit summation over repeated indices is assumed and $i \neq j$.

Following the notation of Ref.~\cite{Ivanov:2011ae}, we reserve the symbol ``$\ast$", as in $\mathcal{G}^\ast$, to indicate the presence of GCP elements in a symmetry group.  For instance, the group $\mathbb{Z}_2^\ast$ generates the trivial CP transformation. This notation can be generalised unambiguously to both cyclic and non-cyclic composite groups (\textit{e.g.}, $D_4^\ast$), denoting an algebraic structure containing both purely linear and conjugate-linear elements.

Let us consider a composite GCP symmetry group, $\mathcal{G}^\ast = \langle U^\mathrm{GCP}_i, U^\mathrm{GCP}_j \rangle$. To make things more clear we introduce the notation $\mathcal{J}$, where any single GCP transformation is explicitly factored as $U^\mathrm{GCP} = \mathcal{U}\,\mathcal{J}$, where $\mathcal{U}$ is a unitary matrix and $\mathcal{J}$ denotes the trivial CP transformation, mapping scalar fields to their conjugates, $\mathcal{J}: h_i \mapsto h_i^\ast$. Specifically, the multiplication rule for two GCP elements is given by:
\begin{equation}\label{Eq:Separate_2_GCP}
U^\mathrm{GCP}_i U^\mathrm{GCP}_j  = \left(\mathcal{U}_i \mathcal{J}\right)\left(\mathcal{U}_j \mathcal{J}\right) = \mathcal{U}_i \,\mathcal{U}_j^\ast,
\end{equation}
due to the commutation relation $\mathcal{J}\,\mathcal{U} = \mathcal{U}^\ast \mathcal{J}$, and $\mathcal{J}^2 = \mathcal{I}$. Consequently, this cleanly demonstrates that the full group will consist of both HF elements (like the product $\mathcal{U}_i \,\mathcal{U}_j^\ast$) and GCP elements (like the generators themselves).  The full group $\mathcal{G}^\ast$ is thus structurally composed of this purely linear HF subgroup and its conjugate-linear GCP coset.

The $V_\mathbb{R}$ potentials add no fundamentally new structures: they can always be obtained from $V_\mathbb{C}$ by imposing HF invariance and enforcing real couplings through a trivial CP. Moreover, applying a GCP transformation to $V_\mathbb{C}$ may generate complex phases in the scalar potential; these phases could be basis-dependent (but not always, as in the CP4 model~\cite{Ivanov:2015mwl,Ferreira:2017tvy,Ivanov:2018ime}) and therefore unphysical. Some can be removed by unitary redefinitions of the scalar fields. In particular, if only a single complex phase is present, it can always be rotated away, yielding a real scalar potential without loss of generality.

\subsection[\texorpdfstring{$O(2) \times U(1)$}{O(2) x U(1)}]{\boldmath$O(2) \times U(1)$}

 Let us consider the general real potential, $V_\mathbb{R}$, further constrained by requiring invariance under a specific $\mathbb{Z}_4$ representation (one can verify this by checking for the eigenvalues):
\begin{equation}
g_1 = \left(
\begin{array}{ccc}
 \frac{i}{\sqrt{3}} & \sqrt{\frac{2}{3}} e^{-\frac{5 i \pi}{8}} & 0 \\
 \sqrt{\frac{2}{3}} e^{-\frac{3 i \pi }{8}} & -\frac{i}{\sqrt{3}} & 0 \\
 0 & 0 & 1 \\
\end{array}
\right)_\mathrm{HF}.
\end{equation}
The scalar potential invariant under this transformation is given by
\begin{equation}
\begin{aligned}
V ={}& \mu^2_{11}(h_{11} + h_{22}) + \mu_{33}^2 h_{33} \\
& + \lambda_{1111} (h_{11}^2 + h_{22}^2) + \lambda_{3333} h_{33}^2 + \lambda_{1221} h_{12}h_{21} + \lambda_{1331} (h_{13}h_{31} + h_{23}h_{32}) \\
& + \lambda_{1122} h_{11}h_{22} + \lambda_{1133} (h_{11} + h_{22}) h_{33} + \lambda_{1212} (h_{12}^2 + h_{21}^2) \\
& +  \lambda_{1112} (h_{12} + h_{21})(h_{22} - h_{11}),
\end{aligned}
\end{equation}
with couplings related by
\begin{subequations}
\begin{align}
\lambda_{1221} ={}& 2 \lambda_{1111} + \sqrt{2 + \sqrt{2}} \lambda_{1112} - \lambda_{1122},\\
\lambda_{1212} ={}& - \lambda_{1112} \sin (\pi/8).
\end{align}
\end{subequations}

We note that by imposing this particular $\mathbb{Z}_4$ symmetry on the most general complex potential would yield a $\mathbb{Z}_4$-symmetric one, retaining complex phases. However, imposing this symmetry on $V_\mathbb{R}$, the $\mathbb{Z}_4$ is no longer realisable, we end up with an enlarged symmetry.

The scalar potential is also invariant under 
\begin{align}
g_2 ={} & \begin{pmatrix}
0 & 1 & 0 \\
-1 & 0 & 0 \\
0 & 0 & 1
\end{pmatrix}_\mathrm{HF}.
\end{align}

The group $\mathcal{G} = \left\langle g_1,\,g_2 \right\rangle$ satisfies: (i) $g_1^4 = g_2^4 = e$, (ii) $g_1^2 = g_2^2$ and (iii) $g_2^{-1} g_1^2 g_2 = g_1^2$. From the second relation we identify $\left\langle g_1 \right\rangle \cap \left\langle g_2 \right\rangle = \left\langle e,\,g_1^2 \right\rangle \cong \mathbb{Z}_2$, and the third relation shows that this $\mathbb{Z}_2$ is central. Moreover, there exists no integer $n$ such that $( g_1 g_2)^n=e$. This indicates that $\mathcal{G}$ is an example of a finitely generated infinite group.

In the basis of $\mathcal{G} = \left\langle g_1 g_2,\, g_1 \right\rangle$, the group under consideration is of the form $\mathcal{G} = \mathbb{Z} \rtimes \mathbb{Z}_2$. Actually, we are considering a non-trivial central extension of the infinite dihedral group by $\mathbb{Z}_2$, \textit{i.e.}, a double-cover of $D_\infty$, an infinite dicyclic $Dic_\infty$ group. Modding out the common central subgroup we get $Dic_\infty / \mathbb{Z}_2 \cong D_\infty$.

Actually, the scalar potential associated with this symmetry is also invariant under the $U(1)_{h_3}$ group. Consequently, $D_\infty$ is effectively extended by this continuous symmetry, resulting in an enlarged group structure---$O(2) \times U(1)$.

After performing a basis transformation
\begin{equation}
\begin{pmatrix}
h_1 \\
h_2 \\
h_3
\end{pmatrix} = \begin{pmatrix}
e^\frac{3 i \pi}{4} \cos \theta & e^\frac{i \pi}{4} \sin \theta & 0 \\
e^\frac{3 i \pi}{4} \sin \theta & -e^\frac{i \pi}{4} \cos \theta & 0 \\
0 & 0 & 1
\end{pmatrix} \begin{pmatrix}
h_1^\prime \\
h_2^\prime \\
h_3^\prime
\end{pmatrix},
\end{equation}
where 
\begin{equation}
\theta = \frac{1}{4} \left[ \arctan\left( 2 \sqrt{2 + \sqrt{2}} \right)  - \pi \right],
\end{equation}
we arrive at the $O(2) \times U(1)$ invariant potential:
\begin{equation}
\begin{aligned}
V={}& \mu^2_{11}(h_{11} + h_{22})  + \mu^2_{33} h_{33} \\
& + \lambda_{1111} (h_{11}^2 + h_{22}^2) + \lambda_{3333} h_{33}^2 + \lambda_{1221} h_{12}h_{21} + \lambda_{1331} (h_{13}h_{31} + h_{23}h_{32}) \\
& + \lambda_{1122} h_{11}h_{22} + \lambda_{1133} (h_{11} + h_{22}) h_{33}.
\end{aligned}
\end{equation}

Equivalently, we can define $O(2) \times U(1)$ in terms of the generators
\begin{equation}
\left\langle \begin{pmatrix}
0 & e^{i \theta_1} & 0 \\
e^{i \theta_2} & 0 & 0 \\
0 & 0 & 1
\end{pmatrix},\, \begin{pmatrix}
e^{i \theta_1} & 0 & 0 \\
0 & e^{i \theta_2} & 0 \\
0 & 0 & 1
\end{pmatrix} \right\rangle.
\end{equation}
To be precise we have :
\begin{equation}
\left[ U(1) \times U(1) \right] \rtimes \mathbb{Z}_2 \cong U(1)_\mathrm{center} \times \left[ SO(2) \rtimes \mathbb{Z}_2 \right] \cong U(1) \times O(2),
\end{equation}
where $ U(1)_\mathrm{center}$ is invariant under the swap generated by $\mathbb{Z}_2$.

\subsection[The \texorpdfstring{$S_3$}{S3}-based cases ]{The \boldmath$S_3$-based cases}\label{Sec:G2}

Consider a generator representing even permutations (in terms of $\mathcal{U}$),
\begin{equation}\label{Eq:G2_g1}
g_1 = \begin{pmatrix}
0 & 1 & 0\\
0 & 0 & 1\\
1 & 0 & 0
\end{pmatrix}_\ast,
\end{equation}
applied as a GCP transformation. We get $g_1^3=e$, but the full group is of dimension six, since we need to consider $(U^\mathrm{GCP})^{2n} = \mathrm{diag}\left( \mathcal{U} \mathcal{U}^\ast, (\mathcal{U} \mathcal{U}^\ast)^\ast \right)^n = \mathcal{I}$, for $n=3$. Applying $\mathcal{U} = g_1$ as a GCP transformation yields a potential with real couplings. In contrast, requiring invariance under the action of $\left\langle g_1 \right\rangle_\mathrm{HF}$ results in the $\mathbb{Z}_3$-symmetric 3HDM.

The $S_3$-symmetric (we recall that $S_3 \cong \mathbb{Z}_3 \rtimes \mathbb{Z}_2$) scalar potential in the reducible-triplet framework~\cite{Derman:1978rx,Derman:1979nf} can be presented as ($\phi_{ij} = \phi_i^\dagger \phi_j$)
{\interdisplaylinepenalty=1000
\begin{align}\label{Eq:S3_Derman_redtr}
V ={}& -\lambda  \sum_i \phi_{ii} + \frac{1}{2} \gamma \sum_{i<j}  \left( \phi_{ij} + \mathrm{h.c.}\right) \\
{}&  +A \sum_i \phi_{ii}^2+\sum_{i<j} \left[ C \phi_{ii}  \phi_{jj}  + \bar{C}\phi_{ij}\phi_{ji} + \frac{1}{2}D\left( \phi_{ij}^2 + \mathrm{h.c.} \right)
\right] +\frac{1}{2}\sum_{i\neq j} \left(   E_1 \phi_{ii}  \phi_{ij}  + \mathrm{h.c.}\right) \nonumber\\
&\quad + \frac{1}{2}\sum_{i \neq j \neq k \neq i ,j<k} \left(  E_2 \phi_{ij} \phi_{ki} +  E_3 \phi_{ii}  \phi_{kj} + E_4  \phi_{ij} \phi_{ik} + \mathrm{h.c.} 
\right).\nonumber
\end{align}}
Due to the underlying $S_3$ symmetry not all of the coefficients can be complex. The two complex couplings are $E_1$ and $E_4$; one degree of freedom is redundant~\cite{Kuncinas:2023ycz}, for example, either $E_1$ or $E_4$ can be made real.

Since we are only interested in the even permutations applied as a GCP transformation, and not the full $S_3$ (in the current context we are considering $\left\langle g_1 \right\rangle_\mathrm{HF} \rtimes \mathbb{Z}_2^\ast$), it is to be expected that some additional terms may be allowed. Namely, the
\begin{equation*}
\frac{1}{2}\sum_{i\neq j} \left(   E_1 \phi_{ii}  \phi_{ij}  + \mathrm{h.c.}\right)
\end{equation*}
term gets split into two parts (before we had $E_1 = E_1' = E_1''$),
\begin{equation}
\frac{1}{2}\sum_i \left(   E_1' \phi_{ii}  \phi_{i i'}  +  E_1'' \phi_{i'i'} \phi_{i i'}   + \mathrm{h.c.}\right),
\end{equation}
where $i'=(i\mod 3)+1$. There are no other modifications with respect to the scalar potential of eq.~\eqref{Eq:S3_Derman_redtr}. 

By going into the basis of the irreducible representation of $S_3$ and further on performing an additional basis transformation,
\begin{equation}
\begin{aligned}
\begin{pmatrix}
\phi_1 \\
\phi_2 \\
\phi_3
\end{pmatrix} =&{} \left(
\begin{array}{ccc}
 \frac{1}{\sqrt{2}} & \frac{1}{\sqrt{6}} & \frac{1}{\sqrt{3}} \\
 -\frac{1}{\sqrt{2}} & \frac{1}{\sqrt{6}} & \frac{1}{\sqrt{3}} \\
 0 & -\sqrt{\frac{2}{3}} & \frac{1}{\sqrt{3}} \\
\end{array}
\right)  \left(
\begin{array}{ccc}
 \frac{1}{\sqrt{2}} & \frac{1}{\sqrt{2}} & 0 \\
 \frac{i}{\sqrt{2}} & -\frac{i}{\sqrt{2}} & 0 \\
 0 & 0 & 1 \\
\end{array}
\right) \begin{pmatrix}
h_1 \\
h_2 \\
h_3
\end{pmatrix}\\
={}& 
\frac{1}{\sqrt{3}} \left(
\begin{array}{ccc}
 e^{i \pi / 6} & e^{-i \pi / 6} & 1 \\
 -e^{-i \pi / 6} &  -e^{i \pi / 6} & 1 \\
 -i & i & 1
\end{array}
\right) \begin{pmatrix}
h_1 \\
h_2 \\
h_3
\end{pmatrix},
\end{aligned}
\end{equation}
the scalar potential gets modified,
\begin{equation}\label{Eq:V_S3pr_N}
\begin{aligned}
V ={}& \mu^2_{11}(h_{11} + h_{22})  + \mu^2_{33} h_{33} \\
& + \lambda_{1111} (h_{11}^2 + h_{22}^2) + \lambda_{3333} h_{33}^2 + \lambda_{1221} h_{12}h_{21} + \lambda_{1331} (h_{13}h_{31} + h_{23}h_{32}) \\
& + \lambda_{1122} h_{11}h_{22} + \lambda_{1133} (h_{11} + h_{22}) h_{33} \\
& +  \lambda_{1323} (h_{13}h_{23} + h_{31}h_{32}) + \left\lbrace \lambda_{1213} h_{12}(h_{13} + h_{32}) + \mathrm{h.c.} \right\rbrace.
\end{aligned}
\end{equation}
Next, one can rotate the $h_1$ and $h_2$ doublets by opposite phases (chosen to leave the $\lambda_{1323}$ term invariant), with the rotation angle proportional to the argument of $\lambda_{1213}$. In this way, the $\lambda_{1213}$ term becomes real.

In the $S_3$-symmetric potential the last line of eq.~\eqref{Eq:V_S3pr_N} would read:
\begin{equation}\label{Eq:S3_pm}
\left\lbrace \lambda_{1323} h_{13}h_{23} +  \lambda_{1213}(h_{12}h_{13} \pm h_{21}h_{23}) + \mathrm{h.c.} \right\rbrace,
\end{equation}
where one of the phases can be rotated away. The origin of the ``$\pm$" sign between the two terms in $\lambda_{1213}$ is discussed in Appendix~\ref{App:S3_sign}.

A more direct derivation of the scalar potential in eq.~\eqref{Eq:V_S3pr_N} is achieved by imposing invariance under the single GCP generator:
\begin{equation}
g_2 = \begin{pmatrix}
0 & e^{-\frac{2 i \pi}{3}} & 0 \\
e^\frac{2 i \pi}{3} & 0 & 0 \\
0 & 0 & 1
\end{pmatrix}_\ast,
\end{equation}
which spans a symmetry group of dimension six. Within the full $6\times6$ $H$ space, the powers of $\left\langle g_{2} \right\rangle_\ast$ generate: (i) $S_2$ acting on $\{h_1,\, h_2\}$ as an HF transformations, (ii) a diagonal $\mathbb{Z}_3$ as an HF transformation and (iii) a trivial GCP transformation. Consequently, $\left\langle g_{2} \right\rangle_\ast$ encapsulates $S_3$ (same elements) realised at the levels of $U^\mathrm{HF}$ and $U^\mathrm{GCP}$, and hence the full symmetry group of the model is $S_3 \times \mathbb{Z}_2^\ast$~\cite{Ivanov:2012fp,Ivanov:2014doa}.

In Ref.~\cite{Bree:2024edl} the scalar potential in question was referred to as CPc ($S_3 \times \mathrm{GCP}_{\theta = \pi}$). However, it was presented with redundant couplings. A basis change from the scalar potential of eq.~(35) of Ref.~\cite{Bree:2024edl} to the one of eq.~\eqref{Eq:V_S3pr_N} in this work can be performed via a basis transformation $\Phi = \mathcal{R}_{(\frac{\pi}{4},\, 0,\, \frac{\pi}{4},\, \frac{\pi}{2})}h$. 

\subsection{CP4}

Let us consider two different symmetry transformations, corresponding to CP4:
\begin{subequations}
\begin{align}
g_1 = \begin{pmatrix}
0 & -i & 0 \\
i & 0  & 0 \\
0 & 0 & 1
\end{pmatrix}_\ast, \label{Eq:CP4_gen_pm_i}\\
g_2 = \begin{pmatrix}
0 & -1 & 0 \\
1 & 0  & 0 \\
0 & 0 & 1
\end{pmatrix}_\ast \label{Eq:CP4_gen_pm_1}.
\end{align}
\end{subequations}

While $g_1^2 = g_2^4 = e$, one should keep in mind that we are dealing with the GCP transformations. When the generator is represented by a complex matrix, the presence of complex conjugation effectively doubles the action of the generator. In this sense, a complex generator of order $n$ in the unitary subgroup corresponds to a GCP generator of order $2n$. Consequently, the generators $g_1$ and $g_2$ are different representations of $\mathbb{Z}_4^\ast$.

The same potential in different bases is given by:
\begin{subequations}\label{Eq:V_CP4}
\begin{align}
\begin{split}
V_{g_1} ={}& \mu^2_{11}(h_{11} + h_{22})  + \mu^2_{33} h_{33} \\
& + \lambda_{1111} (h_{11}^2 + h_{22}^2) + \lambda_{3333} h_{33}^2 + \lambda_{1221} h_{12}h_{21} + \lambda_{1331} (h_{13}h_{31} + h_{23}h_{32}) \\
& + \lambda_{1122} h_{11}h_{22} + \lambda_{1133} (h_{11} + h_{22}) h_{33} + \lambda_{1323} (h_{13}h_{23} + h_{31}h_{32})\\
& + \left\lbrace \lambda_{1212} h_{12}^2 + \lambda_{1313}(h_{13}^2-h_{32}^2)  - \lambda_{1112} h_{12}(h_{11} - h_{22}) + \mathrm{h.c.}\right\rbrace,\\
\end{split}\\
\begin{split}
V_{g_2} ={}& \mu^2_{11}(h_{11} + h_{22})  + \mu^2_{33} h_{33} \\
& + \lambda_{1111} (h_{11}^2 + h_{22}^2) + \lambda_{3333} h_{33}^2 + \lambda_{1221} h_{12}h_{21} + \lambda_{1331} (h_{13}h_{31} + h_{23}h_{32}) \\
& + \lambda_{1122} h_{11}h_{22} + \lambda_{1133} (h_{11} + h_{22}) h_{33} + i\lambda_{1323} (h_{13}h_{23} - h_{31}h_{32})\\
& + \left\lbrace \lambda_{1212} h_{12}^2 + \lambda_{1313}(h_{13}^2+h_{32}^2)  - \lambda_{1112} h_{12}(h_{11} - h_{22}) + \mathrm{h.c.}\right\rbrace.\\
\end{split}
\end{align}
\end{subequations}

This structure was identified in Refs.~\cite{Ivanov:2011ae,Ivanov:2014doa} and later dubbed CP4, since $(g_n g_n^\ast)^2=e$, in Ref.~\cite{Ivanov:2015mwl}. The defining characteristic of the CP4 3HDM is that, despite the model being explicitly CP-conserving, there is no basis in which all the potential coefficients are real.

One might notice that all possible 2HDM terms (for $h_1$ and $h_2$) are included in the above scalar potentials $V_{g_1}$ and $V_{g_2}$, which, in turn, indicates the existence of redundant couplings; in the 2HDM one can utilise the $SU(2)$ freedom to re-define 14 parameters to 11 physical ones. Different approaches to simplifying the CP4 scalar potentials were followed in Refs.~\cite{Ferreira:2017tvy,Haber:2018iwr}. There is actually also a $\mathbb{Z}_2$ HF transformation, which can be expressed as $h_3\mapsto -h_3$ in the above bases.

In Ref.~\cite{Ivanov:2018ime}, see eq.~(A1), the CP4-invariant 3HDM was described by a one-parameter family of generators---different bases of the same scalar potential being related by conjugation of a single matrix,
\begin{equation}\label{Eq:CP4_param_g}
g_3 = \begin{pmatrix}
0 & 1 & 0 \\
-1 & 0 & 0 \\
0 & 0 & e^{i \beta}
\end{pmatrix}_\ast,
\end{equation}
for any fixed value of the $\beta$ phase. Then, some phases of the quartic couplings will be fixed in terms of the $\beta$ value. We suggest an alternative parameterisation given by a family of generators:
\begin{equation}
g_4 = \begin{pmatrix}
0 & 1 & 0 \\
-1 & 0  & 0 \\
0 & 0 & e^{2 i \pi / N}
\end{pmatrix}_\ast,\text{ for } N \in \mathbb{Z} \backslash \{0\}.
\end{equation}
which one arrives at naturally, considering an $N^\mathrm{th}$ root of unity.

Finally, we want to state that we did not succeed in finding a basis, following our approach, in which there would be no redundant couplings present.

\subsection[The \texorpdfstring{$D_4$}{D4}-based cases ]{The \boldmath$D_4$-based cases}\label{Sec:D4_GCP}

Now, suppose we extended $\mathbb{Z}_4^\ast$ by an additional $\mathbb{Z}_2^\ast$. Let us consider invariance under
\begin{equation}\label{Eq:GCP_CP4_Z2_Z2}
\left\langle
\begin{pmatrix}
0 & -i & 0\\
i & 0 & 0\\
0 & 0 & 1
\end{pmatrix},\,
\begin{pmatrix}
-1 & 0 & 0\\
0 & -1 & 0\\
0 & 0 & 1
\end{pmatrix} \right\rangle_\ast.
\end{equation}
In this case the scalar potential is given by:
\begin{gather}
\begin{aligned}
V ={}& \mu^2_{11}(h_{11} + h_{22})  + \mu^2_{33} h_{33} \\
& + \lambda_{1111} (h_{11}^2 + h_{22}^2) + \lambda_{3333} h_{33}^2 + \lambda_{1221} h_{12}h_{21} + \lambda_{1331} (h_{13}h_{31} + h_{23}h_{32}) \\
& + \lambda_{1122} h_{11}h_{22} + \lambda_{1133} (h_{11} + h_{22}) h_{33} + \lambda_{1212} (h_{12}^2 + h_{21}^2) \\
& + \lambda_{1112} (h_{12} + h_{21})(h_{22} - h_{11}) + \lambda_{1313} (h_{13}^2 - h_{23}^2 + \mathrm{h.c.})+\lambda_{1323} (h_{13}h_{23} + \mathrm{h.c.} ).
\end{aligned}
\label{Eq:V_GCP_CP4_Z2_Z2}
\raisetag{40pt}
\end{gather}
It can be viewed as applying two consecutive GCP transformations: CP2 and CP4; see Refs.~\cite{Haber:2018iwr,Ivanov:2018ime} for a discussion. 

After performing a basis rotation of $\mathcal{R}_{(\frac{\pi}{4},\, \pi,\, -\frac{\pi}{4},\, \frac{\pi}{2})}$ the scalar potential becomes:
\begin{equation}\label{Eq:V_Z2_Z2_GCP_1}
\begin{aligned}
V ={}& \mu^2_{11}(h_{11} + h_{22})  + \mu^2_{33} h_{33} \\
& + \lambda_{1111} (h_{11}^2 + h_{22}^2) + \lambda_{3333} h_{33}^2 + \lambda_{1221} h_{12}h_{21} + \lambda_{1331} (h_{13}h_{31} + h_{23}h_{32}) \\
& + \lambda_{1122} h_{11}h_{22} + \lambda_{1133} (h_{11} + h_{22}) h_{33}\\
& + \left\lbrace \lambda_{1212} h_{12}^2 +  \lambda_{1313} (h_{13}^2 + h_{32}^2) + \mathrm{h.c.} \right\rbrace.
\end{aligned}
\end{equation}
In this form four real quartic couplings were exchanged for two complex couplings. Without loss of generality one of these complex phases can be rotated away. Equivalently, we could have performed the $\mathcal{R}_{(\alpha,\, 0,\, \theta,\, 0)}$ basis transformation starting with the scalar potential of eq.~\eqref{Eq:V_GCP_CP4_Z2_Z2}. Another aspect to note is that the scalar potential resembles the $D_4$-symmetric one of eq.~\eqref{Eq:D4_g1}, apart from the fact that the terms multiplied by $\lambda_{1313}$ are grouped in another way, as $(h_{13}^2 + h_{32}^2)$ rather than $(h_{13}^2 + h_{23}^2)$ in the $D_4$-symmetric potential.

The scalar potential of eq.~\eqref{Eq:V_Z2_Z2_GCP_1} is invariant under
\begin{equation}
(\mathbb{Z}_2 \times \mathbb{Z}_2) \rtimes \mathbb{Z}_2^\ast = \left\langle \begin{pmatrix}
-1 & 0 & 0\\
0 & 1 & 0\\
0 & 0 & 1
\end{pmatrix}_\mathrm{HF},\, 
\begin{pmatrix}
1 & 0 & 0\\
0 & -1 & 0\\
0 & 0 & 1
\end{pmatrix}_\mathrm{HF},\, 
\begin{pmatrix}
0 & 1 & 0\\
1 & 0 & 0\\
0 & 0 & 1
\end{pmatrix}_\ast \right\rangle,
\end{equation}
or equivalently under
\begin{equation}
(\mathbb{Z}_2 \times \mathbb{Z}_2) \rtimes \mathbb{Z}_2^\ast = \left\langle \begin{pmatrix}
0 & -i & 0\\
i & 0 & 0\\
0 & 0 & 1
\end{pmatrix}_\mathrm{HF},\, 
\begin{pmatrix}
0 & i & 0\\
-i & 0 & 0\\
0 & 0 & 1
\end{pmatrix}_\mathrm{HF},\, 
\begin{pmatrix}
1 & 0 & 0\\
0 & 1 & 0\\
0 & 0 & 1
\end{pmatrix}_\ast \right\rangle.
\end{equation}
In both cases the $D_4$ group is spanned. For $D_4$ there exists a special isomorphism \mbox{$(\mathbb{Z}_2 \times \mathbb{Z}_2) \rtimes \mathbb{Z}_2 \cong D_4$}, while in general we have $D_n \cong \mathbb{Z}_n \rtimes \mathbb{Z}_2$. One can apply
\begin{equation}
D_4 = \left\langle \begin{pmatrix}
0 & -1 & 0\\
1 & 0 & 0\\
0 & 0 & 1
\end{pmatrix},\, 
\begin{pmatrix}
0 & 1 & 0\\
1 & 0 & 0\\
0 & 0 & 1
\end{pmatrix}\right\rangle_\ast,
\end{equation}
as a GCP transformation to arrive at the potential of eq.~\eqref{Eq:V_Z2_Z2_GCP_1}. 

At this point we would like to stress that there exists a $V_4 \times \mathbb{Z}_2^\ast$ symmetric model,
\begin{equation}
\begin{aligned}
V_{V_4} ={}& \sum_i \mu_{ii}^2 h_{ii}\\
& + \sum_i \lambda_{iiii} h_{ii}^2 + \sum_{i<j} \lambda_{iijj} h_{ii} h_{jj} + \sum_{i<j} \lambda_{ijji} h_{ij} h_{ji} + \sum_{i<j} \left\lbrace \lambda_{ijij} h_{ij}^2 + \mathrm{h.c.}\right\rbrace,
\end{aligned} 
\end{equation}
with $\lambda_{ijij} \in \mathbb{R}$, as required by the invariance under $\mathbb{Z}_2^\ast$. In the bilinear formalism the pattern of the eigenvalues for $V_4 \times \mathbb{Z}_2^\ast$ is $8_1$, while for $V_4 \rtimes \mathbb{Z}_2^\ast$ it is $2_2 4_1$.

Our initial assumption was to consider invariance under the symmetry transformations of eq.~\eqref{Eq:GCP_CP4_Z2_Z2}, applied as GCP transformations, as a starting point. One may wonder what is the structure behind the GCP transformations
\begin{equation}
\left\langle
\begin{pmatrix}
0 & -1 & 0\\
1 & 0 & 0\\
0 & 0 & 1
\end{pmatrix},\,
\begin{pmatrix}
-1 & 0 & 0\\
0 & -1 & 0\\
0 & 0 & 1
\end{pmatrix} \right\rangle_\ast.
\end{equation}
Equivalently it can be presented as $\mathbb{Z}_4 \times \mathbb{Z}_2^\ast$. In this case the scalar potential is given by:
\begin{equation}\label{Eq:D4_Z2cp}
\begin{aligned}
V ={}& \mu^2_{11}(h_{11} + h_{22})  + \mu^2_{33} h_{33} \\
& + \lambda_{1111} (h_{11}^2 + h_{22}^2) + \lambda_{3333} h_{33}^2 + \lambda_{1221} h_{12}h_{21} + \lambda_{1331} (h_{13}h_{31} + h_{23}h_{32}) \\
& + \lambda_{1122} h_{11}h_{22} + \lambda_{1133} (h_{11} + h_{22}) h_{33} + \lambda_{1212} (h_{12}^2 + h_{21}^2) \\
& + \lambda_{1112} (h_{12} + h_{21})(h_{22} - h_{11}) + \lambda_{1313} (h_{13}^2 + h_{23}^2 + \mathrm{h.c.}).
\end{aligned}
\end{equation}
After performing a basis change $\mathcal{R}_{( \frac{\pi}{4},\, 0,\, \frac{\pi}{4},\, \frac{\pi}{2} )}$ we end up with a different representation of $D_4$, where the potential in the new basis looks like the potential of eq.~\eqref{Eq:D4_g2}, where now $\lambda_{1212}^\prime \in \mathbb{C}$ and $\lambda_{1323}^\prime \in \mathbb{R}$. A basis change can then eliminate the complex phase.

While this section may initially seem somewhat redundant or like an unnecessary exercise, what we have actually learned is that requiring the invariance of the scalar potential under different representations of the same symmetry group can lead to redundant couplings in one basis, while not in another. Furthermore, in some cases, by considering the same set of elements applied as HF or GCP transformations, we can relatively easily conclude whether there is explicit CP violation, provided we can recognise symmetries in a basis-independent way~\cite{deMedeirosVarzielas:2019rrp}.

\subsection[The \texorpdfstring{$U(1) \circ V_4$}{U(1) ○ V4}-based case]{The \boldmath$U(1) \circ V_4$-based case}

Forcing invariance under
\begin{equation}
[U(1) \times \mathbb{Z}_2] \rtimes \mathbb{Z}_4^\ast = \left\langle \begin{pmatrix}
e^{i \theta} & 0 & 0\\
0 & e^{i \theta} & 0\\
0 & 0 & 1
\end{pmatrix}_\mathrm{HF},\,
\begin{pmatrix}
-1 & 0 & 0\\
0 & 1 & 0\\
0 & 0 & 1
\end{pmatrix}_\mathrm{HF},\,
\begin{pmatrix}
0 & -1 & 0\\
1 & 0 & 0\\
0 & 0 & 1
\end{pmatrix}_\ast \right\rangle.
\end{equation}
results in the scalar potential
\begin{equation}
\begin{aligned}\label{Eq:V_l1212_C}
V ={}& \mu^2_{11}(h_{11} + h_{22})  + \mu^2_{33} h_{33} \\
& + \lambda_{1111} (h_{11}^2 + h_{22}^2) + \lambda_{3333} h_{33}^2 + \lambda_{1221} h_{12}h_{21} + \lambda_{1331} (h_{13}h_{31} + h_{23}h_{32}) \\
& + \lambda_{1122} h_{11}h_{22} + \lambda_{1133} (h_{11} + h_{22}) h_{33} + \{\lambda_{1212} h_{12}^2 + \mathrm{h.c.}\}.
\end{aligned}
\end{equation}

Requiring invariance under
\begin{equation}
 [U(1) \times \mathbb{Z}_4]  \rtimes \mathbb{Z}_2^\ast  = \left\langle \begin{pmatrix}
e^{i \theta} & 0 & 0\\
0 & e^{i \theta} & 0\\
0 & 0 & 1
\end{pmatrix}_\mathrm{HF},\,
\begin{pmatrix}
0 & -1 & 0\\
1 & 0 & 0\\
0 & 0 & 1
\end{pmatrix}_\mathrm{HF},\,
\begin{pmatrix}
1 & 0 & 0\\
0 & 1 & 0\\
0 & 0 & 1
\end{pmatrix}_\ast \right\rangle,
\end{equation}
yields
\begin{equation}\label{Eq:U1_V4_CP}
\begin{aligned}
V ={}& \mu^2_{11}(h_{11} + h_{22})  + \mu^2_{33} h_{33} \\
& + \lambda_{1111} (h_{11}^2 + h_{22}^2) + \lambda_{3333} h_{33}^2 + \lambda_{1221} h_{12}h_{21} + \lambda_{1331} (h_{13}h_{31} + h_{23}h_{32}) \\
& + \lambda_{1122} h_{11}h_{22} + \lambda_{1133} (h_{11} + h_{22}) h_{33} + \lambda_{1212} (h_{12}^2 + h_{21}^2) \\
& +  \lambda_{1112} (h_{12} + h_{21})(h_{22} - h_{11}).
\end{aligned}
\end{equation}

Applying the basis transformation $\mathcal{R}_{(\frac{\pi}{4},\, -\frac{\pi}{4},\, \frac{\pi}{4},\, 0)}$ to the scalar potential given by eq.~\eqref{Eq:U1_V4_CP} one recovers the form of eq.~\eqref{Eq:V_l1212_C}.

Provided the semi-direct product actions are chosen appropriately, we have $([U(1) \times \mathbb{Z}_2] \rtimes \mathbb{Z}_4^\ast)/ \mathbb{Z}_2^\mathrm{cent} \cong ([U(1) \times \mathbb{Z}_4]  \rtimes \mathbb{Z}_2^\ast)/ \mathbb{Z}_2^\mathrm{cent}$, where $ U(1) \cap\mathbb{Z}_4 \cong \mathbb{Z}_2^\mathrm{cent}$ and the central factor group is $\mathbb{Z}_2^\mathrm{cent} = \left\langle \mathrm{diag}(-1,-1,1) \right\rangle_\mathrm{HF}$. Equivalently, the symmetry can be written as
\begin{equation}
([U(1)\times D_4]/\mathbb Z_2)\rtimes\mathbb Z_2^\ast \cong [U(1)\times V_4]\rtimes\mathbb Z_2^\ast.
\end{equation}

The scalar potential in eq.~\eqref{Eq:V_l1212_C} closely resembles the $U(1)\circ V_4$-symmetric potential of eq.~\eqref{Eq:V_U1_D4}; the only formal difference is that $\lambda_{1212}\in\mathbb C$ in eq.~\eqref{Eq:V_l1212_C}. Since $\lambda_{1212}$ is the sole coupling sensitive to the relative phase between $h_1$ and $h_2$, its phase can be removed by a relative rephasing of $h_1$ and $h_2$. Hence $\arg(\lambda_{1212})$ is a basis artefact rather than an independent physical parameter, and the GCP extension does not produce a physically distinct scalar potential: after removing redundant parameters we recover the potential of $U(1)\circ V_4$.

In the GCP realisation leading to eq.~\eqref{Eq:V_l1212_C}, the cohomological choices available in the HF classification of $U(1)\circ V_4$ are fixed and the structure is realised through a specific finite lift $\mathcal G=D_4$ (\textit{cf.} eq.~\eqref{Eq:Def_tilde_G_true}). By rotating away the redundant phase (setting $\lambda_{1212} \in \mathbb{R}$) one can extend the underlying symmetry to admit an additional GCP transformation $\mathbb{Z}_2^\ast$, so that the scalar potential simultaneously realises both HF and GCP symmetries.

In the discussed setup, CP acts as a non-trivial automorphism of the HF symmetry:
\begin{equation}
(\mathrm{CP})~ \mathrm{diag}(e^{i \theta},\,e^{i \theta},\,1)~(\mathrm{CP})^{-1} = \mathrm{diag}(e^{-i \theta},\,e^{-i \theta},\,1).
\end{equation}
Accordingly, the full symmetry is the projective semi-direct product
\begin{equation}
[U(1)\times V_4]\rtimes_\varphi \mathbb Z_2^\ast \qquad \varphi:
\begin{cases}
U(\theta) \mapsto U(-\theta),\\
g \mapsto g,\quad \forall g\in V_4 .
\end{cases}
\end{equation}

By minimising the scalar potential we must satisfy eq.~\eqref{Eq:Min_U1_D4}. We see that when neither $v_1$ nor $v_2$ vanishes, we must require $ v_2 = \pm  v_1$, or else the symmetry of the scalar potential gets increased to $O(2) \times U(1)$, since this equation relates quartic couplings. There is the possibility of having a relative phase between the vacuum expectation values (vevs) of $h_1$ and $h_2$ which cannot be rotated away while leaving the potential invariant. Therefore, CP can be spontaneously violated.

Let us analyse the scalar potential from a slightly different point of view. Suppose that we promote the parameterised generator of eq.~\eqref{Eq:CP4_param_g}, which led us to the CP4 model, to a continuous one,
\begin{equation}\label{Eq:U1_Z2_comp}
g_1 = \begin{pmatrix}
0 & -e^{i \theta} & 0\\
e^{i \theta} & 0 & 0\\
0 & 0 & 1
\end{pmatrix}.
\end{equation}
From the HF transformation, $\left\langle g_1\right\rangle_\mathrm{HF}$,  point of view we are looking at the $U(1) \times \mathbb{Z}_2$-symmetric scalar potential in a representation where it turns out that some of the couplings must be purely imaginary, $\{\lambda_{1112},\, \lambda_{1332},\, \lambda_{1233}\} \in i \mathbb{R}$. The couplings of the potential can be re-arranged by a basis transformation of $\mathcal{R}_{(\frac{\pi}{4},\, 0,\, \frac{\pi}{4},\,0)}$, allowing all to be real.

Invariance under the GCP transformation $\left\langle g_1 \right\rangle_\ast$ yields the following scalar potential:
\begin{equation}\label{Eq:V_U1_Z2st_8C}
\begin{aligned}
V ={}& \mu^2_{11}(h_{11} + h_{22})  + \mu^2_{33} h_{33} \\
& + \lambda_{1111} (h_{11}^2 + h_{22}^2) + \lambda_{3333} h_{33}^2 + \lambda_{1221} h_{12}h_{21} + \lambda_{1331} (h_{13}h_{31} + h_{23}h_{32}) \\
& + \lambda_{1122} h_{11}h_{22} + \lambda_{1133} (h_{11} + h_{22}) h_{33}  \\
& + \left\lbrace \lambda_{1212} h_{12}^2  + \lambda_{1112} h_{12} (h_{22} - h_{11}) + \mathrm{h.c.} \right\rbrace.
\end{aligned}
\end{equation}

While the scalar potential is not globally $U(2)$ invariant, the $U(2)$ transformations acting on the $\{h_1,h_2\}$ subspace do not generate additional independent monomials, rather they only mix coefficients. This basis freedom allows us to perform a rotation $\mathcal R_{(\alpha,0,\theta,0)}$ that removes either $\lambda_{1212}$ or $\lambda_{1112}$. Choosing $\lambda_{1112}$ as the redundant coupling yields the scalar potential of eq.~\eqref{Eq:V_l1212_C}. Starting from $\left\langle g_1 \right\rangle_\ast$, we are considering \mbox{$U(1) \times \mathrm{CP4}$}. However, CP4 is defined up to a fixed phase, see eq.~\eqref{Eq:CP4_param_g}. Therefore, it is more appropriate to describe this case by $[U(1) \times \mathrm{CP4}]/ \mathbb{Z}_2^\mathrm{cent}$. Modulo the central phases, dropping the factor in eq.~\eqref{Eq:CP4_param_g}, elements of the set realise a central extension of $V_4$ by a finite cyclic subgroup of $U(1)$.

\subsection[\texorpdfstring{$O(2)$}{O(2)}]{\boldmath$O(2)$}

In Ref.~\cite{Bree:2024edl} a GCP symmetry generated by 
\begin{equation}
g_1 = \begin{pmatrix}
\cos \theta & \sin \theta & 0\\
-\sin \theta & \cos \theta & 0\\
0 & 0 & 1
\end{pmatrix}_\ast.
\end{equation}
was discussed and named CPd or $O(2) \times \mathrm{CP}$. Requiring invariance under $g_1$ results in:
\begin{equation}\label{Eq:V_SO2_Z2}
\begin{aligned}
V ={}& \mu^2_{11}(h_{11} + h_{22})  + \mu^2_{33} h_{33} \\
& + \lambda_{1111} (h_{11}^2 + h_{22}^2) + \lambda_{3333} h_{33}^2 + \lambda_{1221} h_{12}h_{21} + \lambda_{1331} (h_{13}h_{31} + h_{23}h_{32}) \\
& + \lambda_{1122} h_{11}h_{22} + \lambda_{1133} (h_{11} + h_{22}) h_{33}\\
& + \Lambda(h_{12}^2 + h_{21}^2) + \lambda_{1313} (h_{13}^2 + h_{23}^2 + \mathrm{h.c.}),
\end{aligned}
\end{equation}
where
\begin{equation}
\Lambda = \frac{1}{2} (2\lambda_{1111} - \lambda_{1221} - \lambda_{1122}).
\end{equation}

If a rotation is implemented as multiplication by a complex phase on each complex field (the usual $U(1)$ re-phasing), we get another representation of $O(2)$, given by $U(1) \rtimes \mathbb{Z}_2$.

In this case the scalar potential takes the form
\begin{equation}\label{Eq:V_U1_Z2}
\begin{aligned}
V ={}& \mu^2_{11}(h_{11} + h_{22})  + \mu^2_{33} h_{33} \\
& + \lambda_{1111} (h_{11}^2 + h_{22}^2) + \lambda_{3333} h_{33}^2 + \lambda_{1221} h_{12}h_{21} + \lambda_{1331} (h_{13}h_{31} + h_{23}h_{32}) \\
& + \lambda_{1122} h_{11}h_{22} + \lambda_{1133} (h_{11} + h_{22}) h_{33}\\
& + \lambda_{1323} (h_{13}h_{23} + h_{31}h_{32}).
\end{aligned}
\end{equation}
The two representations of $O(2)$ are connected via a basis change of $\mathcal{R}_{(0,\, -\frac{\pi}{4},\, \frac{\pi}{4},\, \frac{\pi}{2})}$.

The scalar potentials presented in eq.~\eqref{Eq:V_SO2_Z2} and eq.~\eqref{Eq:V_U1_Z2} are  invariant under $O(2) \cong SO(2) \rtimes \mathbb{Z}_2 \cong U(1) \rtimes \mathbb{Z}_2$. We end up with a real potential, which coincides with what we would have obtained, had we started from $V_\mathbb{R}$. In fact, the full symmetry can be expressed as $O(2) \times \mathbb{Z}_2^\ast$. This notation is consistent with the choice of Ref.~\cite{Bree:2024edl} $O(2) \times \mathrm{CP}$.
 
However, requiring invariance under the HF transformation of $O(2)$, results in a scalar potential with a single phase-sensitive coupling, $\lambda_{1313} \in \mathbb{C}$ in the case of eq.~\eqref{Eq:V_SO2_Z2} or $\lambda_{1323} \in \mathbb{C}$ in the case of eq.~\eqref{Eq:V_U1_Z2}. This phase can be rotated away. Therefore, the HF $O(2)$ symmetry is automatically extended to the $O(2) \times \mathbb{Z}_2^\ast$ one.

\subsection[\texorpdfstring{$\Delta(54)/ \mathbb{Z}_3$}{Δ(54)/Z3}]{\boldmath$\Delta(54)/ \mathbb{Z}_3$}\label{Sec:Delta54_Z3}

The final case we want to discuss is $\Delta(54) \cong (\mathbb{Z}_3 \times \mathbb{Z}_3) \rtimes S_3$; strictly speaking, in the context of 3HDMs it is more appropriate to refer to it as $\Delta(54) / \mathbb{Z}_3$ due to its non-trivial intersection with $U(1)_Y$. Our aim is to discuss the constraints of $\Delta(54)$ on the $V_\mathbb{R}$ space. While this might at first appear to be of little relevance from our point of view, the structure of the eigenvalues in the bilinear space actually differs from that of the $\Delta(54)/\mathbb{Z}_3$-symmetric potential when applied to $V_\mathbb{C}$. We assume that the phases of the complex couplings are free parameters.

The scalar potential
\begin{equation}\label{Eq:V_Delta_54}
\begin{aligned}
V ={}& \mu_{11}^2 h_{ii}\\
&+ \lambda_{1111} h_{ii}^2 + \lambda_{1221} h_{ij}h_{ji} + \lambda_{1122} h_{ii}h_{jj}\\
& + \left\lbrace \lambda_{1323} (h_{13}h_{23} + h_{21}h_{31} + h_{32}h_{12}) + \mathrm{h.c.} \right\rbrace,
\end{aligned}
\end{equation}
can be generated by an (minimal) HF transformation
\begin{equation}\label{Eq:Z3_rt_Z3}
\mathbb{Z}_3 \rtimes \mathbb{Z}_3 = \left\langle  
\begin{pmatrix}
1 & 0 & 0 \\
0 & \omega & 0\\
0 & 0 & \omega^2
\end{pmatrix},\,
\begin{pmatrix}
0 & 1 & 0 \\
0 & 0 & 1 \\
1 & 0 & 0
\end{pmatrix}
\right\rangle_\mathrm{HF},
\end{equation}
where $\omega = e^{2i \pi/3}$.

Requiring invariance under $\mathbb{Z}_3 \rtimes \mathbb{Z}_3$, the pattern of eigenvalues in the bilinear formalism is $4_2$. By promoting the latter generator to the GCP transformation we are led to $\mathbb{Z}_3 \rtimes \mathbb{Z}_6^\ast$, where $\mathbb{Z}_6^\ast$ can be presented as $\mathbb{Z}_3 \times \mathbb{Z}_2^\ast$, but the form is consistent with eq.~\eqref{Eq:Z3_rt_Z3}. This case amounts to setting $\lambda_{1323} \in \mathbb{R}$, and the pattern of eigenvalues becomes $1_4 2_2$, see Ref.~\cite{deMedeirosVarzielas:2019rrp}. 

In both cases there is an additional $\mathbb{Z}_2$ symmetry realised as $h_i \leftrightarrow h_j$. Therefore, the structure is automatically extended to the dihedral-like group \mbox{$\Delta(6n^2) \cong (\mathbb{Z}_n \times \mathbb{Z}_n) \rtimes S_3$} for $n=3$; see Ref.~\cite{Fairbairn:1964sga}. There is a non-trivial intersection of the center given by $Z(\Delta(54)) = \left\langle \omega\,\mathcal{I} \right\rangle = \mathbb{Z}_3 \leq U(1)_Y$. Consequently, in the language of NHDMs, we are dealing with $\Delta(54)/\mathbb{Z}_3$. For explicit formulations in the context of 3HDMs, see Refs.~\cite{Ivanov:2012fp,Ivanov:2014doa}.

We note that $\Delta(54)/\mathbb{Z}_3$ corresponds to the $\mathsf{SmallGrp}$ identifier $[54,13]/\mathbb{Z}_3 \cong [18,4] $. There are several ways of presenting this group: $\mathbb{Z}_3 \rtimes S_3$ or the generalised dihedral group formed from $\mathbb{Z}_3 \times \mathbb{Z}_3$, $D(E_9) \cong E_9 \rtimes \mathbb{Z}_2$. 

Next, we arrive at the real scalar potential of eq.~\eqref{Eq:V_Delta_54} by requiring invariance under:
\begin{equation} \label{Eq:Delta_246x3}
\left\langle  \begin{pmatrix}
1 & 0 & 0 \\
0 & \omega^{-1} & 0 \\
0 & 0 & 1
\end{pmatrix},\, \begin{pmatrix}
0 & \omega & 0 \\
0 & 0 & \omega \\
1 & 0 & 0
\end{pmatrix},\, \frac{i}{\sqrt{3}} \begin{pmatrix}
1 & 1 & 1 \\
1 & \omega & \omega^2 \\
1 & \omega^2 & \omega
\end{pmatrix} \right\rangle_\ast.
\end{equation}

The first two generators yield the $S_3$-symmetric scalar potential, while imposing all three transformations leads to the $\Delta(54)/\mathbb{Z}_3$ applied to $V_\mathbb{R}$; see Refs.~\cite{Kuncinas:2023ycz, Doring:2024kdg} for the path from $S_3$ to $\Delta(54)$. The underlying structure can be presented as $(\Delta(54)/\mathbb{Z}_3) \rtimes \mathbb{Z}_2^\ast$, and it corresponds to the $\mathsf{SmallGrp}$ identifier $[36,13]$. From the group-theoretical point of view the quotient group can be presented as $\mathbb{Z}_2 \times \mathbb{Z}_3 \rtimes S_3$.

\section{Summary of HF and GCP transformations}\label{Sec:GT_GCP_summary}

Let us here summarise the cases discussed so far. Table~\ref{Table:Indep_Copuplings_Cases} lists the realisable symmetry groups together with the numbers of independent bilinear and quartic couplings (complex couplings are counted as a single entry), while Table~\ref{Table:Bilinear_Patterns_Cases} shows the associated bilinear eigenvalue patterns. The GCP extensions of the HF symmetries take the form $G\rtimes\mathbb{Z}_2^\ast$ or $G\times\mathbb{Z}_2^\ast$, for example $S_3\times\mathbb{Z}_2^\ast$ or $\mathbb{Z}_3\rtimes\mathbb{Z}_2^\ast$. The GCP  cases not being listed in the tables require some explanations.

In the GCP columns we only list the non-trivial realisations, \textit{i.e.}, we exclude cases where the application of the HF symmetry to the most general complex potential already leads to a real scalar potential. An example is the $\mathbb{Z}_4$ symmetry, which corresponds to a real potential, therefore this potential is already $\mathbb{Z}_4 \rtimes \mathbb{Z}_2^\ast$ symmetric. Apart from that, we exclude cases where the GCP transformation amounts to setting the complex couplings of the potential to real values; the only exception to this rule in our listing is the appearance of the $(\Delta(54)/ \mathbb{Z}_3) \rtimes \mathbb{Z}_2^\ast$ case in Table~\ref{Table:Indep_Copuplings_Cases}, since in the bilinear space $\Delta(54)/ \mathbb{Z}_3$ and $(\Delta(54)/ \mathbb{Z}_3) \rtimes \mathbb{Z}_2^\ast$ do not share the same pattern, see Table~\ref{Table:Bilinear_Patterns_Cases}. 
 
In addition to the HF realisations of the GCP transformations we include the CP2 and CP4 cases, respectively $\mathbb{Z}_2^\ast$ and $\mathbb{Z}_4^\ast$.

Table~\ref{Table:Indep_R_Copuplings_Cases} supplements Table~\ref{Table:Indep_Copuplings_Cases} by listing upper bounds on the number of independent real parameters for each symmetry. These account for the residual continuous basis freedom that leaves the symmetry manifest. By fixing the basis to the manifest form, we identify a reduced set of parameters sufficient to define the physics of that symmetry. For instance, the $\mathbb Z_2$-symmetric potential retains its form under basis transformations belonging to \mbox{$(U(2)\times U(1))/U(1)_Y$}, implying four redundant real parameters. As further reductions may occur upon imposing stationary-point equations, the values quoted represent upper bounds. See Appendix~\ref{App:Redundancies} for more details.

{{\renewcommand{\arraystretch}{1.05}
\begin{table}[H]
\caption{Realisable symmetries in 3HDMs along with the number of couplings. The first column lists the total number of distinct bilinear and quartic coupling constants, where each complex coupling is counted as a single entry. The $U(1) \circ V_4$ entry (projectively $U(1) \times V_4$) is given as a quotient group; the full structure is provided in eq.~\eqref{Eq:Def_tilde_G_true}.}
\label{Table:Indep_Copuplings_Cases}
\begin{center}
\begin{tabular}{|c|c|c|c|} \hline\hline
\begin{tabular}[l]{@{}c@{}@{}} Independent \\couplings\end{tabular} & HF discrete & HF continuous & GCP \\ \hline \hline
1 + 2 &  & $SU(3)$ &\\ \hline
1 + 3 & $\Sigma (36)$ & $SO(3)$, $\left[ U(1) \times U(1)\right] \rtimes S_3$ & \\ \hline 
1 + 4 &  $S_4$, $A_4$, $\Delta(54)/ \mathbb{Z}_3$ &   & $(\Delta(54)/ \mathbb{Z}_3) \rtimes \mathbb{Z}_2^\ast$\\ \hline
2 + 5 &  & $U(2)$  & \\ \hline
2 + 6 &  & $O(2) \times U(1)$  &  \\ \hline
2 + 7 &  & $U(1) \circ V_4$, $O(2)$  &   \\ \hline
2 + 8 & $S_3$, $D_4$ &  & $(\mathbb{Z}_2 \times \mathbb{Z}_2) \rtimes \mathbb{Z}_2^\ast$ \\ \hline
2 + 9 &  &  & $\mathbb{Z}_4^\ast$ (CP4)  \\ \hline
3 + 9 &  & $U(1) \times U(1)$  &  \\ \hline
3 +10 &  & $U(1) \times \mathbb{Z}_2$, $U(1)_1$  &  \\ \hline
3 +11 & $\mathbb{Z}_4$ &   &  \\ \hline
3 +12 & $\mathbb{Z}_2 \times \mathbb{Z}_2$, $\mathbb{Z}_3$ &   &  \\ \hline
4 +14 &  & $U(1)_2$  &  \\ \hline
4 +17 & $\mathbb{Z}_2$ &   &  \\ \hline
6 +27 &  &   & $\mathbb{Z}_2^\ast$ (CP2) \\ \hline
\end{tabular}\vspace*{-8pt}
\end{center}
\end{table}}

{{\renewcommand{\arraystretch}{1.05}
\begin{table}[H]
\caption{Same symmetries as in Table~\ref{Table:Indep_Copuplings_Cases}, shown with corresponding eigenvalue patterns ($v_m$ indicates $v$ eigenvalues, each with multiplicity $m$) in the bilinear formalism.}
\label{Table:Bilinear_Patterns_Cases}
\begin{center}
\begin{tabular}{|c|c|c|c|} \hline\hline
\begin{tabular}[l]{@{}c@{}@{}} Patterns of\\ eigenvalues\end{tabular} & HF discrete & HF continuous & GCP  \\ \hline \hline
$1_8$ &   & $SU(3)$  &  \\ \hline
$1_6 1_2$ &   & $[U(1) \times U(1)] \rtimes S_3$  &  \\ \hline
$1_5 1_3$ &   & $SO(3)$  &  \\ \hline
$2_4$ & $\Sigma (36)$  &   &  \\ \hline
$1_4 1_3 1_1$ &   & $U(2)$  &  \\ \hline
$1_4 2_2$ &   &   & $(\Delta(54)/ \mathbb{Z}_3) \rtimes \mathbb{Z}_2^\ast$ \\ \hline
$1_4 1_2 2_1$ &   & $O(2) \times U(1)$  &  \\ \hline
$1_4 4_1$ &   & $U(1) \circ V_4$  &  \\ \hline
$2_3 1_2$ & $S_4$, $A_4$  & &    \\ \hline
$4_2$ & $\Delta(54)/ \mathbb{Z}_3$ & & \\ \hline
$3_2 2_1$ &  $S_3$, $\mathbb{Z}_3$ & $O(2)$, $U(1) \times U(1)$, $U(1)_1$  &   \\ \hline
$2_2 4_1$ & $D_4$, $\mathbb{Z}_4$  & $U(1) \times \mathbb{Z}_2$, $U(1)_2$  & $\mathbb{Z}_4^\ast$ (CP4), $(\mathbb{Z}_2 \times \mathbb{Z}_2) \rtimes \mathbb{Z}_2^\ast$  \\ \hline
$8_1$ & $\mathbb{Z}_2 \times \mathbb{Z}_2$, $\mathbb{Z}_2$ &  & $\mathbb{Z}_2^\ast$ (CP2) \\ \hline
\end{tabular}\vspace*{-8pt}
\end{center}
\end{table}}

\renewcommand{\arraystretch}{1.05}
\begin{table}[H]
\caption{Complementary to Table~\ref{Table:Indep_Copuplings_Cases}, giving upper bounds on the number $N$ of independent parameters  for each symmetry. See discussion in Appendix~\ref{App:Redundancies}, specifically Table~\ref{Table:Redundancies}.}
\label{Table:Indep_R_Copuplings_Cases}
\begin{center}
\begin{tabular}{|c|c|c|c|} \hline\hline
\begin{tabular}[l]{@{}c@{}@{}} $~N \leq~$ \end{tabular} & HF discrete & HF continuous & GCP \\ \hline \hline
3  &  & $SU(3)$ & \\ \hline
4  & $\Sigma(36)$ & $SO(3)$, $\big[U(1)\times U(1)\big]\rtimes S_3$ & \\ \hline
5  & $S_4$ &  & $(\Delta(54)/\mathbb{Z}_3)\rtimes \mathbb{Z}_2^\ast$ \\ \hline
6  & $A_4$, $\Delta(54)/\mathbb{Z}_3$ &  & \\ \hline
7  &  & $U(2)$ & \\ \hline
8  &  & $O(2)\times U(1)$ & \\ \hline
9  &  & $O(2)$, $U(1)\circ V_4$ & \\ \hline
10 & $D_4$ &  & $S_3 \times \mathbb{Z}_2^\ast$ \\ \hline
11 & $S_3$ &  & $(\mathbb{Z}_2\times\mathbb{Z}_2)\rtimes \mathbb{Z}_2^\ast$ \\ \hline
12 &  & $U(1)\times U(1)$ & \\ \hline
13 &  & $U(1)_1$, $U(1)\times\mathbb{Z}_2$ &  $\mathbb{Z}_4^\ast$ (CP4) \\ \hline
14 & $\mathbb{Z}_4$ &  & \\ \hline
15 &  &  & $\mathbb{Z}_2\times\mathbb{Z}_2 \times \mathbb{Z}_2^\ast$, $\mathbb{Z}_3 \rtimes \mathbb{Z}_2^\ast$ \\ \hline
16 & $\mathbb{Z}_2\times\mathbb{Z}_2$, $\mathbb{Z}_3$ &  & \\ \hline
17 &  & & $U(1)_2 \rtimes \mathbb{Z}_2^\ast$ \\ \hline
20 &  & & $\mathbb{Z}_2 \times \mathbb{Z}_2^\ast$\\ \hline
21 &  & $U(1)_2$ & \\ \hline
26 & $\mathbb{Z}_2$ &  & \\ \hline
30 &  &  & $\mathbb{Z}_2^\ast$ (CP2) \\ \hline
46 & 3HDM &  &  \\ \hline
\end{tabular}\vspace*{-8pt}
\end{center}
\end{table}

To our knowledge, Tables~\ref{Table:Indep_Copuplings_Cases} and \ref{Table:Bilinear_Patterns_Cases} together provide the most complete overview to date of HF and GCP symmetries admissible in 3HDMs, they can be used as a practical guide when classifying specific models. A robust, basis-independent diagnostic is to compare the bilinear eigenvalue pattern of a given potential with the patterns listed in Table~\ref{Table:Bilinear_Patterns_Cases}. This comparison is often more reliable than attempting to match complex couplings across different bases, since in some bases a single complex parameter may appear as two independent real parameters and thus obscure identification. Nevertheless, bilinear patterns are not unique identifiers: distinct symmetry classes can share the same pattern, and additional checks are therefore usually required (see Refs.~\cite{Ivanov:2018ime,deMedeirosVarzielas:2019rrp} for examples and methods).

Regarding the eigenvalue patterns in Table~\ref{Table:Bilinear_Patterns_Cases}, it is noteworthy that several naively admissible multiplicity patterns do not occur in the bilinear space for the standard HF and GCP implementations: $1_7 1_1$, $1_6 2_1$, $1_5 1_2 1_1$, $1_5 3_1$, $2_3 2_1$, $1_3 2_2 1_1$, $1_3 1_2 3_1$, $1_3 5_1$, and $1_2 6_1$. At first sight, one might suspect that such patterns can never arise; however, some of these patterns do arise in non-standard (see Appendix~\ref{App:Gen_TGOOFy}) or constrained constructions. For instance, in Ref.~\cite{Kuncinas:2024zjq} minimising the $O(2)$-symmetric potential with a vacuum of the form $(\hat v_1 e^{i\sigma},\,\hat v_2,\,0)$ implies \mbox{$2 \lambda_{1111} - \lambda_{1221} - \lambda_{1122} = 0$}, and the resulting bilinear eigenvalue pattern is $1_3 2_2 1_1$. We were unable to relate this algebraic constraint to a symmetry enhancement, but the example shows that eigenvalue patterns alone do not exhaust all structural possibilities of the scalar potential. This observation raises the question of whether there may exist alternative or non-standard ways of implementing symmetries, beyond the conventional HF and GCP transformations considered so far.

At this stage we find it useful to collect some group theory concepts, which are summarised in the following subsection.

\subsection{Some useful identities and definitions}

\paragraph{Abelian groups}

\begin{itemize}
  \item $\mathbb{Z}_n$---cyclic group of order $n$;
  \item $\mathbb{Z}_2\times\mathbb{Z}_2 \cong V_4$---Klein four-group;
  \item $U(1)\times U(1) \cong T^2$---the two-torus.
\end{itemize}

Many of the symmetry groups relevant to our discussion admit a representation as semidirect products. In particular, we have
\begin{equation}
\mathcal{G} = \mathcal{N} \rtimes_\phi \mathcal{H},\quad \mathcal{N} \triangleleft \mathcal{G},~ \phi:\,\mathcal{H} \rightarrow \mathrm{Aut}(\mathcal{N}),
\end{equation}
with $\mathcal{H}$ acting non-trivially on the normal subgroup $\mathcal{N}$. Note that, the definition of the semidirect group depends on the homomorphism $\phi$.  Below we list the observed structures that are relevant for 3HDMs.

\paragraph{Dihedral groups\\}

In general, we have:
\begin{equation}
D_n \cong \mathbb{Z}_n \rtimes \mathbb{Z}_2.
\end{equation}
Special cases that occur in 3HDMs include:
\begin{equation}
S_3 \cong D_3 \cong \mathbb{Z}_3\rtimes\mathbb{Z}_2,\qquad
D_4 \cong \mathbb{Z}_4\rtimes\mathbb{Z}_2 \cong V_4\rtimes\mathbb{Z}_2.
\end{equation}

\paragraph{Symmetric and alternating groups\\}

The symmetric group is a semidirect product of the alternating group $A_n$ (the kernel of the sign homomorphism) and $\mathbb{Z}_2$, generated by any odd permutation:
\begin{equation}
S_n \cong A_n \rtimes \mathbb{Z}_2, \quad \text{for } n \ge 2.
\end{equation}
Relevant specific decompositions for 3HDMs are:
\begin{equation}
S_3 \cong \mathbb{Z}_3\rtimes\mathbb{Z}_2,\qquad
A_4 \cong V_4\rtimes\mathbb{Z}_3,\qquad
S_4 \cong V_4\rtimes S_3.
\end{equation}

\paragraph{Continuous groups\\}

For these groups we have the following isomorphisms and extensions:
\begin{itemize}
  \item $SO(2) \cong U(1)$;  
  \item $O(2)\cong SO(2)\rtimes\mathbb{Z}_2 \cong U(1)\rtimes\mathbb{Z}_2$, where $\mathbb{Z}_2$ acts on $U(1)$ via complex conjugation;  \item $SO(3)\cong SU(2)/\mathbb{Z}_2$;
  \item $U(n)\cong (SU(n)\times U(1))/\mathbb{Z}_n$ (in particular for $n=2,3$).
\end{itemize}

\paragraph{Special non-Abelian discrete groups\\}

There are two distinct $\Delta$-series given by
\begin{subequations}
\begin{align}
\Delta(3n^2) &\cong (\mathbb{Z}_n\times\mathbb{Z}_n)\rtimes \mathbb{Z}_3,\\
\Delta(6n^2) &\cong (\mathbb{Z}_n\times\mathbb{Z}_n)\rtimes S_3,
\end{align}
\end{subequations}
with particular cases
\begin{subequations}
\begin{align}
&\Delta(12)=\Delta(3\times 2^2)\cong A_4,\\
\Delta(6)=\Delta(6\times 1^2)\cong S_3,\qquad
&\Delta(24)=\Delta(6\times 2^2)\cong S_4,\qquad
\Delta(54)=\Delta(6\times 3^2).
\end{align}
\end{subequations}

A further special symmetry in the classification of finite subgroups of $PSU(3)$ belongs to the so-called exceptional series, with the smallest member
\begin{equation}
\Sigma(36) \cong (\mathbb{Z}_3 \times \mathbb{Z}_3)\rtimes \mathbb{Z}_4.
\end{equation}
In $SU(3)$ this group appears with an additional central $\mathbb{Z}_3$ factor, denoted $\Sigma(36\times 3)$,
\begin{equation}\label{Eq:Sigma_36_isom}
\Sigma(36)_{PSU(3)} \cong \Sigma(36\times 3)_{SU(3)} \big/ \mathbb{Z}_3.
\end{equation}

\section{GOOFy-like transformations of 3HDMs}\label{Sec:GOOFY_3HDMs}

Having discussed the conventional HF and GCP transformations, we now turn to a distinct class of field-space operations known as GOOFy transformations. These act on scalar doublets and their conjugates through non-standard transformations that can flip relative signs between fields and invert the sign of kinetic-like terms. Unlike ordinary HF or GCP symmetries, GOOFy transformations do not preserve the canonical kinetic structure, yet they can enforce renormalisation-group (RG) stable relations among parameters of the potential and motivate auxiliary-field limits when promoted to exact symmetries. 

However, at the one-loop level, a conceptual tension exists regarding the validity of the model. While Ref.~\cite{Ferreira:2025ate} argues the $r_0$ invariance is preserved if an extended imaginary scaling acts simultaneously on the fields, spacetime coordinates, momenta, and crucially, the renormalisation scale itself, Ref.~\cite{Pilaftsis:2024uub} demonstrates that within a standard physical framework using dimensional regularisation, finite threshold corrections explicitly violate the invariance. Because inevitable threshold effects spoil the relations when matching scales are properly fixed, we interpret GOOFy operations strictly as algebraic transformations rather than exact dynamical symmetries, viewing them as ``fingerprints'' indicative of a broader, complexified theory.

This interpretation naturally aligns with recent theoretical frameworks developed to classify true RG invariants in the 2HDM. Both the outer automorphism formalism~\cite{deBoer:2026ktb} and the spurion-field approach~\cite{Pilaftsis:2026wyw} outline techniques for understanding how the relations driven by applying GOOFy transformations can technically survive. To maintain an exact RG-invariant surface, the parameter space must simply lack the specific loop interactions needed to regenerate the forbidden operators. While the restricted geometry of the 2HDM accidentally accommodates an isolated $r_0$ parameter parity, these frameworks imply that a standalone parity will be inherently insufficient to protect the potential in expanded parameter spaces.

This limitation becomes immediately apparent in the 3HDM. In the 2HDM limit, the $SU(2)$ algebra lacks a fully symmetric group tensor ($d_{abc} = 0$). This algebraic accident prevents loop diagrams from mixing the mass parameters, artificially protecting the $r_0$ transformation. In contrast, the $SU(3)$ algebra has a non-zero $d_{abc}$ tensor. This, in turn, introduces radiative feedback channels. Consequently, an isolated $r_0$ condition is radiatively destroyed in the generic 3HDM. The present work thus serves as a systematic, group-theoretical exercise to identify models where GOOFy transformations remain useful as organising principles or tree-level limits.

\subsection{The GOOFy framework}

A new kind of global symmetry transformation, not corresponding to any regular symmetry, was recently proposed by Ferreira, Grzadkowski, Ogreid and Osland~\cite{Ferreira:2023dke} for the scalar sector of 2HDMs. It accounts for a region of scalar parameters that is RG invariant to all orders when only scalar and gauge interactions are considered; this might be also expanded to the Yukawa sector. These transformations differ from those previously known by allowing the scalar doublets to transform independently of their Hermitian conjugates. Such transformations have the paradoxical property of leading to RG-invariant fixed points without keeping the kinetic term $(\partial_\mu h_i)^\dagger (\partial^\mu h_i)$ invariant.

In the example of Ref.~\cite{Ferreira:2023dke}, the RG-invariant conditions are
\begin{equation}
m_{11}^2 + m_{22}^2 = 0, \qquad \lambda_1 = \lambda_2, \qquad \lambda_6 = - \lambda_7,
\end{equation}
written in terms of the most common notation for 2HDMs (and not in our notation), and is reminiscent of the CP2-invariant 2HDM. The transformation of scalar fields leading to this property was found to be:
\begin{equation}\label{Eq:r0_2HDM}
\begin{aligned}
& h_1 \mapsto - h_2^*,  \quad &  h_1^\dagger \mapsto  h_2^{\mathrm{T}}, \\
& h_2 \mapsto  h_1^*,         &  h_2^\dagger \mapsto  h_1^{\mathrm{T}}.
\end{aligned}
\end{equation}

This transformation inverts the sign of the kinetic Lagrangian terms, since the fields $h_i$ and their conjugates transform non-trivially under these operations. The scalar covariant derivatives are defined as
\begin{equation}
D^\mu = \partial^\mu + i\frac{g}{2} \sigma_i W_i^\mu + i\frac{g^\prime}{2} B^\mu,
\end{equation}
and invariance of the gauge–interaction terms requires that the gauge fields transform as
\begin{equation}\label{Eq:W_and_B_trans}
W_{1\mu} \rightarrow i W_{1\mu}, \qquad 
W_{2\mu} \rightarrow - i W_{2\mu}, \qquad
W_{3\mu} \rightarrow i W_{3\mu}, \qquad 
B_\mu \rightarrow i B_\mu.
\end{equation}
Clearly, the term $(\partial_\mu h_i)^\dagger (\partial^\mu h_i)$ changes sign and the covariant kinetic terms are therefore not invariant under these transformations. If the transformation of eq.~\eqref{Eq:r0_2HDM} is promoted to an exact symmetry of the theory, the ``inverted'' (scalar) terms must be excluded from the Lagrangian, causing the associated scalar field to lose its kinetic term and become auxiliary.

These kinds of transformations have by now attracted considerable attention (see Refs.~\cite{Haber:2025cbb,Trautner:2025yxz,Ferreira:2025ate} as well as Refs.~\cite{deBoer:2025jhc,Trautner:2025prm,Grzadkowski:2026gkx,Ferreira:2026ccg} for the implications in the Yukawa sector) and are referred to as \emph{GOOFy transformations}. One important implication of these transformations is that they can render certain bilinear terms of the scalar potential odd under the symmetry. Initially, GOOFy transformations were applied as GCP-type transformations, but they were later generalised by Trautner in Ref.~\cite{Trautner:2025yxz} to include HF transformations. The defining criterion for a transformation to be GOOFy became the non-trivial transformation of the kinetic terms.

With the generalisation proposed in Ref.~\cite{Trautner:2025yxz} (we shall refer to these transformations as T-GOOFy, after Trautner, in Section~\ref{Sec:TGOOFY_3HDMs}}, where more details shall be outlined) situations may arise in which some kinetic terms of the form $(\partial_\mu h_i)^\dagger (\partial^\mu h_i)$ remain invariant, while others change sign. This implies that invariance cannot, in general, be imposed simultaneously for all quartic couplings of the fields appearing in the kinetic term, since the transformations given by eq.~\eqref{Eq:W_and_B_trans} only work for doublets that transform in a GOOFy way. In such cases, some of the kinetic terms provide a hard breaking of the GOOFy symmetry.

It was argued in Ref.~\cite{Trautner:2025yxz} that for NHDMs one should expect non-vanishing RG corrections to fixed points at higher orders whenever there is a relative-sign-flipping behaviour of the gauge kinetic terms. In contrast, for cases with global-sign-flipping GOOFy transformations, one should expect all-order stability of the parameter relations resulting from imposing the symmetry. This occurs even though, in the latter case, the kinetic terms are not fully invariant under the GOOFy symmetry. A full proof is expected to appear in a future publication. If one chooses to remove the kinetic terms of the doublets that break the GOOFy transformations, the corresponding scalars will behave as auxiliary fields.

It is interesting to note that the GOOFy transformations are reminiscent of other frameworks, though they are not directly equivalent to any of them.

On one hand, special GOOFy cases when all bilinear terms vanish, align with the Gildener-Weinberg class of models~\cite{Gildener:1976ih}. In these multi-field generalisations of the Coleman-Weinberg mechanism~\cite{Coleman:1973jx}, the tree-level potential is assumed to be classically scale invariant; masses are generated radiatively, while the kinetic terms remain canonical. The scale-invariant 2HDM scenario has been studied in several works, see, for example, Refs.~\cite{Lee:2012jn,Hashino:2015nxa,Lane:2019dbc,Eichten:2022vys}.

On the other hand, the kinetic sign-flip property of the GOOFy transformation mirrors Lee-Wick constructions~\cite{Lee:1969fy,Lee:1970iw} and their modern Lee-Wick Standard Model extensions~\cite{Grinstein:2007mp,Carone:2008iw,Carone:2009nu,Espinosa:2011js,Johansen:2015vyh}. In Lee–Wick theories, scalar modes with opposite-sign kinetic terms arise dynamically from higher-order derivative operators or auxiliary regulator fields to improve ultraviolet behaviour. However, this comes at the cost of potential ghost-like instabilities, requiring special prescriptions or decoupling arguments to preserve unitarity and causality.

\subsection[Transformations of the \texorpdfstring{$SU(2)$}{SU(2)} bilinear singlets]{Transformations of the \boldmath$SU(2)$ bilinear singlets}

In Ref.~\cite{Kuncinas:2024zjq} symmetry-breaking patterns for 3HDMs with continuous symmetries were discussed. There, 2HDMs embedded inside 3HDMs were mentioned. One may therefore envisage a construction consisting of a 2HDM sector accompanied by a separate Standard Model-like scalar sector, without tree-level mixing between the two sectors. A necessary condition for such patterns to emerge is that a specific transformation acts differently on the bilinear and quartic terms of the scalar potential, since we do not want to eliminate the bilinear terms of each independent sector.

A technical aspect must be addressed at this point. We can always eliminate some quadratic couplings by applying symmetries. However, it is not possible to obtain full decoupling between the two sectors, since quartic portals $h_{ii} h_{jj}$ and $|h_{ij}|^2$ are gauge invariant and are allowed to be built for any invariant quadratic couplings. If not present, those would be generated radiatively.

Let us consider an HF-like transformation acting on the $SU(2)$ bilinear singlets. Here we are considering the quadratic terms as independent of each other. We denote this transformation by \(\overline{\mathbb Z}_2\), defined as
\begin{subequations}
\begin{align}
h_{ii} &{}\mapsto - h_{ii},\\
h_{ij} &{}\mapsto - h_{ij},~i \neq j.
\end{align}
\end{subequations}
Under this transformation all $SU(2)$ bilinear singlets are odd. Therefore, the bilinear couplings, $\mu_{ii}^2$ and $\mu_{ij}^2$, are not allowed, whereas all quartic couplings are allowed. Consequently, imposing the $\overline{\mathbb{Z}}_2$ transformation removes the full bilinear sector of the scalar potential, while leaving the quartic structure intact. In this sense, the resulting theory may be viewed as a 3HDM realisation of the Gildener–Weinberg framework. The transformation enforcing $\mu_{ii}^2=\mu_{ij}^2=0$ may thus be interpreted as acting analogously to a global scale transformation on the scalar potential.

Alternatively, we could also think of $\overline{\mathbb{Z}}_2$ acting as a GCP-like transformation, $\overline{\mathbb{Z}}^\ast_2$, under which the $SU(2)$ bilinear singlets transform as
\begin{subequations}
\begin{align}
h_{ii} &{}\mapsto - h_{ii},\\
h_{ij} &{}\mapsto - h_{ji},~i \neq j,
\end{align}
\end{subequations}
which corresponds, except for the sign, to a CP transformation given by $h_{ij} \mapsto h_{ji}$.

In this case, the bilinear part of the potential is given by:
\begin{equation}
V = i \mu_{ij}^2 (h_{ij} - h_{ji}),
\end{equation}
with purely imaginary $i\mu_{ij}^2$ couplings. In contrast, since we can imagine that we are looking at the identity generator acting as a GCP transformation and the quartic potential is even under $\overline{\mathbb{Z}}_{2}^\ast$, we get real couplings, $\lambda_{ijkl} \in \mathbb{R}$.

What about different $\overline{\mathbb{Z}}_{n}$ symmetries? For example, we may consider a $\overline{\mathbb{Z}}_{3}$ transformation acting as
\begin{equation}
h_{ij} \mapsto e^{2 i \pi/3} h_{ij}.
\end{equation}
Since the renormalisable scalar potential contains $SU(2)$ bilinear singlets of multiplicities one and two (bilinear $\mu_{ij}^2$ and quartic $\lambda_{ijkl}$ terms), all terms vanish under $\overline{\mathbb{Z}}_{3}$. However, cubic $SU(2)$ bilinear singlet structures, $h_{ij}h_{kl}h_{mn}$, remain invariant under both $\overline{\mathbb{Z}}_{3}$ and $\overline{\mathbb{Z}}_{3}^\ast$. This observation can be easily extended to any $\overline{\mathbb{Z}}_{n}$ or $\overline{\mathbb{Z}}_{n}^\ast$ transformation.

Up to this point, we have assumed that all $h_{ij}$ singlets transform identically. As an example of a non-uniform transformation, consider the following:
\begin{subequations}\label{Eq:hiip_hijm}
\begin{align}
h_{ii} &{}\mapsto h_{ii},\\
h_{ij} &{}\mapsto - h_{ij},~i \neq j.
\end{align}
\end{subequations}
Invariance results in the bilinear part
\begin{equation}
V = \mu_{ii}^2 \,h_{ii},
\end{equation}
whereas in the quartic potential the $\lambda_{iijk}$ couplings are prohibited. In another basis the quartic potential can be associated with the $\mathbb{Z}_2$-symmetric model. On the other hand, if we switch the signs of the above transformation, \textit{i.e.}, $h_{ii} \mapsto -h_{ii},\, h_{ij} \mapsto h_{ij}$, the bilinear part will be different, while the quartic part remains $\mathbb{Z}_2$-symmetric.

A comprehensive analysis of all possible transformations of this type would be a computationally demanding task since we would be looking at the transformations of the $3^2$ $SU(2)$ bilinear singlets for the 3HDM. In general, we do not need to assume that $h_{ij}$ and $h_{ji}$ transform as conjugates; for instance, if $h_{13} \to i h_{21}$, it is possible to have $h_{31} \to i h_{12}$, with both transformations involving a common factor of ``$i$". Moreover, it is possible to have transformation given by $h_{13} \to h_{23} \to - h_{13}$. These do not violate Hermiticity of the scalar potential.

This naturally raises the question: can these identifications be simplified in line with the methodology developed in the previous sections?

For a scalar doublet $h_i$ transforming under $\mathcal{U}$ as $h_i \mapsto \mathcal{U}_{ij} h_j$, the transformation of $h_{ij}$ will not generally match the standard HF or GCP transformation of eqs.~\eqref{Eq:HF_tr} and \eqref{Eq:GCP_tr} if both $h_i$ and $h_i^\ast$ transform under the same $\mathcal{U}$. In such cases, two different unitary transformations, $\mathcal{U}^{(2)} \neq \left( \mathcal{U}^{(1)} \right)^\ast$, may be considered, as briefly noted but not elaborated upon in Ref.~\cite{Doring:2024kdg}:
\begin{itemize}
\item HF-like transformations,
\begin{equation}\label{Eq:HF_GOOFy_tr}
H \mapsto \begin{pmatrix}
\mathcal{U}^{(1)} & 0 \\
0 & \mathcal{U}^{(2)}
\end{pmatrix} H.
\end{equation}
\item GCP-like transformations,
\begin{equation}\label{Eq:GCP_GOOFy_tr}
H \mapsto \begin{pmatrix}
0 & \mathcal{U}^{(1)}\\
\mathcal{U}^{(2)} & 0
\end{pmatrix} H.
\end{equation}
\end{itemize}
Importantly, not every $h_{ij}$-transformation can be recast into one of these two forms for an arbitrary NHDM; \textit{e.g.}, eqs.~\eqref{Eq:hiip_hijm} serve as a counterexample. The earlier outlined preliminary checks suggest that only the structure of the bilinear terms differs, and only in cases where each $SU(2)$ bilinear singlet maps directly to another bilinear singlet (one-to-one) rather than to a linear combination.

It should be noted that the transformations in question are closely related to eq.~(5.1) of Ref.~\cite{Trautner:2025yxz}, identified independently via a different approach. Since a general prescription for mapping eqs.~\eqref{Eq:HF_GOOFy_tr} and \eqref{Eq:GCP_GOOFy_tr} to the $h_{ij}$-space is lacking, we restrict our attention to the GOOFy transformations explicitly discussed in Ref.~\cite{Trautner:2025yxz}.

Our approach is to analyse each symmetry transformation type (HF, GCP, GOOFy) within its own fixed framework. In many cases, this full symmetry group can be decomposed into a product of a simple GOOFy-like generator and a HF or GCP symmetry. This distinction is crucial, as it highlights that the full symmetry structure cannot always be reduced to a simple GOOFy action combined with a pure HF transformation.

\subsection{GOOFy HF-like transformations}

First, consider symmetry transformations of the form
\begin{equation}\label{Eq:HF_G_mUUs}
H \mapsto U^\mathrm{HF}_\pm H \equiv \begin{pmatrix}
\mathcal{U} & 0 \\
0 & -\mathcal{U}^\ast
\end{pmatrix} H,
\end{equation}
\textit{cf.} eq.~(5.3) of Ref~\cite{Trautner:2025yxz}. In this case the $SU(2)$ bilinear singlets transform as
\begin{equation}
h_{ij} \mapsto -\mathcal{U}_{ik}^\ast \mathcal{U}_{jl} h_{kl}.
\end{equation}
It is clear that these transformations can only generate extra relations among the bilinear terms, without affecting the quartic part. Transformations of the form~\eqref{Eq:HF_G_mUUs} correspond to the GOOFy transformations introduced in Ref.~\cite{Ferreira:2023dke}. 

The choice $\mathcal{U} = \mathcal{I}_3$ corresponds to $\overline{\mathbb{Z}}_2:~h_{ij} \mapsto - h_{ij}$. As noted earlier, this case renders a potential with the most general quartic couplings, while all bilinear terms vanish. This observation identifies an entire family of scalar potentials with $\mu_{ij}^2=0$. This behaviour has been verified explicitly for all HF transformations.

Beyond the $\mu_{ij}^2=0$ family of scalar potentials, there exist other configurations where bilinear terms need not vanish. We now turn to discussing these cases, by categorising potentials based on the patterns of eigenvalues in the bilinear formalism. Here, $V_\mathcal{G}$ will
denote the quartic part of the scalar potential together with the corresponding allowed bilinear terms listed separately:
\begin{itemize}
\item $\mathbf{1_4 1_2 2_1}$:\\
$V_{O(2)_{[SO(2) \rtimes \mathbb{Z}_2]} \times U(1)}$ with $i\mu_{12}^2$; \\
$V_{O(2)_{[U(1) \rtimes \mathbb{Z}_2]} \times U(1)}$ with $\mu_{22}^2=-\mu_{11}^2$.

\item $\mathbf{1_4 4_1}$:\\
$V_{U(1) \circ V_4}$ with $\mu_{22}^2= - \mu_{11}^2$.

\item $\mathbf{3_2 2_1}$:\\
$V_{O(2)_{[SO(2) \rtimes \mathbb{Z}_2]}}$ with $i\mu_{12}^2$;\\
$V_{O(2)_{[U(1) \rtimes \mathbb{Z}_2]}}$ with $\mu_{22}^2=-\mu_{11}^2$;\\
$V_{S_3}$ with $\mu_{22}^2=-\mu_{11}^2$.

\item $\mathbf{2_2 4_1}$:\\
$V_{U(1) \times \mathbb{Z}_2}$ with $\mu_{12}^2$;\\
$V_{\mathbb{Z}_4}$ with $\mu_{12}^2$;\\
$V_{D_4}$ with $\mu_{22}^2 = - \mu_{11}^2$.

\item $\mathbf{8_1}$:\\
$V_{\mathbb{Z}_2 \times \mathbb{Z}_2}$ with $\mu_{12}^2$;\\
$V_{\mathbb{Z}_2}$ with $\{\mu_{12}^2,\, \mu_{13}^2\}$.
\end{itemize}

Since most of the considered symmetry groups are composite, \textit{i.e.}, expressible as products of simpler factors, it is not surprising that they can be decomposed into a product of GOOFy and HF transformations.

These cases are summarised in Table~\ref{Table:GOOFy_HF_cases}, and an explicit list of generators is provided in Appendix~\ref{App:generators}. 
In summary, the models considered here can be characterised by two GOOFy HF-like transformations, $\mathbb{Z}_2$ and $\mathbb{Z}_4$. One could then label these two representations as $\overline{\mathbb{Z}}_{n,\pm}$, where the ``$\pm$" subscript corresponds to $\mathcal{U}^{(2)} = - \left( \mathcal{U}^{(1)} \right)^\ast$, as a mnemonic for the opposite signs of the two blocks of the transformation of $H$.

{{\renewcommand{\arraystretch}{1.6}
\begin{table}[H]
\caption{Symmetry transformations after eq.~\eqref{Eq:HF_G_mUUs} for cases where the bilinear terms are present. In the second column the HF transformation is factored out and provided in the last sub-column, while the first sub-column corresponds to the GOOFy transformation.}
\label{Table:GOOFy_HF_cases}
\begin{center}
\begin{tabular}{|c||c|c|}  \hline\hline 
$\mathcal{G}$  as HF-like GOOFy  & \multicolumn{2}{|c|}{\hspace*{-133pt}$\mathcal{G}_\mathrm{HF-GOOFy} \ast \mathcal{G}_\mathrm{HF}$} \\ \hline  \hline

$O(2)_{[SO(2) \rtimes \mathbb{Z}_2]} \times U(1)$ & $\scriptsize \left\langle \begin{pmatrix}
-1 & 0 & 0\\
0 & 1 & 0\\
0 & 0 & 1
\end{pmatrix}\right\rangle$ & $\scriptsize\left\langle \begin{pmatrix}
\cos \beta & \sin \beta & 0\\
-\sin \beta & \cos \beta & 0\\
0 & 0 & 1
\end{pmatrix},\, \begin{pmatrix}
1 & 0 & 0\\
0 & 1 & 0\\
0 & 0 & e^{i \theta}
\end{pmatrix}\right\rangle$\\ \hline
$O(2)_{[U(1) \rtimes \mathbb{Z}_2]} \times U(1)$ & $\scriptsize\left\langle \begin{pmatrix}
0 & 1 & 0\\
1 & 0 & 0\\
0 & 0 & 1
\end{pmatrix}\right\rangle$ & $\scriptsize\left\langle \begin{pmatrix}
1 & 0 & 0\\
0 & e^{i \theta_1} & 0\\
0 & 0 & e^{i \theta_2}
\end{pmatrix}\right\rangle$ \\ \hline \hline

$U(1) \times V_4$ & $\scriptsize\left\langle \begin{pmatrix}
0 & 1 & 0\\
1 & 0 & 0\\
0 & 0 & 1 
\end{pmatrix}\right\rangle$ & $\scriptsize\left\langle \begin{pmatrix}
-1 & 0 & 0\\
0 & 1 & 0\\
0 & 0 & 1
\end{pmatrix},\,\begin{pmatrix}
1 & 0 & 0\\
0 & 1 & 0\\
0 & 0 & e^{i \theta}
\end{pmatrix}\right\rangle$ \\ \hline \hline

$O(2)_{[SO(2) \rtimes \mathbb{Z}_2]}$ & $\scriptsize\left\langle \begin{pmatrix}
-1 & 0 & 0\\
0 & 1 & 0\\
0 & 0 & 1
\end{pmatrix}\right\rangle$ & $\scriptsize\left\langle \begin{pmatrix}
\cos \beta & \sin \beta & 0\\
-\sin \beta & \cos \beta & 0\\
0 & 0 & 1
\end{pmatrix}\right\rangle$ \\ \hline
$O(2)_{[U(1) \rtimes \mathbb{Z}_2]}$ & $\scriptsize\left\langle \begin{pmatrix}
0 & 1 & 0\\
1 & 0 & 0\\
0 & 0 & 1
\end{pmatrix}\right\rangle$ & $\scriptsize\left\langle \begin{pmatrix}
e^{-i \theta} & 0 & 0\\
0 & e^{i \theta} & 0\\
0 & 0 & 1
\end{pmatrix}\right\rangle$\\ \hline
$S_3$ & $\scriptsize\left\langle \begin{pmatrix}
0 & 1 & 0\\
1 & 0 & 0\\
0 & 0 & 1
\end{pmatrix}\right\rangle$ & $\scriptsize\left\langle \begin{pmatrix}
e^{2 i \pi/3} & 0 & 0\\
0 & e^{-2 i \pi/3} & 0\\
0 & 0 & 1
\end{pmatrix}\right\rangle$\\ \hline\hline

$D_4$ & $\scriptsize\left\langle \begin{pmatrix}
0 & 1 & 0\\
1 & 0 & 0\\
0 & 0 & 1
\end{pmatrix}\right\rangle$ & $\scriptsize\left\langle \begin{pmatrix}
-1 & 0 & 0\\
0 & 1 & 0\\
0 & 0 & 1
\end{pmatrix},\,\begin{pmatrix}
1 & 0 & 0\\
0 & -1 & 0\\
0 & 0 & 1
\end{pmatrix}\right\rangle$\\ \hline
$\overline{\mathbb{Z}}_{4,\pm}$ & $\scriptsize\left\langle \begin{pmatrix}
1 & 0 & 0\\
0 & -1 & 0\\
0 & 0 & i
\end{pmatrix}\right\rangle$ &  \textemdash \\ \hline
$U(1) \times \mathbb{Z}_2$ & $\scriptsize\left\langle \begin{pmatrix}
-1 & 0 & 0\\
0 & 1 & 0\\
0 & 0 & 1
\end{pmatrix}\right\rangle$ & $\scriptsize\left\langle \begin{pmatrix}
1 & 0 & 0\\
0 & 1 & 0\\
0 & 0 & e^{i \theta}
\end{pmatrix}\right\rangle$ \\ \hline\hline

$\mathbb{Z}_2 \times \mathbb{Z}_2$ & $\scriptsize\left\langle \begin{pmatrix}
-1 & 0 & 0\\
0 & 1 & 0\\
0 & 0 & 1
\end{pmatrix}\right\rangle$ & $\scriptsize\left\langle \begin{pmatrix}
1 & 0 & 0\\
0 & -1 & 0\\
0 & 0 & 1
\end{pmatrix}\right\rangle$ \\ \hline
$\overline{\mathbb{Z}}_{2,\pm}$ & $\scriptsize\left\langle \begin{pmatrix}
-1 & 0 & 0\\
0 & 1 & 0\\
0 & 0 & 1
\end{pmatrix}\right\rangle$ &  \textemdash \\ \hline\hline
\end{tabular}
\end{center}
\end{table}}

\subsection{GOOFy GCP-like transformations}

Under a GCP-like transformation of the form
\begin{equation}\label{Eq:GCP_G_mUUs}
H \mapsto U^\mathrm{GCP}_\pm H \equiv \begin{pmatrix}
0 & \mathcal{U} \\
-\mathcal{U}^\ast & 0
\end{pmatrix} H,
\end{equation}
\textit{cf.} eq.~(5.3) of Ref~\cite{Trautner:2025yxz}, the $SU(2)$ bilinear singlets transform as
\begin{equation}
h_{ij} \mapsto -\mathcal{U}_{ik}^\ast \mathcal{U}_{jl} h_{lk}.
\end{equation}

Before proceeding to the discussion of different transformations, let us consider the canonical antisymmetric form
\begin{equation}
\mathcal{K}=\begin{pmatrix}
0 & \mathcal{I}_3 \\
-\mathcal{I}_3 & 0
\end{pmatrix}.
\end{equation}
Linear maps $\mathcal{M}: H \mapsto \mathcal{M} H$ ($\mathcal{M}$ is assumed to be unitary) that preserve the canonical bilinear form $\mathcal{K}$ belong to the complex symplectic group
\begin{equation}
Sp(6,\mathbb{C})=\left\{ \mathcal{M}\in GL(6,\mathbb{C})~\big| ~ \mathcal{M}^\mathrm{T}\, \mathcal{K}\, \mathcal{M} = \mathcal{K}
\right\}.
\end{equation}

Two families of transformations appear naturally as subgroups of $Sp(6,\mathbb{C})$:
\begin{itemize}
\item The usual HF transformations of eq.~\eqref{Eq:H_to_HF}:
\begin{equation}
(U^\mathrm{HF})^\mathrm{T} \, \mathcal{K} \, U^\mathrm{HF} = \mathcal{K};
\end{equation}

\item The GCP-like transformations of eq.~\eqref{Eq:GCP_G_mUUs}:
\begin{equation}
(U^\mathrm{GCP}_\pm)^\mathrm{T} \, \mathcal{K}\, U^\mathrm{GCP}_\pm = \mathcal{K}.
\end{equation}
\end{itemize}
These transformations therefore generate the subgroups of $U(3) \rtimes \mathbb{Z}_2 \leq Sp(6,\mathbb{C})$, where the non-trivial element of $\mathbb{Z}_2$ acts on $U(3)$ by the group automorphism. For $U^\mathrm{GCP}$ and $U^\mathrm{HF}_\pm$ the above $\mathcal{K}$ does not satisfy symplecticity.

Note that for a complex scalar multiplet one finds
\begin{equation}
H^\mathrm{T} \mathcal{K} H = h^T h^* - h^\dagger h = (h^\dagger h)^* - h^\dagger h = 0,
\end{equation}
since $h^\dagger h=\sum_i|h_i|^2\in\mathbb R$. This bilinear does not appear in the Lagrangian because it vanishes identically for commuting scalar fields.

After discussing symplecticity, we proceed to the identification of the transformations given by eq.~\eqref{Eq:GCP_G_mUUs}. Since the HF, GCP, and HF-like GOOFy transformations have already been classified, identifying genuinely new GCP-like GOOFy transformations reduces to finding cases presented by unique generators. Otherwise, the transformation simply falls into one of the previously discussed categories. For example, if we apply the CP2 transformation as the GCP-like GOOFy transformation, we recover exactly the CP2 quartic potential, while, as expected, the bilinear terms vanish. Consequently, we want to check for potentials when at least a single bilinear term remains.

As in the HF-like GOOFy case discussed previously, these transformations can be expressed as products of an HF transformation and a GCP-like GOOFy transformations. This decomposition makes it straightforward to distinguish genuinely new structures from those that simply reproduce the earlier listed cases. The explicit examples obtained through this approach are collected in  Table~\ref{Table:GOOFy_GCP_cases}.

{{\renewcommand{\arraystretch}{1.6}
\begin{table}[H]
\caption{Symmetry transformations according to eq.~\eqref{Eq:GCP_G_mUUs} for cases when the bilinear terms are present. In the second column the HF transformation is factored out.}
\label{Table:GOOFy_GCP_cases}
\begin{center}
\begin{tabular}{|c||c|c|}  \hline\hline 
$\mathcal{G}$ as GCP-like GOOFy  & \multicolumn{2}{|c|}{\hspace*{-107pt}$\mathcal{G}_\mathrm{GCP-GOOFy} \ast \mathcal{G}_\mathrm{HF}$} \\ \hline  \hline

$U(1) \times V_4$ & $\scriptsize\left\langle \begin{pmatrix}
0 & 1 & 0\\
1 & 0 & 0\\
0 & 0 & 1 
\end{pmatrix}\right\rangle$ & $\scriptsize\left\langle \begin{pmatrix}
-1 & 0 & 0\\
0 & 1 & 0\\
0 & 0 & 1
\end{pmatrix},\,\begin{pmatrix}
1 & 0 & 0\\
0 & 1 & 0\\
0 & 0 & e^{i \theta}
\end{pmatrix}\right\rangle$ \\ \hline 
$U(1) \times \mathbb{Z}_2$ & $\scriptsize\left\langle \begin{pmatrix}
0 & 1 & 0\\
-1 & 0 & 0\\
0 & 0 & 1 
\end{pmatrix}\right\rangle$ & $\scriptsize\left\langle \begin{pmatrix}
1 & 0 & 0\\
0 & 1 & 0\\
0 & 0 & e^{i \theta}
\end{pmatrix}\right\rangle$ \\ \hline\hline

$S_3$ & $\scriptsize\left\langle \begin{pmatrix}
0 & 1 & 0\\
1 & 0 & 0\\
0 & 0 & 1 
\end{pmatrix}\right\rangle$ & $\scriptsize\left\langle \begin{pmatrix}
e^{2 i \pi/3} & 0 & 0\\
0 & e^{-2 i \pi/3} & 0\\
0 & 0 & 1
\end{pmatrix}\right\rangle$ \\ \hline\hline

$O(2)_{[SO(2) \rtimes \mathbb{Z}_2]}$ & $\scriptsize\left\langle \begin{pmatrix}
-1 & 0 & 0\\
0 & -1 & 0\\
0 & 0 & 1
\end{pmatrix}\right\rangle$ & $\scriptsize\left\langle \begin{pmatrix}
\cos \beta & \sin \beta & 0\\
-\sin \beta & \cos \beta & 0\\
0 & 0 & 1
\end{pmatrix}\right\rangle$ \\ \hline
$O(2)_{[U(1) \rtimes \mathbb{Z}_2]}$ & $\scriptsize\left\langle \begin{pmatrix}
0 & 1 & 0\\
1 & 0 & 0\\
0 & 0 & 1
\end{pmatrix}\right\rangle$ & $\scriptsize\left\langle \begin{pmatrix}
e^{-i \theta} & 0 & 0\\
0 & e^{i \theta} & 0\\
0 & 0 & 1
\end{pmatrix}\right\rangle$\\ \hline \hline

$U(1)_2$ & $\scriptsize\left\langle \begin{pmatrix}
1 & 0 & 0\\
0 & 1 & 0\\
0 & 0 & -1
\end{pmatrix}\right\rangle$ & $\scriptsize\left\langle \begin{pmatrix}
1 & 0 & 0\\
0 & 1 & 0\\
0 & 0 & e^{i \theta}
\end{pmatrix}\right\rangle$\\ \hline\hline

$\overline{\mathbb{Z}}_{4,\pm}^\ast$ (CP4-like) & $\scriptsize\left\langle \begin{pmatrix}
0 & -1 & 0\\
1 & 0 & 0\\
0 & 0 & 1
\end{pmatrix}\right\rangle$ & \textemdash \\ \hline\hline

$\mathbb{Z}_2$ & $\scriptsize\left\langle \begin{pmatrix}
1 & 0 & 0\\
0 & 1 & 0\\
0 & 0 & -1
\end{pmatrix}\right\rangle$ & $\scriptsize\left\langle \begin{pmatrix}
1 & 0 & 0\\
0 & 1 & 0\\
0 & 0 & -1
\end{pmatrix}\right\rangle$ \\ \hline
$\overline{\mathbb{Z}}_{2,\pm}^\ast$ (CP2-like) & $\scriptsize\left\langle \begin{pmatrix}
1 & 0 & 0\\
0 & 1 & 0\\
0 & 0 & 1
\end{pmatrix}\right\rangle$ &  \textemdash \\ \hline\hline
\end{tabular}
\end{center}
\end{table}}

The unique configurations categorised in terms of the eigenvalue patterns are:
\begin{itemize}
\item $\mathbf{1_4 4_1}$:\\
By requiring invariance under the
\begin{equation*}
\left\langle \begin{pmatrix}
0 & e^{i\theta} & 0\\
e^{i\theta} & 0 & 0\\
0 & 0 & 1
\end{pmatrix}, \begin{pmatrix}
0 & 1 & 0\\
-1 & 0 & 0\\
0 & 0 & 1
\end{pmatrix} \right\rangle_\ast
\end{equation*}
generators we arrive at $V_{([U(1) \times D_4]/ \mathbb{Z}_2) \rtimes \mathbb{Z}_2^\ast}$ along with $\mu_{22}^2 = - \mu_{11}^2$. In another representation,
\begin{equation*}
\left\langle \begin{pmatrix}
0 & -e^{i\theta} & 0\\
e^{i\theta} & 0 & 0\\
0 & 0 & 1
\end{pmatrix}\right\rangle_\ast,
\end{equation*}
we get $V_{([U(1) \times D_4]/ \mathbb{Z}_2) \rtimes \mathbb{Z}_2^\ast}$, see eq.~\eqref{Eq:V_U1_Z2st_8C} for the quartic part, and $\{\mu_{22}^2 = - \mu_{11}^2,\, \mu_{12}^2\}$. In this case, the quartic part contains a redundant coupling, accompanied by an extra bilinear term.

\item $\mathbf{3_2 2_1}$:\\
For $V_{\mathbb{Z}_3 \rtimes \mathbb{Z}_2^\ast}:~ h_1 \mapsto e^{-2i \pi/3} h_2,\, h_2 \mapsto e^{2i \pi/3} h_1$ we get $\mu_{22}^2 = - \mu_{11}^2$;\\
$V_{SO(2) \rtimes \mathbb{Z}_2^\ast}$ with $i(\mu_{12}^2-\mu_{21}^2)$ or $V_{U(1) \rtimes \mathbb{Z}_2^\ast}$  with $\mu_{22}^2=-\mu_{11}^2$.\\

\item $\mathbf{2_2 4_1}$:\\
$V_{U(1)_2 \rtimes \mathbb{Z}_2^\ast}$ with $i\mu_{12}^2$;\\
$V_{\mathbb{Z}_4^\ast}$ with $\{\mu_{22}^2 = - \mu_{11}^2,\, \mu_{12}^2\}$. There are no cases with $\mu_{ij}^2=0$, since otherwise CP4 would reduce to the $D_4$-symmetric model.

\item $\mathbf{8_1}$:\\
$V_{\mathbb{Z}_2 \times \mathbb{Z}_2^\ast}$ with $\mu_{12}^2$;\\
$V_{\mathbb{Z}_2^\ast}$ with $\{i\mu_{12}^2,\, i\mu_{13}^2,\, i\mu_{23}^2\}$, assuming $\left\langle \mathcal{I} \right\rangle_\ast$.

\end{itemize}

An interesting observation is that the $D_4$ group can act as a GCP-like GOOFy transformation, \textit{i.e.}, a non-vanishing bilinear part is allowed. Depending on the choice of $D_4$ representation, the quartic sector can acquire an additional complex coupling.

In summary, the GCP-like GOOFy transformations admit only two unique symmetry groups: $\mathbb{Z}_2$ and $\mathbb{Z}_4$. We can label these as $\overline{\mathbb{Z}}_{n,\pm}^\ast$. Notice that in the HF-like GOOFy case we had also identified two unique symmetry groups: $\overline{\mathbb{Z}}_{2,\pm}$ and $\overline{\mathbb{Z}}_{4,\pm}$. 

\section{T-GOOFy transformations}\label{Sec:TGOOFY_3HDMs}

In the previous section, we considered transformations of $h_i$ and $h_i^\ast$ that differ by an overall minus sign (and complex conjugation). There, the GOOFy transformations of Ref.~\cite{Ferreira:2023dke} were discussed, but for 3HDM, in terms of their action on the $SU(2)$ bilinear singlets, without addressing RG stability; see Refs.~\cite{deBoer:2026ktb,Pilaftsis:2026wyw}. We now turn to a broader class of transformations, which we shall refer to as T-GOOFy transformations, following the transformations identified by Trautner in Ref.~\cite{Trautner:2025yxz} and also presented by eqs.~\eqref{Eq:HF_GOOFy_tr} and \eqref{Eq:GCP_GOOFy_tr}, when additional constraints are applied.

To specify the action of these transformations, we employ the notation
\begin{equation}
\mathcal{G} = \left\langle [\mathcal{U}^{(1)},\, \mathcal{U}^{(2)}] \right\rangle,
\end{equation}
which indicates the subspace of $H$ on which the transformations act. For instance, such a transformation can be represented as a HF-type operation, $H \mapsto \mathrm{diag}(\mathcal{U}^{(1)},\, \mathcal{U}^{(2)})\,H$. However, this notation alone does not distinguish between HF and GCP transformations. We therefore employ a subscript notation consistent with earlier sections, and the first entry, $\mathcal{U}^{(1)}$, corresponds to the upper index of the full transformation acting on the $H$ fields. 

Hermiticity of the Lagrangian imposes a consistency condition on the individual transformations that generate $\mathcal{G}$. Each transformation, constructed from the unitary blocks $\mathcal{U}^{(1)}$ and $\mathcal{U}^{(2)}$, must preserve the Hermitian form of the kinetic term, which effectively requires that the product of a transformation and its Hermitian conjugate yields a diagonal operator with eigenvalues ``$\pm1$". These discrete eigenvalues determine whether the transformation leaves the fields invariant or introduces an overall sign flip, ensuring that all elements of $\mathcal{G}$ act as genuine symmetries of the theory.

As noted previously, identifying all possible symmetries in this framework is considerably more involved, since one must also account for permutations of the generators, which determine the form of the bilinear terms. For both methodological and computational reasons, we restrict our analysis to the classification of distinct quartic terms in the scalar potential. Cases differing only by the restriction $\lambda \in \mathbb{R}$ are omitted, as these correspond to the imposition of a CP2 symmetry, $\mathrm{CP2}:\,\lambda(\mathbb{C}) \mapsto \lambda(\mathbb{R})$.

As before, distinct symmetry classes are identified via the eigenvalue patterns of the bilinear space. Although no additional refinement step is introduced, it is worth noting that under GOOFy transformations certain eigenvalues may be equal in magnitude or vanish entirely. Such features can provide an initial diagnostic for the underlying symmetry and motivate a further subdivision into finer classes.

Many of the cases discussed below can be expressed as $\mathcal{G}_\mathrm{GOOFy} \ast \mathcal{G}_\mathrm{HF\ and/or\ GCP}$. In general, multiple decompositions of a given T-GOOFy transformation are possible: different generator sets can lead to identical quartic potentials while differing in the bilinear sector. Since we focus only on the quartic part, the discussion of free parameters remains limited. In particular, when the scalar potential in the $\{h_1,\,h_2\}$ sector coincides with that of the full 2HDM, an $SU(2)$ transformation can be used to remove redundant degrees of freedom, and we prioritise such two-dimensional group actions in our classification.

\subsection{T-GOOFy HF-like transformations}

We start by covering the HF transformations, which were given by eq.~\eqref{Eq:HF_GOOFy_tr}. For the HF-like transformations we have:
\begin{equation}
h_{ij} \mapsto U^{(2)\,\ast}_{ik} U^{(1)}_{jl} h_{kl}.
\end{equation}

Earlier, we noted that different representations can lead to distinct scalar potentials. For example, consider two different $\mathbb{Z}_2$ representations (these can be viewed as $\mathbb{Z}_2$ acting on the whole $H$ space):
\begin{subequations}
\begin{align}
\mathcal{G}_1 =&{} \left\langle \left[ \mathcal{I}_3,\, \begin{pmatrix}
 1 & 0 & 0 \\
 0 & 1 & 0 \\
 0 & 0 & -1 \\
\end{pmatrix} \right] \right\rangle_\mathrm{HF},\label{Eq:G1_TGOOFY}\\
\mathcal{G}_2 =&{} \left\langle \left[ \mathcal{I}_3,\, \begin{pmatrix}
 -1 & 0 & 0 \\
 0 & -1 & 0 \\
 0 & 0 & 1 \\
\end{pmatrix} \right] \right\rangle_\mathrm{HF}.
\end{align}
\end{subequations}

In both cases we get an identical quartic part:
\begin{equation}
\begin{aligned}
V^4 ={}& \lambda_{1111} h_{11}^2 + \lambda_{2222} h_{22}^2 + \lambda_{3333} h_{33}^2 + \lambda_{1221} h_{12} h_{21} + \lambda_{1122}h_{11}h_{22} \\
&+ \bigg\{ \lambda_{1112} h_{11}h_{12} + \lambda_{1222} h_{12}h_{22} + \lambda_{1323} h_{13}h_{23} \\
&\qquad+ \lambda_{1212} h_{12}^2 + \lambda_{1313} h_{13}^2 + \lambda_{2323} h_{23}^2 + \mathrm{h.c.} \bigg\}.
\end{aligned}
\end{equation}

As for the bilinear parts we have:
\begin{equation}
V^2_{\mathcal{G}_1} = \mu_{11}^2 h_{11} + \mu_{22}^2 h_{22} + \left\lbrace \mu_{12}^2 h_{12} + \mathrm{h.c.} \right\rbrace,
\end{equation}
or 
\begin{equation}
V^2_{\mathcal{G}_2} = \mu_{33}^2 h_{33},
\end{equation}
respectively.

It should be noted that with the help of the $SU(2)$ transformation acting on the $\{h_1,\,h_2\}$ space, like in the 2HDM, it is possible to rotate away one of the couplings. A convenient choice could be to rotate away either $\lambda_{1112}$ or $\lambda_{1222}$.

Now, suppose that on top of either $\mathcal{G}_1$ or $\mathcal{G}_2$ we apply $\mathbb{Z}_2:\,h_1 \mapsto - h_1$. Then, we get
\begin{equation}
\begin{aligned}
V^4 ={}& \lambda_{1111} h_{11}^2 + \lambda_{2222} h_{22}^2 + \lambda_{3333} h_{33}^2 + \lambda_{1221} h_{12} h_{21} + \lambda_{1122}h_{11}h_{22}\\
& + \left\lbrace \lambda_{1212} h_{12}^2 + \lambda_{1313} h_{13}^2 + \lambda_{2323} h_{23}^2 + \mathrm{h.c.} \right\rbrace.
\end{aligned}
\end{equation}
Different choices of bases and of the forms of generators lead to several different possibilities for the bilinear terms, either $\mu_{11}^2 h_{11} + \mu_{22}^2 h_{22}$, or $\mu_{12}^2 h_{12}$, or  $\mu_{33}^2 h_{33}$, or no terms. As a result, the identification of the underlying symmetries becomes a complicated task following our methodology. Hence, as stated earlier, we shall omit discussion of the bilinear terms.

Both of the discussed cases in the bilinear formalism share the $8_1$ pattern of eigenvalues. We can consider yet another $\mathbb{Z}_2$ representation given by 
\begin{equation}\label{Eq:G3_TGOOFy}
\mathcal{G}_3 = \left\langle \left[ \begin{pmatrix}
 1 & 0 & 0 \\
 0 & -1 & 0 \\
 0 & 0 & 1 \\
\end{pmatrix},\, \begin{pmatrix}
 -1 & 0 & 0 \\
 0 & 1 & 0 \\
 0 & 0 & 1 \\
\end{pmatrix}\right] \right\rangle_\mathrm{HF},
\end{equation}
for which,
\begin{equation}
\begin{aligned}
V^4_{\mathcal{G}_3} ={}& \lambda_{1111} h_{11}^2 + \lambda_{2222} h_{22}^2 + \lambda_{3333} h_{33}^2 + \lambda_{1221} h_{12} h_{21} + \lambda_{1122}h_{11}h_{22}\\
& + \left\lbrace \lambda_{1332} h_{13}h_{32} + \lambda_{1233} h_{12}h_{33} + \lambda_{1212} h_{12}^2 + \lambda_{1313} h_{13}^2 + \lambda_{2323} h_{23}^2 + \mathrm{h.c.} \right\rbrace.
\end{aligned}
\end{equation}

Another representation of  $\mathbb{Z}_2$, given by
\begin{equation}
\mathcal{G}_4 = \left\langle \left[ \begin{pmatrix}
 0 & - e^{ 3 i \pi/4} & 0 \\
 e^{i \pi/4} & 0 & 0 \\
 0 & 0 & 1 \\
\end{pmatrix},\, \begin{pmatrix}
 0 & e^{i \pi/4} & 0 \\
 - e^{ 3 i \pi/4} & 0 & 0 \\
 0 & 0 & 1 \\
\end{pmatrix}\right] \right\rangle_\mathrm{HF},
\end{equation}
yields
\begin{equation}
\begin{aligned}
V^4_{\mathcal{G}_4} ={}& \lambda_{1111} (h_{11}^2 - h_{22}^2) + \lambda_{3333} h_{33}^2 + \lambda_{1221} h_{12} h_{21} + \lambda_{1122}h_{11}h_{22} \\
&+ \lambda_{1332} (h_{13}h_{32} + h_{23}h_{31}) + \lambda_{1233} (h_{12} + h_{21})h_{33} + \lambda_{1212} (h_{12}^2+ h_{21}^2) \\
&+ \left\lbrace \lambda_{1323} h_{13}h_{23} + \mathrm{h.c.} \right\rbrace.
\end{aligned}
\end{equation}

There is no unitary transformation between $V^4_{\mathcal{G}_1}$, $V^4_{\mathcal{G}_3}$, and $V^4_{\mathcal{G}_4}$. While the generators resemble different $\mathbb{Z}_2$ representations, it should be noted that $h_i$ and their conjugated fields transform differently, and transformations are not connected by a simple function. In contrast, if we consider how the $h_{ij}$ bilinear singlets transform, we can notice that while for both $\mathcal{G}_1$ and $\mathcal{G}_3$ the $SU(2)$ bilinear singlets transform into themselves after two operations, for $\mathcal{G}_4$ we are looking at six operations. On the other hand, while both $\mathcal{G}_1$ and $\mathcal{G}_3$ seem to correspond to the ``same" $\mathbb{Z}_2$, $\mathcal{G}_1$ and $\mathcal{G}_3$ act on different spaces. To be more precise:
\begin{equation}
\begin{aligned}
\begin{array}{c c c}
 & \mathcal{G}_1 : \vspace{5pt}&  \\ 
h_{11} \mapsto h_{11}, & h_{12} \mapsto h_{12}, & h_{13} \mapsto h_{13}, \\
h_{21} \mapsto h_{21}, & h_{22} \mapsto h_{22}, & h_{23} \mapsto h_{23}, \\
h_{31} \mapsto -h_{31}, & h_{32} \mapsto -h_{32}, & h_{33} \mapsto -h_{33}.
\end{array} \quad
\begin{array}{c c c}
 & \mathcal{G}_3 : \vspace{5pt}&  \\
h_{11} \mapsto -h_{11}, & h_{12} \mapsto h_{12}, & h_{13} \mapsto -h_{13}, \\
h_{21} \mapsto h_{21}, & h_{22} \mapsto -h_{22}, & h_{23} \mapsto h_{23}, \\
h_{31} \mapsto h_{31}, & h_{32} \mapsto -h_{32}, & h_{33} \mapsto h_{33}.
\end{array}
\end{aligned}
\end{equation}

We can also define the following $\mathbb{Z}_4$ generator
\begin{equation}
\mathcal{G}_5 = \left\langle \left[ \begin{pmatrix}
 -i & 0 & 0 \\
 0 & i & 0 \\
 0 & 0 & 1 \\
\end{pmatrix},\, \begin{pmatrix}
 i & 0 & 0 \\
 0 & -i & 0 \\
 0 & 0 & 1 \\
\end{pmatrix}\right] \right\rangle_\mathrm{HF},
\end{equation}
for which we get
\begin{equation}
\begin{aligned}
V^4_{\mathcal{G}_5} ={}& \lambda_{1111} h_{11}^2 + \lambda_{2222} h_{22}^2 + \lambda_{3333} h_{33}^2 + \lambda_{1221} h_{12} h_{21} + \lambda_{1122}h_{11}h_{22}\\
&  + \left\lbrace \lambda_{1323} h_{13}h_{23} + \lambda_{1332} h_{13}h_{32} + \lambda_{1233} h_{12}h_{33} + \lambda_{1212} h_{12}^2 + \mathrm{h.c.} \right\rbrace.
\end{aligned}
\end{equation}

These scalar potentials share the same pattern of eigenvalues, $8_1$. However a new feature emerges: while the above patterns fall into the $8_1$ class, if we take the absolute values of the eigenvalues we get another pattern, $|2_2 4_1|$---six different values. Here, we introduced a new notation inspired by the absolute value.

Potentials other than the ones presented above, sharing the $8_1$ eigenvalues pattern, are subject to an underlying symmetry which can be written as products of T-GOOFy and HF transformations. These are presented in Appendix~\ref{App:HF_TGOOFy}.

The only other unique case is given by the representation of $\mathbb{Z}_6$,
\begin{equation}
\mathcal{G}_6 = \left\langle \left[ \begin{pmatrix}
 0 & 1 & 0 \\
 0 & 0 & 1 \\
 1 & 0 & 0 \\
\end{pmatrix},\, \begin{pmatrix}
 0 & -1 & 0 \\
 0 & 0 & 1 \\
 1 & 0 & 0 \\
\end{pmatrix}\right] \right\rangle_\mathrm{HF}.
\end{equation}
Note that in the conventional HF transformations $\mathbb{Z}_n$, with $n \geq 5$, results in either $U(1)_1$ or $U(1)_2$, based on how the symmetry charges are implemented.

Invariance under this group results in the following scalar potential:
\begin{equation}
\begin{aligned}
V^4_{\mathcal{G}_6} ={}& \lambda_{1111} (h_{11}^2 + h_{22}^2 + h_{33}^2) + \lambda_{1122} ( - h_{11} h_{22} + h_{11} h_{33} + h_{22} h_{33}) \\
& + \lambda_{1221}(-h_{12}h_{21} + h_{13}h_{31} +h_{23}h_{32}) + \left\lbrace \lambda_{1212}(-h_{12}^2 +h_{13}^2 -h_{23}^2 ) + \mathrm{h.c.} \right\rbrace.
\end{aligned}
\end{equation}
For this case the patterns of eigenvalues are $2_2 4_1$ and $|2_3 2_1|$.

For the above discussed models it does not make much sense to discuss CP violation since the bilinear part of the scalar potential remains largely unexplored.

One may wonder whether GOOFy transformations could provide a natural mechanism for constructing a 2HDM-like potential enlarged by an additional, non-interacting doublet. Consider $U(1) \times \mathbb{Z}_2$ applied as a GOOFy transformation ($ \mathbb{Z}_2$ will be identified as $\mathcal{G}_1$ of  eq.~\eqref{Eq:G1_TGOOFY}):
\begin{equation}
H \mapsto \begin{pmatrix} {\cal U}^{(1)} & 0\\ 0 & {\cal U}^{(2)} \end{pmatrix}H, \quad
{\cal U}^{(1)} =\begin{pmatrix}1 & 0 & 0 \\ 0 & 1 & 0 \\0 & 0 & e^{i\theta}\end{pmatrix},  \quad
{\cal U}^{(2)} =\begin{pmatrix}1 & 0 & 0 \\ 0 & 1 & 0 \\0 & 0 & -e^{-i\theta}\end{pmatrix}, 
\end{equation}
or equivalently
\begin{equation}
H \mapsto \text{diag}\{1,1,e^{i\theta},1,1,-e^{-i\theta}\}H.
\end{equation}
This transformation can be split into two combined transformations: (HF-like GOOFy)  $\mathcal{G}_1$ and the HF $U(1)_{h_3}$, which can be presented as  $\mathcal{G}_1 \ast [U(1)]_{h_3}$:
\begin{subequations}
\begin{alignat}{2}
&\mathcal{G}_1:& H &\mapsto \mathrm{diag}\{1, 1, 1, 1, 1, -1\} H, \\
&U(1)_{h_3}: \quad & H &\mapsto \mathrm{diag}\{1, 1, e^{i\theta}, 1, 1, e^{-i\theta}\} H.
\end{alignat}
\end{subequations}

The scalar potential invariant under $\mathcal{G}_1 \ast [U(1)]_{h_3}$ is given by:
\begin{equation}
\begin{aligned}
V ={}& \mu_{11}^2 h_{11} + \mu_{22}^2 h_{22} + \{ \mu_{12}^2 h_{12} + \mathrm{h.c.}\} \\
& + \lambda_{1111} h_{11}^2 + \lambda_{2222} h_{22}^2 + \lambda_{1221} h_{12} h_{21} + \lambda_{1122}h_{11}h_{22} \\ & + \bigg\{ \lambda_{1112} h_{11}h_{12} + \lambda_{1222} h_{12}h_{22} + \lambda_{1212} h_{12}^2  + \mathrm{h.c.} \bigg\}  + \lambda_{3333} h_{33}^2\\
={}& V_\mathrm{2HDM} + \lambda_{3333} h_{33}^2.
\end{aligned}
\end{equation}
This potential successfully decouples $h_3$, replicating the structure of \mbox{$V_\mathrm{2HDM} + V(h_3)$} at tree level, but not once the higher-order corrections are introduced. However, a crucial limitation emerges in the bilinear sector: while allowing all bilinear 2HDM couplings, the symmetry forbids the $\mu_{33}^2$ coupling. There are no quartic terms mixing $\{h_1,\, h_2\}$ with $h_3$ since there is no $\mu_{33}^2$ coupling present. In practice, the only option to achieve decoupling of $h_3$ at tree level is by considering a $\mathbb{Z}_2$ transformation, either \mbox{$\mathcal{G}_1:\,\mu_{33}^2=0$} or $\mathcal{G}_2:\,\{\mu_{11}^2, \mu_{22}^2,\, \mu_{12}^2\}=0$. 

Because the GOOFy transformation acts non-trivially on the kinetic sector, promoting it to an exact symmetry results in a non-standard dynamics for $h_3$ in the case of \mbox{$\mathcal{G}_1 \ast [U(1)]_{h_3}$}. Concretely, the GOOFY invariance either induces an overall minus sign in the $h_3$ kinetic Lagrangian, giving rise to a ghost-like state, or it forces the kinetic coefficient to vanish, in which case $h_3$ is no longer a propagating degree of freedom but instead becomes an auxiliary field. Both possibilities have immediate consequences for radiative stability. Introducing the $\mu_{33}^2$ term can stabilise the physical mass scale of the $h_3$ states; however, it does not turn $h_3$ into a conventional scalar doublet.

This example, nevertheless, provides a basis for identifying other 2HDM-like potentials, presented in Appendix~\ref{App:Gen_TGOOFy}. In the bilinear formalism the off-diagonal/mixed bilinear terms ($i \neq j$), which encode the relative orientation and phase between doublets, are:
\begin{equation*}
\frac{1}{2} \left( h_{ij} + h_{ji} \right), \text{  and  } -\frac{1}{2} \left( i  h_{ij} - i h_{ji} \right).
\end{equation*}
The key, basis-independent signature of such a decoupled structure is the pattern of eigenvalues in the bilinear formalism: since $\{h_1,\,h_2\}$ and $h_3$ do not mix, at least four eigenvalues must vanish. This provides a simple diagnostic tool---any 3HDM potential exhibiting a null space of this dimension falls into the 2HDM-like category.

\subsection{T-GOOFy GCP-like transformations}

Now, we move to the GCP transformations, which were given by eq.~\eqref{Eq:GCP_GOOFy_tr}. For this case, the $SU(2)$ bilinear singlets transform as
\begin{equation}\label{eq:TGOOFY_GCP_singlets}
h_{ij} \mapsto U^{(2)\,\ast}_{ik} U^{(1)}_{jl} h_{lk}.
\end{equation}

In total, we managed to identify two unique (not a product of other groups) symmetries. First, consider a $\mathbb{Z}_4$ representation
\begin{equation}\label{Eq:G11_GCP_gen}
\mathcal{G}_7^\ast = \left\langle \left[ \begin{pmatrix}
 0 & -1 & 0 \\
 1 & 0 & 0 \\
 0 & 0 & i \\
\end{pmatrix},\, \begin{pmatrix}
 0 & -1 & 0 \\
 1 & 0 & 0 \\
 0 & 0 & -i \\
\end{pmatrix}\right] \right\rangle_\ast.
\end{equation}
The $\mathcal{G}_7$ structure is connected to $\mathcal{G}_3$ of eq.~\eqref{Eq:G3_TGOOFy} via a basis change of $\mathcal{R}_{(\frac{\pi}{4},\, -\frac{\pi}{4},\, -\frac{\pi}{4},\, 0)}$. However, utilising the $\mathcal{G}_3^\ast$ group (we recall that we are considering GCP transformations)  would result in a scalar potential with three purely imaginary couplings, $\{\lambda_{1112},\, \lambda_{1222},\, \lambda_{1323}\}$. On the other hand, $\mathcal{G}_7^\ast$ results in some redundant couplings. While the two cases are connected via a unitary basis transformation, it should be noted that we consider \textit{only} the quartic part of the potential. As in the previous subsection, while the choice of $\mathcal{G}_3^\ast$ or $\mathcal{G}_7^\ast$ would render an identical quartic part of the scalar potential, the bilinear terms will be different. For $\mathcal{G}_3^\ast$ we get \mbox{$V =\mu_{12}^2 (h_{12} + h_{21}) + \mu_{33}^2$}, while for $\mathcal{G}_7^\ast$ it is $V = \mu_{11}^2 (h_{11} + h_{22})$. Moreover, $\mathcal{G}_3$ and $\mathcal{G}_7$ are not the only two generators that result in a different bilinear part of the potential.

For $\mathcal{G}_7^\ast$ we get the following scalar potential:
\begin{equation}
\begin{aligned}
V^4_{\mathcal{G}_7^\ast} ={}& \lambda_{1111} (h_{11}^2 + h_{22}^2) + \lambda_{3333} h_{33}^2 + \lambda_{1221} h_{12} h_{21} + \lambda_{1122}h_{11}h_{22}\\
&  + \lambda_{1133} (h_{22} - h_{11})h_{33} +  \lambda_{1331} (h_{13}h_{31} - h_{23}h_{32}) + \lambda_{1323} (h_{13}h_{23} + h_{31}h_{32})\\
& + \bigg\{ \lambda_{1212} h_{12}^2 + \lambda_{2331} h_{23} h_{31} + \lambda_{1112} (h_{11} - h_{22})h_{21}\\
& \qquad+  \lambda_{1233} h_{12}h_{33} + \lambda_{1313} (h_{13}^2 - h_{32}^2) + \mathrm{h.c.} \bigg\}.
\end{aligned}
\end{equation}
In terms of the eigenvalues pattern we are considering the $8_1$, $|2_2 4_1|$, class of the potentials.

The presence of phase-sensitive couplings in the scalar potential might seem at odds with a GCP-like symmetry. However, taking into consideration that the underlying symmetry is $\mathbb{Z}_4^\ast$, it should be of no surprise that these couplings are allowed. 

Another unique group is given by another $\mathbb{Z}_4$ representation:
\begin{equation}
\mathcal{G}_8 = \left\langle \left[ \begin{pmatrix}
 1 & 0 & 0 \\
 0 & 1 & 0 \\
 0 & 0 & 1 \\
\end{pmatrix},\, \begin{pmatrix}
 0 & -i & 0 \\
 i & 0 & 0 \\
 0 & 0 & 1 \\
\end{pmatrix}\right] \right\rangle_\ast.
\end{equation}
In this case the scalar potential is:
\begin{equation}
\begin{aligned}
V^4_{\mathcal{G}_8^\ast} ={}& \lambda_{1111} (h_{11}^2 + h_{22}^2 - h_{12}^2 - h_{21}^2) + \lambda_{3333} h_{33}^2 + \lambda_{1122} (h_{11}h_{22} + h_{12} h_{21}) \\
& + \lambda_{1112} (h_{22} - h_{11})(h_{12} + h_{21}) + \lambda_{1323} (h_{13}h_{23} + h_{31}h_{32})\\
& + \lambda_{1313} ( h_{13}^2 - h_{23}^2 + \mathrm{h.c.}).
\end{aligned}
\end{equation}
In the bilinear formalism it corresponds to the $2_2 4_1$ and $|1_4 1_2 2_1|$ pattern.

All other T-GOOFy GCP-like scalar potentials can be derived starting from these two. The complete set of scalar potentials is presented in Appendix~\ref{App:GCP_TGOOFy}.

\section{Summary}\label{Sec:Summary}

The set of realisable symmetries in NHDMs is substantially richer and more intricate than often assumed. A ``brute-force'' approach, although less elegant than purely group-theoretical methods, has been instrumental in systematically classifying symmetries: by explicitly constructing and comparing scalar potentials.

Our analysis began with a detailed re-examination of conventional HF and GCP transformations in 3HDMs. By systematically constructing the corresponding scalar potentials, we have compiled what we believe to be the most complete picture to date, summarised in Tables~\ref{Table:Indep_Copuplings_Cases}--\ref{Table:Indep_R_Copuplings_Cases}. These tables serve as a practical diagnostic reference for model builders, providing the number of independent couplings and basis-invariant eigenvalue patterns associated with each symmetry. Although we have not exhaustively scanned all subgroups of $U(3)$, our results remain broadly consistent with previous classifications and reveal no evidence for additional symmetries beyond those identified here.

Another direction of our paper was to consider GOOFy transformations, which act asymmetrically on scalar doublets and their conjugates. Such transformations challenge the conventional paradigm of symmetry imposition and naturally give rise to new families of scalar potentials. Due to the complexity of the analysis, we restricted our attention to the quartic sector, providing a first systematic study of 3HDM scalar potentials invariant under the GOOFy-type transformations. We found that certain representations in the combined space of doublets and their conjugates can selectively eliminate bilinear terms while leaving the quartic sector invariant. This feature cannot be attributed merely to a basis choice alone and suggests new directions for model building, possibly offering novel approaches to hierarchy~\cite{deBoer:2025jhc,Trautner:2025prm,Grzadkowski:2026gkx,Ferreira:2026ccg} and flavour problems. We have not attempted to propose a formal naming convention for such symmetries; instead, GOOFy transformations have been denoted generically as $\mathcal{G}_n$.

One of the central lessons from our analysis is the observation of the critical roles played by the choice of basis and of representation in the identification of the symmetries.
Furthermore, invariance under a given set of generators does not necessarily capture the full underlying symmetry.

While a complete classification of 3HDMs, invariant under all possible transformations, remains an ambitious goal, $e.g.$, we have not considered custodial symmetries, our work provides a transparent framework and lays the ground for several research directions. The new scalar potentials identified here warrant detailed phenomenological study. Extending the GOOFy transformations to the fermion and gauge sectors is an important next step, with the potential for distinctive collider signatures and cosmological implications. Furthermore, extending the work of Ref.~\cite{deMedeirosVarzielas:2019rrp} in order to recognise potentials invariant under GOOFy transformations, when written in a particular basis would be useful to gain a deeper understanding of these theories.

\section*{Acknowledgements}

It is a pleasure to thank Darius Jur\v ciukonis for discussions on group theory and for help with the implementation of the $\mathsf{GAP}$ software. We are also very grateful to Andreas Trautner for enlightening discussions on  properties of generalisations of GOOFy transformations. We thank Luís Lavoura for questions which led us to clarify some aspects.

The work of AK and MNR was partially supported by Funda\c c\~ ao para a Ci\^ encia e a Tecnologia (FCT), Portugal through the projects UIDB/00777/2020, UIDP/00777/2020, UID/00777/2025 \href{https://doi.org/10.54499/UID/00777/2025}{(https://doi.org/10.54499/UID/00777/2025)}, UID/PRR/777/2025,\linebreak  UID/PRR2/777/2025, CERN/FIS-PAR/0002/2021 and 2024.02004.CERN, which are partially funded through POCTI (FEDER), COMPETE, QREN and the EU. Furthermore, the work of AK has been supported by the research studentships for doctoral students BL186/2024\_IST-ID within the framework of the project UIDB/00777/2020 as well as BL325/2024\_IST-ID within the framework of the project 2024.02004.CERN, and also by the Spanish grants PID2023-147306NB-I00, CNS2024-154524 and CEX2023-001292-S (MICIU/AEI/10.13039/501100011033), as well as CIPROM/2021/054 (Generalitat Valenciana). PO is supported in part by the Research Council of Norway. We thank the University of Bergen and CFTP/IST/University of Lisbon as well as the CERN Theory Department, where collaboration visits took place.

\appendix

\section{Redundant couplings}\label{App:Redundancies}

When analysing the scalar potential of an NHDM, symmetric under a group $\mathcal H \subset U(N)$, it is important to distinguish between physical couplings and redundant ones. Redundant couplings are artefacts of the basis choice: they appear in the scalar potential but can be removed by a unitary basis transformation altering neither any  physical prediction nor the explicit form of the imposed symmetry $ \mathcal H$. From the group-theoretical point of view, identification of these redundancies follows from the orbit-stabiliser theorem.

For the most general NHDM, the scalar potential preserves its form under the full group of unitary basis transformations $U(N)$. However, the overall $U(1)_Y$ acts trivially on the potential. Consequently, one can always utilise the remaining $SU(N)$ freedom to remove $\dim(SU(N)) = N^2 - 1$ redundant real degrees of freedom from the general potential; note that a complex coupling can be presented in terms of two real ones.

When a specific symmetry $\mathcal H$ is imposed, we essentially select a particular embedding of the symmetry generators in the $N$-dimensional scalar space. The set of continuous basis transformations that preserve this orientation is the centraliser of $\mathcal H$ in $U(N)$:
\begin{equation}
    C_{U(N)}(\mathcal H) = \left\lbrace U \in U(N) \mid U\,h=h\,U,\quad \forall\,h\in \mathcal H \right\rbrace.
\end{equation}
Intuitively, $C_{U(N)}(\mathcal H)$ represents the residual freedom, which can be utilised to re-write the fields while keeping the explicit form of the symmetry manifest.

However, not all degrees of the centraliser correspond to redundant directions in the parameter space. If $ \mathcal H$ itself contains continuous factors (\textit{e.g.}, a global $U(1)$ symmetry), the potential is invariant under these transformations by construction, and hence they do not correspond to redundant directions that can be used to rotate away redundant parameters. In order to identify redundancies, we must subtract the genuine continuous symmetries, as well as the the overall $U(1)_Y$:
\begin{equation}\label{Eq:redundant_dofs}
    N_{\mathrm{redundant}} = \dim (C_{U(N)}( \mathcal H)) - \dim ( \mathcal H_{\mathrm{continuous}}) - 1,
\end{equation}
where the final term accounts for $U(1)_Y$, and $\mathcal H_{\mathrm{continuous}}$ denotes the global continuous transformations of the scalar potential.

This counting is a direct application of the orbit–stabiliser theorem for the action of the (effective) basis-transformation group on the space of scalar potentials. The orbit corresponds to the continuous re-parameterisation directions in the space generated by basis transformations, while the stabiliser is the subgroup of transformations that leave a given potential invariant. For a potential invariant under $\mathcal H$, this stabiliser coincides with the continuous symmetry group of the potential itself.

There is a compelling conceptual analogy between this counting and Goldstone's theorem. In spontaneous symmetry breaking, massless Goldstone bosons arise from broken generators, corresponding to flat directions along the vacuum orbit $\mathcal G/ \mathcal H$. Similarly, in model building, redundant parameters correspond to flat directions in the parameter space. Just as a Goldstone boson represents a fluctuation moving along a symmetry orbit (changing the parameterisation, not the energy), a redundant parameter represents a change in couplings that moves along a basis-equivalence orbit (changing the couplings of the scalar potential, neither the form nor the physics). Note that, this is purely a formal analogy, the redundancies discussed here are non-physical re-parameterisation freedoms.

To implement eq.~\eqref{Eq:redundant_dofs} without relying on explicit angle parametrisations, one can work at the level of the Lie algebra, following the outlined technique:
\begin{enumerate}
\item Let $A \in \mathfrak{u}(N)$ (using $N^2$ real variables) be a general skew-Hermitian generator and $U = e^A \in U (N)$ its exponentiation. For $j < k$, set $A_{jk} = a_{jk} + i b_{jk}$, and for the diagonal entries set $A_{jj} = i c_j$, where $\{a,\, b,\, c\} \in \mathbb{R}$; 
\item For every HF  (the case of GCP requires some further explanation) generator $h \in \mathcal H$ impose $[A,h] = 0$ (equivalently, $A\, h - h\, A =0$) in terms of a linear system for the variables $\{a,\,b,\, c\}$;
\item Compute the dimension of the real null-space (it shows which infinitesimal basis changes preserve the form of the imposed symmetry) of the resulting $U$ matrix. This yields $\dim C_{U(N)}(\mathcal H)$. Apply eq.~\eqref{Eq:redundant_dofs} to find $N_{\mathrm{redundant}}$.
\end{enumerate}

This technique extends in a straightforward way to the GCP transformations, with a crucial distinction arising from their inherent inclusion of complex conjugation. A GCP transformation acts as $h \mapsto \mathcal U h^\ast$, involving complex conjugation. Consequently, under a basis transformation $h \mapsto U h$ we have:
\begin{equation}
\mathcal U \rightarrow  U\, \mathcal U \, U^\mathrm{T}, \quad U=e^A, \quad A  \in \mathfrak{u}(N).
\end{equation}
We expand to first order in an infinitesimal parameter $\epsilon$, $U = \mathcal I + \epsilon A + \mathcal O(\epsilon^2)$. This yields:
\begin{equation}
(\mathcal I + \epsilon A)\,\mathcal U\,(\mathcal I + \epsilon A^{\mathrm T})
= \mathcal U+ \epsilon\,(A\, \mathcal U + \mathcal U\, A^{\mathrm T}) + \mathcal O(\epsilon^2).
\end{equation}
Requiring invariance, $\delta \mathcal U = 0$, yields the congruence condition
\begin{equation}
A\,\mathcal U + \mathcal U\,A^{\mathrm T} = 0.
\end{equation}

By using this constraint in step 2 of the aforementioned algorithm, the same null-space procedure yields the correct count of redundant parameters for GCP-invariant potentials. Note, however, that while $U(1)_Y$ always commutes with HF transformations, it does not satisfy the linearised congruence condition for GCP transformations. Consequently, for potentials constrained solely by GCP symmetries, the hypercharge generator is automatically excluded from the null-space.

We summarise cases which have not-trivial residual continuous basis freedom in Table~\ref{Table:Redundancies}. These cases list the full residual freedom, rather than a particular centraliser.

{{\renewcommand{\arraystretch}{1.35}
\begin{table}[H]
\caption{Realisable 3HDM symmetries with non-trivial residual continuous basis transformation freedom. The third column indicates how many real degrees of freedom can be rotated away. Note that, for the $O(2)$-symmetric case the centraliser is given by $U(1)^2$, while for the CP4 case the quotient indicates that one degree is fixed due to the form of the $\lambda_{1323}$ coupling, see potentials provided by eqs.~\eqref{Eq:V_CP4}. }
\label{Table:Redundancies}
\begin{center}
\begin{tabular}{|c|c|c|} \hline\hline
\begin{tabular}[l]{@{}c@{}@{}} Underlying  symmetry \end{tabular} & \begin{tabular}[l]{@{}c@{}@{}} Residual  freedom \end{tabular} & \begin{tabular}[l]{@{}c@{}@{}} Redundant  degrees \end{tabular} \\ \hline \hline
$\{ e \}$ (3HDM) & $U(3)$ & $9 - 1 = 8$ \\ \hline
$\mathbb{Z}_2$ & $U(2) \times U(1)$ & $5-1=4$ \\ \hline
$U(1)_2$ & $U(2) \times U(1)$ & $5-1-1=3$\\  \hline
$\mathbb{Z}_2 \times \mathbb{Z}_2$, $\mathbb{Z}_3$, $\mathbb{Z}_4$ & $U(1)^3$ & $3-1=2$ \\ \hline
$U(1)_1$, $U(1) \times \mathbb{Z}_2$ & $U(1)^3$ & $3-1-1=1$ \\ \hline
$S_3$, $D_4 $ & $U(1)^2$ & $ 2-1=1$\\  \hline
$O(2)$ & \begin{tabular}[l]{@{}c@{}@{}} $O(2) \times U(1)$,  $\,C_{U(3)} (O(2)) \cong U(1)^2$ \end{tabular} &$ 2 - 1 = 1$ \\ \hline\hline
$\mathbb{Z}_2^\ast$ (CP2) & $SO(3)$ & $3$ \\ \hline
$\mathbb{Z}_4^\ast$ (CP4) & $SU(2)/U(1)$ & $2$ \\ \hline\hline
$(\mathbb{Z}_2 \times \mathbb{Z}_2) \rtimes \mathbb{Z}_2^\ast$ & $U(1)^2$ & $2 - 1 = 1$ \\ \hline
\end{tabular}\vspace*{-8pt}
\end{center}
\end{table}}

\section{The potential with \texorpdfstring{\boldmath$U(1) \circ V_4$}{U(1)◦V4} symmetry}\label{App:U1_D4}

In Ref.~\cite{Kuncinas:2024zjq} we discussed a new potential, starting from the $U(1) \times \mathbb{Z}_2$-symmetric one,
\begin{equation}
\begin{aligned}
V_{U(1) \times \mathbb{Z}_2} ={}& \sum_i \mu_{ii}^2 h_{ii} + \sum_i \lambda_{iiii} h_{ii}^2 + \sum_{i<j} \lambda_{iijj} h_{ii} h_{jj} + \sum_{i<j} \lambda_{ijji} h_{ij} h_{ji}\\
& + \left\lbrace \lambda_{1212} h_{12}^2 + \mathrm{h.c.} \right\rbrace,
\end{aligned}
\end{equation}
and acting upon it with an additional $\mathbb{Z}_2:\,h_1 \leftrightarrow h_2$. The resulting potential is then:
\begin{equation}\label{Eq:V_U1_D4}
\begin{aligned}
V_{\widetilde{\mathcal{G}}} =& \mu_{11}^2 (h_{11} + h_{22}) + \mu_{33}^2 h_{33}\\
& + \lambda_{1111} (h_{11}^2 + h_{22}^2) + \lambda_{3333} h_{33}^2 + \lambda_{1122} h_{11} h_{22} + \lambda_{1133} (h_{11} h_{33} + h_{22} h_{33})\\
& + \lambda_{1221} h_{12} h_{21} + \lambda_{1331} (h_{13} h_{31} + h_{23} h_{32}) + \lambda_{1212} (h_{12}^2 + h_{21}^2).
\end{aligned}
\end{equation}
It is invariant under the symmetry transformations spanned by 
\begin{equation}\label{Eq:Gen_U1_D4}
{\widetilde{\mathcal{G}}} = \begin{aligned}
 \left\langle g_1,\, g_2,\, g_3 \right\rangle_\mathrm{HF} = 
\left\langle 
\begin{pmatrix}
e^{i \alpha} & 0 & 0\\
0 & e^{i \alpha} & 0\\
0 & 0 & 1
\end{pmatrix},\,
\begin{pmatrix}
-1 & 0 & 0\\
0 & 1 & 0\\
0 & 0 & 1
\end{pmatrix},\,
\begin{pmatrix}
0 & 1 & 0\\
1 & 0 & 0\\
0 & 0 & 1
\end{pmatrix} \right\rangle_\mathrm{HF}.
\end{aligned}
\end{equation}
The subset of generators $\{g_2,\, g_3\}$ spans a group of order eight, $ \mathbb{Z}_4 \rtimes \mathbb{Z}_2  \cong D_4$, which made us conclude that the underlying symmetry could be denoted by $U(1) \times D_4$.

In this appendix, we shall argue why $U(1) \times D_4$ does not reflect the true structure and why the symmetry of the potential should better be referred to by the short-hand notation $U(1) \circ V_4$, representing the class of central products of $V_4$ by a finite cyclic subgroup of $U(1)$, as explained below.

\subsection{Refining structures}

First of all, we note that the center of $D_4=\left\langle g_2,\, g_3 \right\rangle$ is given by $Z(D_4) \cong \mathbb{Z}_2$, where the center of a group $\mathcal{G}$ is defined by
\begin{equation}
Z(\mathcal{G}) = \{z \in \mathcal{G} \mid zg = gz, \, \forall \, g \in \mathcal{G}\}.
\end{equation}
The center of $D_4$ overlaps non-trivially with $U(1)$, since $ U(1) \cap D_4 = Z(D_4)$, namely the common subgroup $Z(D_4) \cong \mathbb{Z}_2$ acting on the $\{h_1,\,h_2\}$ doublets. To avoid double-counting the common $ \mathbb{Z}_2$, one should consider a quotient group. In particular we have
\begin{equation}
D_4/Z(D_4) \cong \mathbb{Z}_2 \times \mathbb{Z}_2 \cong V_4,
\end{equation}
while the quotient $U(1)/\mathbb{Z}_2$ is isomorphic to $U(1)$ itself.

It is worth noting that although $\{g_2,\, g_3\}$ resemble two $\mathbb{Z}_2$-type generators, within $D_4$ they generate a non-Abelian subgroup, while the quotient group $D_4/Z(D_4)$ is Abelian. Concretely, the quotient identifies rotations and reflections that differ only by the central $-\mathcal{I}_2$ (acting on $\{h_1,\,h_2\})$, so the resulting cosets commute.

We emphasise that the isomorphism \(D_4/Z(D_4)\cong V_4\) is an abstract group isomorphism. Concretely, in the original field representation the coset representatives are implemented by matrices that may differ by central elements (phases) belonging to the factored subgroup; hence the quotient corresponds to the action on fields modulo these central phases (a projective action). Accordingly, while we may continue to label convenient coset representatives by \(\{g_2,g_3\}\), one should keep in mind that the equality \(D_4/Z(D_4)=V_4\) refers to the abstract multiplication table of cosets. To realise the quotient as literal matrices acting on the fields one must choose explicit representatives, and products of representatives are then defined up to multiplication by elements of the factored center.

In Ref.~\cite{Kuncinas:2024zjq} we also noted that $Q_8$ (order 8) and the Pauli group $\mathcal{P}_1$ (order 16) are also symmetries of the scalar potential given by eq.~\eqref{Eq:V_U1_D4}. Let us consider the $\mathcal{P}_1$ group, which can be presented in terms of:
\begin{equation}
\mathcal{P}_1 = \left\langle 
\begin{pmatrix}
-1 & 0 & 0\\
0 & 1 & 0\\
0 & 0 & 1
\end{pmatrix},\,
\begin{pmatrix}
0 & 1 & 0\\
1 & 0 & 0\\
0 & 0 & 1
\end{pmatrix},\,
\begin{pmatrix}
0 & -i & 0\\
i & 0 & 0 \\
0 & 0 & 1
\end{pmatrix} \right\rangle_\mathrm{HF}.
\end{equation}
Its center is given by
\begin{equation}
Z(\mathcal{P}_1) = \{ \pm\mathcal{I}_2, \pm i\mathcal{I}_2 \} \cong \mathbb{Z}_4 \leq U(1).
\end{equation}
Again the intersection with $U(1)$ is non-trivial, and factoring it out results in:
\begin{equation}
\mathcal{P}_1 / Z(\mathcal{P}_1) \;\cong\; V_4.
\end{equation}

While starting from any of $ \mathcal{G} = \{D_4,\, Q_8,\, \mathcal{P}_1\}$, we are led to consider a quotient of the direct product along the common center
\begin{equation}\label{Eq:wt_G_U1_V4}
\widetilde{\mathcal{G}} \cong \left[ U(1) \times \mathcal{G} \right] / Z(\mathcal{G}) \cong U(1) \times V_4, 
\end{equation}
since the center $Z(\mathcal{G})$ is identified with a subgroup of $U(1)$.

\subsection{Identifying the underlying structure}

So far we mentioned several cases $\mathcal{G} = \{ D_4,\, Q_8,\, \mathcal{P}_1 \}$, all of which satisfy the structure of eq.~\eqref{Eq:wt_G_U1_V4} and yield the scalar potential of eq.~\eqref{Eq:V_U1_D4}. A natural question arises: what is the general structure of $\mathcal{G}$, that admits the same scalar potential. 

As a matter of fact, we are considering a central product $U(1) \circ_{Z(\mathcal{G})} \mathcal{G}$ (for example, $U(1) \circ_{\mathbb{Z}_2} D_4$) formed by identifying the subgroup of the $n^\mathrm{th}$ roots of unity in $U(1)$ with the center $Z(\mathcal{G}) \cong \mathbb{Z}_n$. In general, we are considering the following structure:
\begin{equation}\label{Eq:Def_tilde_G}
\widetilde{\mathcal{G}} \cong U(1) \circ_{Z(\mathcal{G})} \mathcal{G} \quad \text{with} \quad Z(\mathcal{G}) \cong \mathbb{Z}_n, ~ \mathcal{G}/Z(\mathcal{G}) \cong V_4,
\end{equation}
which ensures that the redundant $\mathbb{Z}_n$ (literally contained in $U(1)$) is not double-counted. One can identify an admissible $\mathcal{G}$ by considering an exact sequence
\begin{equation}
1 \longrightarrow Z(\mathcal{G}) \longrightarrow \mathcal{G} \longrightarrow V_4 \longrightarrow 1,
\end{equation}
where we require the kernel $Z(\mathcal G)$ to be central in $\mathcal G$. Such central extensions are classified (up to equivalence) by the second cohomology group,
\begin{equation}
H^2(V_4, \mathbb{Z}_n) \cong \mathrm{Hom}(\mathbb{Z}_2, \mathbb{Z}_n) \oplus \mathbb{Z}_{\mathrm{gcd}(2,n)},
\end{equation}
see Refs.~\cite{Brown1982,Karpilovsky1993,Weibel1994}. Note that, the cohomology classification may also produce extensions in which the normal subgroup is not exactly the center of $\mathcal{G}$. In such cases one must check explicitly that the subgroup is central and of the desired form $\mathbb{Z}_n$.

A useful consequence is that $H^2(V_4,\mathbb Z_n)$ is non-trivial only when $n$ is even. Thus, non-trivial finite lifts (like $D_4,\,Q_8,\,\mathcal P_1$) occur for even $n$. For odd $n$ the extension splits, $\mathcal G\cong\mathbb Z_n \times V_4$, and the scalar potential is automatically enlarged to the $U(1)\times U(1)$-symmetric 3HDM.

Having identified that $n$ should be even, we can further refine the structure. Writing $n=2^k m$ with $m$ odd, the odd factor contributes trivially to cohomology, so any admissible finite lift factorises as $\mathcal G \cong \mathbb Z_m \times \mathcal G_{2^k}$, where $\mathcal G_{2^k}$ denotes a (not necessarily unique) central extension of $V_4$ by $\mathbb Z_{2^k}$. Here the subscript $2^k$ labels the order of the central subgroup in the extension, rather than the order of the group itself. Concretely, for $k=1$ ($n=2$) one obtains the familiar non-split lifts $D_4$ or $Q_8$, while for $k=2$ ($n=4$) one recovers $\mathcal P_1$. 

At the continuous level, the full HF symmetry fits into the exact sequence
\begin{equation}
1\longrightarrow U(1)\longrightarrow\widetilde{\mathcal G}\longrightarrow V_4\longrightarrow 1,
\end{equation}
so the identity component of $\widetilde{\mathcal G}$ is $U(1)$ and the discrete quotient is $V_4$. When the extension class is non-trivial (even $n$), the lifts of the non-trivial elements of $V_4$ commute only up to a central phase in $U(1)$, \textit{i.e.}, their commutators lie in the identified cyclic subgroup of $U(1)$. When the intersection is trivial ($n$ odd), the extension splits and one obtains $\widetilde{\mathcal{G}} \cong U(1)\times V_4$, which, actually, results in the $U(1) \times \mathbb{Z}_2$-symmetric scalar potential due to the overall $U(1)$ symmetry.

To sum up: when the intersection is non-trivial (equivalently, when $n$ is even) the underlying structure can be described by the central product
\begin{equation}\label{Eq:Def_tilde_G_true}
\widetilde{\mathcal{G}} \cong U(1) \circ_{Z(\mathcal{G})} \mathcal{G} \quad \text{with} \quad Z(\mathcal{G}) \cong \mathbb{Z}_n, ~ \mathcal{G}/Z(\mathcal{G}) \cong V_4, ~ n \in 2\mathbb{Z}.
\end{equation}
Group-cohomologically, non-trivial extension classes are measured by \(H^2(V_4,\mathbb Z_n)\); at the continuous level the relevant obstruction to splitting is contained in $H^2(V_4,U(1))$ (which contains \(\mathbb Z_2\)). When the extension is non-trivial, the lifts of the non-trivial elements of $V_4$ commute only up to a central phase in $U(1)$, \textit{i.e.}, their commutators lie in the identified cyclic subgroup of $U(1)$.

\subsection[Explicit presence of \texorpdfstring{$V_4$}{V4}]{Explicit presence of \boldmath$V_4$}

In the previous subsections we identified that the structure of the potential permits the $V_4$ symmetry. While $V_4$ enters the scalar potential as a quotient group, $\mathcal{G}/Z(\mathcal{G}) \cong V_4$, we would like to elaborate on how it emerges.

We start by observing that due to $\mathbb{Z}_2:\, h_1 \leftrightarrow h_2$ and $U(1)$ acting on the $\{h_1,\, h_2\}$ pair, the $h_3$ doublet plays the role of a spectator: it does not mix with the others. Since at the level of the scalar potential $h_1$ and $h_2$ are indistinguishable (we are not considering vevs at this point), we can group them together into
\begin{equation}
\Phi = \begin{pmatrix}
h_1 \\
h_2
\end{pmatrix}.
\end{equation}
As a result, we are considering a mapping $V(h_1,\, h_2,\, h_3) \to V(\Phi,\, h_3)$, leading to
\begin{equation}
\begin{aligned}
V(\Phi,\, h_3) ={}& \mu_{11}^2 \Phi^\dagger \Phi + \mu_{33}^2 h_{33}\\
& + \lambda_a (\Phi^\dagger \Phi)^2  + \lambda_b (\Phi^\dagger \sigma_3 \Phi)^2 + \lambda_{1221} | \Phi^\dagger \sigma_+ \Phi |^2  + \lambda_{1212} \left( \left( \Phi^\dagger \sigma_+ \Phi \right)^2 + \mathrm{h.c.} \right)\\
& + \lambda_{1133} \Phi^\dagger \Phi h_{33} + \lambda_{1331} | \Phi^\dagger h_3 |^2 + \lambda_{3333} h_{33}^2,
\end{aligned}
\end{equation}
where
\begin{subequations}
\begin{align}
\lambda_a ={}& \frac{1}{2} \lambda_{1111} + \frac{1}{4} \lambda_{1122},\\
\lambda_b ={}& \frac{1}{2} \lambda_{1111} - \frac{1}{4} \lambda_{1122},
\end{align}
\end{subequations}
and we have
\begin{subequations}
\begin{align}
& \text{Pauli matrices:  } \sigma_1 = \begin{pmatrix}
0 & 1 \\
1 & 0
\end{pmatrix},~ \sigma_2 = \begin{pmatrix}
0 & -i \\
i & 0
\end{pmatrix},~ \sigma_3 = \begin{pmatrix}
1 & 0 \\
0 & -1
\end{pmatrix},\\
& \text{Ladder operator:  } \sigma_+ = \frac{1}{2}\left( \sigma_1 + i \sigma_2 \right) = \begin{pmatrix}
0 & 1\\
0 & 0
\end{pmatrix}.
\end{align}
\end{subequations}

The doublet $\Phi$ transforms under the global family group as:
\begin{equation}
\Phi \mapsto U \Phi, \quad U \in U(2).
\end{equation}
The potential is invariant when $ V(\Phi) = V(U\Phi)$.

Starting from the maximal $U(2)$ family symmetry of the kinetic term, various couplings reduce the group as follows (we are not considering $h_3$ at this step):
\begin{equation}\label{Eq:V_P1_Phi_h3}
U(2) \xrightarrow{\lambda_b} [U(1)_{h_1} \times U(1)_{h_2}]/U(1) \xrightarrow{\lambda_{1212}} U(1) \times \mathbb{Z}_2,			  
\end{equation}
where $\mathbb{Z}_2$ acts as a sign flip. As a result, the generators in the $\{h_1,\,h_2\}$ space can be represented by the coset representatives $\sigma_1$ and $\sigma_3$. Note that, the action of $\sigma_3 \sigma_1$ is identical to that of $\sigma_1 \sigma_3$ due to the global $U(1)$ symmetry acting on $\{h_1,\, h_2\}$, so while the group is spanned by $\mathcal{P}_1 = \left\langle \sigma_1,\, \sigma_2,\, \sigma_3 \right\rangle$, in reality we are considering $\mathcal{P}_1 / \mathbb{Z}_4 \cong V_4$, where $\mathbb{Z}_4$ is contained in the  global $U(1)$. This ensures the action on the potential is Abelian, even if the matrix representation is non-Abelian.

\subsection{Identifying other continuous symmetries}

In general, $\mathcal{G}$ of eq.~\eqref{Eq:Def_tilde_G_true} can be an arbitrarily large finite group. The structure of $\widetilde{\mathcal{G}}$ may suggest a continuous symmetry other than $U(1)$---a sequence of increasingly fine discrete symmetry subgroups $\mathbb{Z}_n$ with $n\to\infty$ can approximate a $U(1)$ at the level of the potential. Moreover, we point to the result of Ref.~\cite{Ivanov:2011ae}: the highest admissible cyclic factor for 3HDMs is $\mathbb{Z}_4$, and any $\mathbb{Z}_{n>4}$-invariant scalar potential cannot be distinguished from a $U(1)$-invariant one.

According to the Goldstone theorem, the number of massless Goldstone bosons equals the number of broken continuous generators,\textit{ i.e.}, the difference between the dimensions of the continuous parts of $\mathcal G$ and of the unbroken subgroup $\mathcal H$.

The physical spectrum of the $U(1) \times D_4$-symmetric 3HDM was analysed in Ref.~\cite{Kuncinas:2024zjq}. For a vacuum of the form $(v_1,\, v_2,\, v_3)$, with $v_i$ all different,  one must satisfy the condition:
\begin{equation}\label{Eq:Min_U1_D4}
\left( 2 \lambda_{1111} - \lambda_{1122} - 2 \lambda_{1212} - \lambda_{1221} \right) v_1 v_2 (v_1^2 - v_2^2) = 0.
\end{equation}
Requiring the vanishing of the first factor in eq.~\eqref{Eq:Min_U1_D4} corresponds to enhancing the symmetry of the potential to that of $O(2) \times U(1)$, which, however, is broken by the vacuum. We recall that $O(2) \cong SO(2) \rtimes \mathbb{Z}_2 \cong U(1) \rtimes \mathbb{Z}_2$. An equivalent condition corresponds to setting $\lambda_{1212}=0$, this can be easily identified by observing the symmetry breaking pattern of eq.~\eqref{Eq:V_P1_Phi_h3}. Therefore, we have two different bases for the same symmetry, $O(2) \times U(1)$:
\begin{subequations}
\begin{align}
\begin{split}\label{Eq:V_SO2_Z2_U1}
V_{\left[ SO(2) \rtimes \mathbb{Z}_2 \right] \times U(1) } ={}& \mu_{11}^2 (h_{11} + h_{22}) + \mu_{33}^2 h_{33}\\
& + \lambda_{1111} (h_{11}^2 + h_{22}^2) + \lambda_{3333} h_{33}^2 + \lambda_{1122} h_{11} h_{22}+ \lambda_{1133} (h_{11} h_{33} + h_{22} h_{33})\\
&   + \lambda_{1221} h_{12} h_{21}  + \lambda_{1331} (h_{13} h_{31} + h_{23} h_{32})\\
& + \frac{1}{2}\left( 2\lambda_{1111} - \lambda_{1122}  - \lambda_{1221}\right) \left( h_{12}^2 + h_{21}^2 \right),
\end{split}\\
\begin{split}
V_{\left[ U(1) \rtimes \mathbb{Z}_2 \right] \times U(1) } ={}& \mu_{11}^2 (h_{11} + h_{22}) + \mu_{33}^2 h_{33}\\
& + \lambda_{1111} (h_{11}^2 + h_{22}^2) + \lambda_{3333} h_{33}^2 + \lambda_{1122} h_{11} h_{22}\\
& + \lambda_{1133} (h_{11} h_{33} + h_{22} h_{33}) + \lambda_{1221} h_{12} h_{21} + \lambda_{1331} (h_{13} h_{31} + h_{23} h_{32}).
\end{split}
\end{align}
\end{subequations}

By solving the stationary-point equations we get several different solutions~\cite{Kuncinas:2024zjq}:
\begin{itemize}
\item {\boldmath${(v_1,\, v_2,\, v_3)}$}: one arrives at $V_{O(2) \times U(1)}$ broken by the vacuum, there are 3 massless neutral states;
\item {\boldmath${(v_1,\, v_1,\, v_3)}$}: there are 2 massless neutral states;
\item {\boldmath${(v_1,\, v_2,\, 0)}$}: one arrives at $V_{O(2) \times U(1)}$ broken by the vacuum, there are 2 massless neutral states, furthermore the neutral states from $h_3$ are mass degenerate;
\item {\boldmath${(v/\sqrt{2},\, v/\sqrt{2},\, 0)}$}: the neutral states from $h_3$ are mass degenerate;
\item {\boldmath${(0,\, v_2,\, v_3)}$}: there are 2 massless neutral states.
\end{itemize}
In the above list, one of the massless neutral states is associated with the would-be neutral Goldstone boson.

The only peculiar case is the ${(v_1,\, v_2,\, 0)}$ implementation. While the $U(1)$ symmetry is not broken, $\left\langle h_3 \right\rangle = 0$, there is an additional massless state present. This is easy to understand by considering the scalar potential $V(\Phi, h_3)$. For $\left\langle h_3 \right\rangle = 0$, there is no mixing between the neutral fields of $\{ h_1,\, h_2\}$ and $h_3$. Therefore, invariance of $\Phi$ under $U(1) \rtimes \mathbb{Z}_2 \cong O(2)$ creates a degenerate manifold vacuum, which is broken by the ${(v_1,\, v_2,\, 0)}$ implementation. Notice that for the ${(v/\sqrt{2},\, v/\sqrt{2},\, 0)}$ implementation there are no new massless states. This signals that there is an accidental continuous symmetry.

In the bilinear formalism~\cite{Nishi:2006tg,Maniatis:2006fs,Nishi:2007nh,Maniatis:2007vn,Ivanov:2010ww,Maniatis:2014oza} the presence of the accidental symmetry is manifest. In this formalism the scalar potential can be written as:
\begin{equation}
V = \Lambda_0 r_0^2 + L_i r_0 r_i + \Lambda_{ij} r_i r_j,
\end{equation}
where $r_i$ are the gauge-invariant bilinear combinations in the Gell-Mann basis. To ensure a basis-independent treatment of the symmetries of the model, we stick to the notation of Ref.~\cite{deMedeirosVarzielas:2019rrp}. Specifically, the quartic sector of the potential given by eq.~\eqref{Eq:V_U1_D4} is (see Ref.~\cite{Kuncinas:2024zjq}):
\begin{equation}
\begin{aligned}
\Lambda ={}&  \begin{pmatrix}
\Lambda_0 & L_i \\
L_i^\mathrm{T} & \Lambda_{ij}
\end{pmatrix}\\
={}&
\scriptstyle\begin{pmatrix}
 \lambda_a & 0 & 0 & 0 & 0 & 0 & 0 & 0 & \lambda_b \\
 0 & \lambda_{1221}+2\lambda_{1212} & 0 & 0 & 0 & 0 & 0 & 0 & 0 \\
 0 & 0 & \lambda_{1221}-2\lambda_{1212} & 0 & 0 & 0 & 0 & 0 & 0 \\
 0 & 0 & 0 & 2 \lambda_{1111}-\lambda_{1122} & 0 & 0 & 0 & 0 & 0 \\
 0 & 0 & 0 & 0 & \lambda_{1331} & 0 & 0 & 0 & 0 \\
 0 & 0 & 0 & 0 & 0 & \lambda_{1331} & 0 & 0 & 0 \\
 0 & 0 & 0 & 0 & 0 & 0 & \lambda_{1331} & 0 & 0 \\
 0 & 0 & 0 & 0 & 0 & 0 & 0 & \lambda_{1331} & 0 \\
 \lambda_b & 0 & 0 & 0 & 0 & 0 & 0 & 0 & \lambda_c
\end{pmatrix},
\end{aligned}
\end{equation}
where
\begin{subequations}
\begin{align}
\lambda_a ={}& \frac{1}{3} \left( 2 \lambda_{1111} + \lambda_{1122} + 2 \lambda_{1133} + \lambda_{3333} \right),\\
\lambda_b ={}& \frac{1}{3} \left( 2 \lambda_{1111} + \lambda_{1122} - \lambda_{1133} - 2\lambda_{3333} \right),\\
\lambda_c ={}& \frac{1}{3} \left( 2 \lambda_{1111} + \lambda_{1122} - 4 \lambda_{1133} + 4\lambda_{3333} \right).
\end{align}
\end{subequations}

The quartic coupling matrix $\Lambda_{ij}$ is diagonal, with the subspace $\lambda_{1331} (r_4^2 + r_5^2 + r_6^2 + r_7^2)$ completely decoupled from the remaining bilinears and the $L_i$ vector. Therefore this subspace forms a 4D sphere. As a consequence, the presence of an accidental $SO(4)$ symmetry is manifest. Note that, this $SO(4)$ does not act linearly on the scalar doublets.

\subsection[Final comments on \texorpdfstring{$U(1) \circ V_4$}{V4}]{Final comments on  \boldmath$U(1) \circ V_4$}

Let us recall that the scalar potential of eq.~\eqref{Eq:V_U1_D4} can be generated by eq.~\eqref{Eq:Def_tilde_G_true}:
\begin{equation*}
\widetilde{\mathcal{G}} \cong U(1) \circ_{Z(\mathcal{G})} \mathcal{G} \quad \text{with} \quad Z(\mathcal{G}) \cong \mathbb{Z}_n, ~ \mathcal{G}/Z(\mathcal{G}) \cong V_4, ~ n \in 2\mathbb{Z},
\end{equation*}

We start by picking four transformations acting on the scalar triplet, whose cosets correspond to the elements of the quotient group $\mathcal G/U(1)$. A convenient choice is to require invariance under
\begin{equation}
\left\lbrace
\mathcal{I}_3,\;
s=\begin{pmatrix}
0 & 1 & 0\\
1 & 0 & 0\\
0 & 0 & 1
\end{pmatrix},\;
t=\begin{pmatrix}
-1 & 0 & 0\\
0 & 1 & 0\\
0 & 0 & 1
\end{pmatrix},\;
u=\begin{pmatrix}
0 & -1 & 0\\
1 & 0 & 0 \\
0 & 0 & 1
\end{pmatrix}
\right\rbrace .
\end{equation}
The group laws are understood modulo the common $U(1)= \left\langle \mathrm{diag}(e^{i \alpha},\,e^{i \alpha},\,1  ) \right\rangle
$:
\begin{equation}
s^2 = t^2 = u^2 = e,\qquad st = ts = u .
\end{equation}
These relations define the quotient group $V_4$.

Different finite groups $\mathcal{G}$ provide different lifts of the same effective HF symmetry $V_4$, whose centers are absorbed by the common $U(1)$. As a concrete example, one may consider the $D_4$ lift
\begin{equation}
U(1) \circ_{D_4} V_4 \cong (U(1) \times D_4) / \mathbb{Z}_2^\mathrm{cent},
\end{equation}
where $\mathbb Z_2^{\mathrm{ cent}}$ identifies the $\pi$ rotation in $D_4$ with $\mathrm{diag}(-1,\,-1,\,1)\in U(1)$.

The scalar potential is insensitive to the particular choice of lift, it is only sensitive to the projective action
\begin{equation*}
(U(1)\times\mathcal G)/Z(\mathcal G) \cong U(1) \times V_4,
\end{equation*}
so that different admissible $\mathcal G$ in eq.~\eqref{Eq:Def_tilde_G_true} lead to the same potential. Although different $\mathcal G$ are constructed as extensions of $V_4$ by a central subgroup $\mathbb Z_n \leq U(1)$, this $U(1)$ is not the gauged $U(1)_Y$. Thus, $U(1)\times V_4$ can be viewed as a projective group, considering equivalence classes $[\mathcal G]$ under multiplication by central phases of $U(1)$, while the full underlying symmetry is captured by the central product $U(1)\circ V_4$.

The specific choice of $\mathcal G$ is not fixed by the scalar sector alone. It becomes physically meaningful once the Yukawa Lagrangian is introduced and the representation under which fermions transform is chosen. The scalar potential defines an equivalence class $[\mathcal G]$ of admissible lifts, while the Yukawa sector may select a particular representative within this equivalence class.

Throughout this work we use the abbreviated notation
\begin{equation*}
U(1)\circ V_4
\end{equation*}
to designate the class of central products (or, equivalently, central extensions) of $V_4$ by a finite cyclic subgroup of $U(1)$. Once a specific admissible lift $\mathcal G$ is fixed, the resulting HF symmetry is realised as a direct product $U(1)\times V_4$.

\section{Two different choices of the \texorpdfstring{\boldmath$S_3$}{S3} representations}\label{App:S3_sign}

We devote this appendix to the examination of two possible irreducible representations of the $S_3$ group, commenting on the sign ambiguity ($\pm$) of eq.~\eqref{Eq:S3_pm}.

The irreducible $S_3$ representations are given by $\mathbf{1} \oplus \mathbf{1^\prime} \oplus \mathbf{2}$. The most general renormalisable $S_3$-invariant potential in the singlet-doublet representation, $(h_S)_\mathbf{1} \oplus (h_1~h_2)^\mathrm{T}_{~~\mathbf{2}}$, can be written as~\cite{Pakvasa:1977in,Kubo:2004ps,Teshima:2012cg,Das:2014fea}:
\begin{equation}\label{Eq:V_S3_3HDM}
\begin{aligned}
V ={}& \mu_{11}^2 \left( h_{11} + h_{22} \right) + \mu_{SS}^2 h_{SS}\\
{}&+ \lambda_1 \left( h_{11} + h_{22} \right)^2 + \lambda_2 \left( h_{12} - h_{21} \right)^2+ \lambda_3 \left[ \left( h_{11} - h_{22} \right)^2 + \left( h_{12} + h_{21} \right)^2 \right] \\
&+ \left\{ \lambda_4 \left[ h_{S1} \left( h_{12} + h_{21} \right) + h_{S2} \left( h_{11} - h_{22} \right)\right] +  \mathrm{h.c.} \right\}+ \lambda_5  h_{SS}  \left( h_{11} + h_{22} \right)\\
& + \lambda_6 \left( h_{1S}  h_{S1} + h_{2S} h_{S2} \right)+ \left\{ \lambda_7 \left( h_{S1}^2 + h_{S2}^2  \right) +  \mathrm{h.c.} \right\} + \lambda_8 h_{SS}^2.
\end{aligned}
\end{equation}
Another possibility would be to consider the pseudosinglet-doublet representation~\cite{Emmanuel-Costa:2016vej}, $(h_A)_\mathbf{1^\prime} \oplus (h_1~h_2)^\mathrm{T}_{~~\mathbf{2}}$. This representation would yield an equivalent scalar potential up to the $\lambda_4$ coupling:
\begin{subequations}
\begin{align}
\mathbf{1} \oplus \mathbf{2}:~& \left\lbrace \lambda_4 \left[ h_{S1} \left( h_{12} + h_{21} \right) + h_{S2} \left( h_{11} - h_{22} \right)\right] +  \mathrm{h.c.} \right\rbrace, \\
\mathbf{1}^\prime \oplus \mathbf{2}:~& \left\lbrace \lambda_4^\prime \left[ h_{A2} \left( h_{12} + h_{21} \right) - h_{A1} \left( h_{11} - h_{22} \right)\right] +  \mathrm{h.c.}  \right\rbrace.
\end{align}
\end{subequations}
The scalar potentials are connected by a substitution of $h_S \leftrightarrow h_A$ along with an interchange of $h_1 \leftrightarrow h_2$. Consequently, the two representations are equivalent (at least in the scalar sector). As a result, the ``$\pm$" signs of eq.~\eqref{Eq:S3_pm} correspond to: ``-" for $\mathbf{1} \oplus \mathbf{2}$, while ``+" is for the $\mathbf{1^\prime} \oplus \mathbf{2}$ representation of $S_3$.

The scalar potential discussed in Section~\ref{Sec:G2} would then be a combination of both $\mathbf{1} \oplus \mathbf{2}$ and $\mathbf{1^\prime} \oplus \mathbf{2}$ up to the $\lambda_4$ term:
\begin{align}
V ={}& \mu_{11}^2 \left( h_{11} + h_{22} \right) + \mu_{33}^2 h_{33}, \nonumber \\
& + \lambda_1 \left( h_{11} + h_{22} \right)^2 + \lambda_2 \left( h_{12} - h_{21} \right)^2+ \lambda_3 \left[ \left( h_{11} - h_{22} \right)^2 + \left( h_{12} + h_{21} \right)^2 \right] \nonumber\\
& +  \left\lbrace \lambda_4 \left[ h_{31} \left( h_{12} + h_{21} \right) + h_{32} \left( h_{11} - h_{22} \right)  \right] +  \mathrm{h.c.} \right\rbrace \\
& + \left\lbrace \lambda_4^\prime \left[ h_{32} \left( h_{12} + h_{21} \right) - h_{31} \left( h_{11} - h_{22} \right) \right] +  \mathrm{h.c.}\right\rbrace\nonumber\\
& + \lambda_5    \left( h_{11} + h_{22} \right)h_{33} + \lambda_6 \left( h_{13}  h_{31} + h_{23} h_{32} \right)+ \left\lbrace \lambda_7 \left( h_{13}^2 + h_{23}^2   \right) +  \mathrm{h.c.} \right\rbrace + \lambda_8 h_{33}^2,\nonumber
\end{align}
where $h_3$ plays the role of both $h_S$ and $h_A$. In this form, the potential is identical to the CPc ($S_3 \times \mathrm{GCP}_{\theta = \pi}$) of Ref.~\cite{Bree:2024edl}. However, a redundant quartic term is present. 

\section{List of generators}\label{App:generators}

For the symmetry transformations given by eq.~\eqref{Eq:HF_G_mUUs} we define the following (not unique) generators, which are grouped according to the patterns of eigenvalues (the list of generators is also applicable to HF transformations, but not the discussion of the bilinear terms):

\begin{itemize}
\item $\mathbf{1_4 1_2 2_1}$:\\
$\circ V_{O(2)\times U(1)}$:\\

For generators
\begin{equation}
\left\langle \begin{pmatrix}
-\cos \theta & \sin \theta & 0\\
\sin \theta & \cos \theta & 0\\
0 & 0 & 1
\end{pmatrix}, \begin{pmatrix}
-1 & 0 & 0\\
0 & 1 & 0\\
0 & 0 & e^{i\theta}
\end{pmatrix} \right\rangle,
\end{equation}
we get $V_{O(2)_{[SO(2) \rtimes \mathbb{Z}_2]} \times U(1)}$ with $i\mu_{12}^2$. While for generators
\begin{equation}
\left\langle \begin{pmatrix}
-\cos \theta & \sin \theta & 0\\
\sin \theta & \cos \theta & 0\\
0 & 0 & 1
\end{pmatrix}, \begin{pmatrix}
1 & 0 & 0\\
0 & 1 & 0\\
0 & 0 & e^{i\theta}
\end{pmatrix} \right\rangle,
\end{equation}
we get $V_{O(2)_{[SO(2) \rtimes \mathbb{Z}_2]} \times U(1)}$ with $\mu_{ij}^2=0$.

These can be presented in another representation. For generator
\begin{equation}
\left\langle \begin{pmatrix}
0 & e^{i \theta_1} & 0 \\
e^{i \theta_2} & 0 & 0 \\
0 & 0 & 1
\end{pmatrix} \right\rangle,
\end{equation}
we get $V_{O(2)_{[U(1) \rtimes \mathbb{Z}_2]} \times U(1)}$ with $\mu_{22}^2=-\mu_{11}^2$. And for
\begin{equation}
\left\langle \begin{pmatrix}
0 & e^{i \theta_1} & 0 \\
e^{i \theta_2} & 0 & 0 \\
0 & 0 & 1
\end{pmatrix}, \begin{pmatrix}
e^{i\theta} & 0 & 0\\
0 & 1 & 0\\
0 & 0 & 1
\end{pmatrix} \right\rangle,
\end{equation}
we end up with $V_{O(2)_{[U(1) \rtimes \mathbb{Z}_2]} \times U(1)}$ and $\mu_{ij}^2=0$.

\item $\mathbf{1_4 4_1}$:\\

$\circ V_{U(1)\circ V_4}$:\\
For generators
\begin{equation}
\left\langle \begin{pmatrix}
0 & e^{i\theta} & 0\\
e^{i\theta} & 0 & 0\\
0 & 0 & 1
\end{pmatrix}, \begin{pmatrix}
0 & 1 & 0\\
-1 & 0 & 0\\
0 & 0 & 1
\end{pmatrix} \right\rangle,
\end{equation}
we get $V_{U(1) \circ V_4}$ with $\mu_{22}^2= - \mu_{11}^2$. While for 
\begin{equation}
\left\langle \begin{pmatrix}
-1 & 0 & 0\\
0 & 1 & 0\\
0 & 0 & 1
\end{pmatrix}, \begin{pmatrix}
0 & 1 & 0\\
-1 & 0 & 0\\
0 & 0 & 1
\end{pmatrix}, \begin{pmatrix}
1 & 0 & 0\\
0 & 1 & 0\\
0 & 0 & e^{i\theta}
\end{pmatrix} \right\rangle,
\end{equation}
we get $V_{U(1) \circ V_4}$ with $\mu_{ij}^2=0$. In both cases we have $\lambda_{1212} \in \mathbb{R}$.

\item $\mathbf{3_2 2_1}$:\\

$\circ V_{O(2)}$:\\
For generators
\begin{equation}
\left\langle \begin{pmatrix}
-\cos \theta & \sin \theta & 0\\
\sin \theta & \cos \theta & 0\\
0 & 0 & 1
\end{pmatrix} \right\rangle,
\end{equation}
we get $V_{O(2)_{[SO(2) \rtimes \mathbb{Z}_2]}}$ with $i\mu_{12}^2$, while for generators
\begin{equation}
\left\langle \begin{pmatrix}
-\cos \theta & \sin \theta & 0\\
\sin \theta & \cos \theta & 0\\
0 & 0 & 1
\end{pmatrix}, \begin{pmatrix}
1 & 0 & 0\\
0 & 1 & 0\\
0 & 0 & 1
\end{pmatrix} \right\rangle,
\end{equation}
we get $V_{O(2)_{[SO(2) \rtimes \mathbb{Z}_2]}}$ with $\mu_{ij}^2=0$.

These can be presented as generators in another representation, given by
\begin{equation}
\left\langle \begin{pmatrix}
0 & e^{-i \theta} & 0 \\
e^{i \theta} & 0 & 0 \\
0 & 0 & 1
\end{pmatrix} \right\rangle,
\end{equation}
yielding $V_{O(2)_{[U(1) \rtimes \mathbb{Z}_2]}}$ with $\mu_{22}^2=-\mu_{11}^2$. Also, generators 
\begin{equation}
\left\langle \begin{pmatrix}
0 & e^{-i \theta} & 0 \\
e^{i \theta} & 0 & 0 \\
0 & 0 & 1
\end{pmatrix}, \begin{pmatrix}
1 & 0 & 0\\
0 & 1 & 0\\
0 & 0 & 1
\end{pmatrix} \right\rangle,
\end{equation}
result in $V_{O(2)_{[U(1) \rtimes \mathbb{Z}_2]} }$ with $\mu_{ij}^2=0$.

$\circ V_{S_3}$:\\

For generators
\begin{equation}
\left\langle \begin{pmatrix}
0 & 1 & 0\\
1 & 0 & 0\\
0 & 0 & 1
\end{pmatrix}, \begin{pmatrix}
e^{2 i \pi/3} & 0 & 0\\
0 & e^{-2 i \pi/3} & 0\\
0 & 0 & 1
\end{pmatrix} \right\rangle,
\end{equation}
we get $V_{S_3}$ with $\mu_{ij}^2=0$, while for generators
\begin{equation}
\left\langle \begin{pmatrix}
0 & 1 & 0\\
1 & 0 & 0\\
0 & 0 & 1
\end{pmatrix}, \begin{pmatrix}
0 & e^{-2 i \pi/3} & 0\\
e^{2 i \pi/3} & 0 & 0\\
0 & 0 & 1
\end{pmatrix} \right\rangle,
\end{equation}
we get $V_{S_3}$ with $\mu_{22}^2=\mu_{11}^2$.

\item $\mathbf{2_2 4_1}$:\\

$\circ V_{U(1) \times \mathbb{Z}_2}$:\\

For generators
\begin{equation}
\left\langle \begin{pmatrix}
-1 & 0 & 0\\
0 & 1 & 0\\
0 & 0 & 1
\end{pmatrix}, \begin{pmatrix}
1 & 0 & 0\\
0 & 1 & 0\\
0 & 0 & e^{i \theta}
\end{pmatrix} \right\rangle,
\end{equation}
we get $V_{U(1) \times \mathbb{Z}_2}$ with $\mu_{ij}^2=0$. While for 
\begin{equation}
\left\langle \begin{pmatrix}
1 & 0 & 0\\
0 & -1 & 0\\
0 & 0 & e^{i \theta}
\end{pmatrix}\right\rangle,
\end{equation}
we get $V_{U(1) \times \mathbb{Z}_2}$ with $\mu_{12}^2$.

\newpage
$\circ V_{\mathbb{Z}_4}$:\\

For
\begin{equation}
\left\langle \begin{pmatrix}
-1 & 0 & 0\\
0 & -1 & 0\\
0 & 0 & 1
\end{pmatrix}, \begin{pmatrix}
0 & 1 & 0\\
-1 & 0 & 0\\
0 & 0 & 1
\end{pmatrix} \right\rangle,
\end{equation}
we get $V_{\mathbb{Z}_4}$ with $\mu_{ij}^2=0$. While for
\begin{equation}
\left\langle \begin{pmatrix}
1 & 0 & 0\\
0 & -1 & 0\\
0 & 0 & i
\end{pmatrix}\right\rangle,
\end{equation}
we get $V_{\mathbb{Z}_4}$ with $\mu_{12}^2$.

$\circ V_{D_4}$:\\

For generators
\begin{equation}
\left\langle \begin{pmatrix}
0 & 1 & 0\\
1 & 0 & 0\\
0 & 0 & 1
\end{pmatrix}, \begin{pmatrix}
0 & 1 & 0\\
-1 & 0 & 0\\
0 & 0 & 1
\end{pmatrix}, \begin{pmatrix}
1 & 0 & 0\\
0 & 1 & 0\\
0 & 0 & 1
\end{pmatrix} \right\rangle,
\end{equation}
we get $V_{D_4}$ with $\mu_{ij}^2=0$. While for
\begin{equation}
\left\langle \begin{pmatrix}
0 & 1 & 0\\
1 & 0 & 0\\
0 & 0 & 1
\end{pmatrix}, \begin{pmatrix}
0 & 1 & 0\\
-1 & 0 & 0\\
0 & 0 & 1
\end{pmatrix}\right\rangle,
\end{equation}
we end up with $V_{D_4}$ with $\mu_{22}^2=-\mu_{11}^2$.

\item $\mathbf{8_1}$:\\

$\circ V_{\mathbb{Z}_2 \times \mathbb{Z}_2}$:\\
For
\begin{equation}
\left\langle \begin{pmatrix}
-1 & 0 & 0\\
0 & 1 & 0\\
0 & 0 & 1
\end{pmatrix}, \begin{pmatrix}
1 & 0 & 0\\
0 & -1 & 0\\
0 & 0 & 1
\end{pmatrix} \right\rangle,
\end{equation}
we get $V_{\mathbb{Z}_2 \times \mathbb{Z}_2}$ with $\mu_{12}^2$, while for 
\begin{equation}
\left\langle \begin{pmatrix}
-1 & 0 & 0\\
0 & 1 & 0\\
0 & 0 & 1
\end{pmatrix}, \begin{pmatrix}
1 & 0 & 0\\
0 & -1 & 0\\
0 & 0 & 1
\end{pmatrix},  \begin{pmatrix}
1 & 0 & 0\\
0 & 1 & 0\\
0 & 0 & 1
\end{pmatrix} \right\rangle,
\end{equation}
we get $V_{\mathbb{Z}_2 \times \mathbb{Z}_2}$ with $\mu_{ij}^2=0$. 

$\circ V_{\mathbb{Z}_2}$:\\
For generators
\begin{equation}
\left\langle \begin{pmatrix}
-1 & 0 & 0\\
0 & 1 & 0\\
0 & 0 & 1
\end{pmatrix} \right\rangle,
\end{equation}
we get $V_{\mathbb{Z}_2}$ with $\{\mu_{12}^2,\, \mu_{13}^2\}$. For 
\begin{equation}
\left\langle \begin{pmatrix}
-1 & 0 & 0\\
0 & 1 & 0\\
0 & 0 & 1
\end{pmatrix}, \begin{pmatrix}
1 & 0 & 0\\
0 & 1 & 0\\
0 & 0 & 1
\end{pmatrix} \right\rangle,
\end{equation}
we get $V_{\mathbb{Z}_2}$ with $\mu_{ij}^2=0$.

\end{itemize}

\section{Different T-GOOFy-symmetric potentials}\label{App:Gen_TGOOFy}

Here, we collect different quartic potentials, which can be expressed as $\mathcal{G}_\mathrm{GOOFy} \ast \mathcal{G}_\text{HF\,/\,GCP}$. The form of the underlying symmetry groups can be deceiving since several different transformations can lead to identical quartic potentials.

\subsection{HF-like invariant potentials}\label{App:HF_TGOOFy}

Categorised in terms of the eigenvalues patterns, different identified scalar potentials are:
\begin{itemize}

\item $\mathbf{1_8}$

$\circ$ For $\mathcal{G}_1 \ast [SU(3)]$ we get $V=0$.

\item $\mathbf{1_7 1_1}$

$\circ$ For $\mathcal{G}_1 \ast [U(2)]_{(h_2{-}h_3)}$ we get
\begin{equation}
\begin{aligned}
V ={}& \lambda_{1111} h_{11}^2.
\end{aligned}
\end{equation}

\item $\mathbf{1_6 1_2}$

This pattern of eigenvalues is unique in terms of the classification presented in Table~\ref{Table:Bilinear_Patterns_Cases}.

$\circ$ For $\mathcal{G}_1 \ast [[U(1) \times U(1)] \rtimes S_3]$ we get
\begin{equation}
\begin{aligned}
V ={}& \lambda_{1111} (h_{11}^2 + h_{22}^2 + h_{33}^2).
\end{aligned}
\end{equation}

\item $\mathbf{1_6 2_1}$

This pattern of eigenvalues is unique in terms of the classification presented in Table~\ref{Table:Bilinear_Patterns_Cases}.

$\circ$ For $\mathcal{G}_1 \ast [O(2) \times U(1)]_{(h_2{-}h_3)}$ (as $U(1) \rtimes \mathbb{Z}_2$) we get 
\begin{equation}
\begin{aligned}
V ={}& \lambda_{1111} h_{11}^2 + \lambda_{2222} (h_{22}^2 + h_{33}^2).
\end{aligned}
\end{equation}

$\circ$ For $[\mathcal{G}_1 \times \mathcal{G}_1] \ast [U(1) \times U(1)]$ we get 
\begin{equation}
\begin{aligned}
V ={}& \lambda_{1111} h_{11}^2 + \lambda_{2222} h_{22}^2 + \lambda_{3333} h_{33}^2.
\end{aligned}
\end{equation}

$\circ$ For $\mathcal{G}_4 \ast [O(2) \times U(1)]_{(h_2{-}h_3)}$ (as $U(1) \rtimes \mathbb{Z}_2$) we get
\begin{equation}
\begin{aligned}
V ={}& \lambda_{1111} (h_{11}^2 - h_{22}^2 - h_{33}^2).
\end{aligned}
\end{equation}

\item $\mathbf{1_5 1_3},~\mathbf{|1_8|}$

$\circ$ For $\mathcal{G}_1 \ast SO(3)$ we get
\begin{equation}
\begin{aligned}
V ={}& \lambda_{1111} (h_{11}^2 + h_{22}^2 + h_{33}^2 + h_{12}^2 + h_{21}^2 + h_{13}^2 + h_{31}^2 + h_{23}^2 + h_{32}^2).
\end{aligned}
\end{equation}

\item $\mathbf{1_4 1_3 1_1}$

\begin{itemize}

 \item $\mathbf{|1_7 1_1|}$
 
$\circ$ For $\mathcal{G}_{6} \ast SU(2)$ we get
\begin{equation}
\begin{aligned}
V ={}& \lambda_{1111} (h_{11} + h_{22} - h_{33})^2 \\
& + \lambda_{1122} ( -h_{12}h_{21} + h_{13}h_{31} +h_{23}h_{32} + h_{11} h_{22} - h_{11} h_{33} - h_{22} h_{33}).
\end{aligned}
\end{equation}

 \item $\mathbf{|1_4 1_3 1_1|}$
 
$\circ$ For $\mathcal{G}_1 \ast SU(2)$ we get
\begin{equation}
\begin{aligned}
V ={}& \lambda_{1111} (h_{11} + h_{22})^2 + \lambda_{3333} h_{33}^2 + \lambda_{1221}( h_{12}h_{21} - h_{11}h_{22} ) .
\end{aligned}
\end{equation}

$\circ$ For $\mathcal{G}_4 \ast SU(2)$ we get
\begin{equation}
\begin{aligned}
V ={}& \lambda_{1122}( h_{11}h_{22} - h_{12}h_{21}) + \lambda_{3333} h_{33}^2.
\end{aligned}
\end{equation}
 
\end{itemize}

\item $\mathbf{1_4 1_2 2_1}$

\begin{itemize}

\item  $\mathbf{|1_6 2_1|}$

$\circ$ For $\mathcal{G}_6 \ast [O(2) \times U(1)]$ (as $U(1) \rtimes \mathbb{Z}_2$) we get
\begin{equation}
\begin{aligned}
V ={}& \lambda_{1111} (h_{11}^2 + h_{22}^2 + h_{33}^2) + \lambda_{1122} ( - h_{11} h_{22} + h_{11} h_{33} + h_{22} h_{33}) \\
& + \lambda_{1221}(-h_{12}h_{21} + h_{13}h_{31} +h_{23}h_{32}).
\end{aligned}
\end{equation}

\item  $\mathbf{|1_4 1_3 1_1|}$

$\circ$ For $\mathcal{G}_1 \ast [O(2) \times U(1)]_{(h_2{-}h_3)}$ we get
\begin{equation}
\begin{aligned}
V ={}& \lambda_{1111} h_{11}^2 + \lambda_{2222} (h_{22}^2 + h_{33}^2 + h_{23}^2 + h_{32}^2).
\end{aligned}
\end{equation}

\item  $\mathbf{|1_4 1_2 2_1|}$

$\circ$ For $\mathcal{G}_1 \ast [U(1) \times U(1)]$ we get
\begin{equation}
\begin{aligned}
V ={}& \lambda_{1111} h_{11}^2 + \lambda_{2222} h_{22}^2 + \lambda_{3333} h_{33}^2 + \lambda_{1221} h_{12} h_{21} + \lambda_{1122}h_{11}h_{22}.
\end{aligned}
\end{equation}

$\circ$ For $\mathcal{G}_1 \ast [O(2) \times U(1)]$ we get
\begin{equation}
\begin{aligned}
V ={}& \lambda_{1111} (h_{11}^2 +  h_{22}^2) + \lambda_{3333} h_{33}^2 + \lambda_{1221} h_{12} h_{21} + \lambda_{1122}h_{11}h_{22}.
\end{aligned}
\end{equation}

$\circ$ For $\mathcal{G}_4 \ast [U(1) \times U(1)]$ we get
\begin{equation}
\begin{aligned}
V ={}& \lambda_{1111} (h_{11}^2 -  h_{22}^2) + \lambda_{3333} h_{33}^2 + \lambda_{1221} h_{12} h_{21} + \lambda_{1122}h_{11}h_{22}.
\end{aligned}
\end{equation}

$\circ$ For $\mathcal{G}_4 \ast [O(2) \times U(1)]$ (as $U(1) \rtimes \mathbb{Z}_2$) we get
\begin{equation}
\begin{aligned}
V ={}& \lambda_{1122} h_{11}h_{22} + \lambda_{1221} h_{12}h_{21} +  \lambda_{3333} h_{33}^2.
\end{aligned}
\end{equation}

\end{itemize}

\item $\mathbf{1_4 4_1}$

\begin{itemize}

\item  $\mathbf{|1_4 1_2 2_1|}$

$\circ$ For $\mathcal{G}_1 \ast [U(1) \circ V_4]_{(h_2{-}h_3)}$ we get
\begin{equation}
\begin{aligned}
V ={}& \lambda_{1111} h_{11}^2  + \lambda_{3333}(h_{22}^2 + h_{33}^2)  + \lambda_{2323} (h_{23}^2 + h_{32}^2).
\end{aligned}
\end{equation}

$\circ$ For $[ \mathcal{G}_1 \times \mathcal{G}_1 ] \ast [U(1) \times \mathbb{Z}_2]$ we get
\begin{equation}
\begin{aligned}
V ={}& \lambda_{1111} h_{11}^2 + \lambda_{2222} h_{22}^2 + \lambda_{3333} h_{33}^2 +  \{ \lambda_{1212} h_{12}^2 + \mathrm{h.c.} \}.
\end{aligned}
\end{equation}

$\circ$ For $[\mathcal{G}_{1} \times \mathcal{G}_4] \ast [U(1) \times \mathbb{Z}_2]_{(h_2{-}h_3)}$ we get
\begin{equation}
\begin{aligned}
V ={}& \lambda_{1111} (h_{11}^2 - h_{22}^2) + \lambda_{3333} h_{33}^2 + \lambda_{1212} (h_{12}^2 + h_{21}^2).
\end{aligned}
\end{equation}

\item  $\mathbf{|1_4 4_1|}$

$\circ$ For $\mathcal{G}_1 \ast [U(1)]_{h_3}$ we get
\begin{equation}
\begin{aligned}
V ={}& \lambda_{1111} h_{11}^2 + \lambda_{2222} h_{22}^2 + \lambda_{3333} h_{33}^2 + \lambda_{1221} h_{12} h_{21} + \lambda_{1122}h_{11}h_{22} \\ & + \bigg\{ \lambda_{1112} h_{11}h_{12} + \lambda_{1222} h_{12}h_{22} + \lambda_{1212} h_{12}^2  + \mathrm{h.c.} \bigg\}.
\end{aligned}
\end{equation}

$\circ$ For $\mathcal{G}_1 \ast [U(1) \times \mathbb{Z}_2]$ we get
\begin{equation}
\begin{aligned}
V ={}& \lambda_{1111} h_{11}^2 + \lambda_{2222} h_{22}^2 + \lambda_{3333} h_{33}^2 + \lambda_{1221} h_{12} h_{21} + \lambda_{1122}h_{11}h_{22} \\ & + \bigg\{ \lambda_{1212} h_{12}^2  + \mathrm{h.c.} \bigg\}.
\end{aligned}
\end{equation}

$\circ$ For $\mathcal{G}_1 \ast [U(1) \circ V_4]$ we get
\begin{equation}
\begin{aligned}
V ={}& \lambda_{1111} (h_{11}^2 + h_{22}^2) + \lambda_{3333} h_{33}^2 + \lambda_{1221} h_{12} h_{21} + \lambda_{1122}h_{11}h_{22} \\ & +  \lambda_{1212} (h_{12}^2 + h_{21}^2).
\end{aligned}
\end{equation}

$\circ$ For $\mathcal{G}_4 \ast [U(1) \times \mathbb{Z}_2]$ we get
\begin{equation}
\begin{aligned}
V ={}& \lambda_{1111} (h_{11}^2 - h_{22}^2) + \lambda_{3333} h_{33}^2 + \lambda_{1221} h_{12} h_{21} + \lambda_{1122}h_{11}h_{22}\\ 
& + \lambda_{1212} (h_{12}^2 + h_{21}^2).
\end{aligned}
\end{equation}

$\circ$ For $\mathcal{G}_4 \ast [U(1) \circ V_4]$ we get
\begin{equation}
\begin{aligned}
V ={}& \lambda_{3333} h_{33}^2 + \lambda_{1221} h_{12} h_{21} + \lambda_{1122}h_{11}h_{22} +  \lambda_{1212} (h_{12}^2 + h_{21}^2).
\end{aligned}
\end{equation}

\end{itemize}

\item $\mathbf{2_3 1_2},~\mathbf{|1_6 1_2|}$

$\circ$ For $\mathcal{G}_1 \ast A_4$ we get
\begin{equation}
\begin{aligned}
V ={}& \lambda_{1111} (h_{11}^2 + h_{22}^2 + h_{33}^2) +  \left\lbrace\lambda_{1212}( h_{12}^2 + h_{13}^2 + h_{23}^2 )  + \mathrm{h.c.} \right\rbrace.
\end{aligned}
\end{equation}

$\circ$ For $\mathcal{G}_1 \ast S_4$ we get
\begin{equation}
\begin{aligned}
V ={}& \lambda_{1111} (h_{11}^2 + h_{22}^2 + h_{33}^2) + \lambda_{1212}( h_{12}^2 + h_{13}^2 + h_{23}^2  + \mathrm{h.c.} ).
\end{aligned}
\end{equation}

\item $\mathbf{3_2 2_1}$

\begin{itemize}

\item $\mathbf{|1_4 3_1 1_1|}$

$\circ$ For $\mathcal{G}_{6} \ast SO(2)$ we get
\begin{equation}
\begin{aligned}
V ={}& \lambda_{1111} (h_{11} + h_{22} - h_{33})^2\\
& + \lambda_{1122} ( - h_{11} h_{22} + h_{11} h_{33} + h_{22} h_{33}-h_{12}h_{21} + h_{13}h_{31} +h_{23}h_{32})\\
& + \lambda_{1212}(h_{12}^2 +h_{13}^2 +h_{23}^2 - 2h_{11}h_{22} +2 h_{11}h_{33} + 2 h_{22}h_{33}).
\end{aligned}
\end{equation}

\item $\mathbf{|1_4 1_2 2_1|}$

$\circ$ For $\mathcal{G}_1 \ast U(1)_1$ we get
\begin{equation}
\begin{aligned}
V ={}& \lambda_{1111} h_{11}^2 + \lambda_{2222} h_{22}^2 + \lambda_{3333} h_{33}^2 + \lambda_{1122} h_{11} h_{22} + \lambda_{1221} h_{12} h_{21}\\
&  + \left\lbrace  \lambda_{1323}h_{13}h_{23}  + \mathrm{h.c.} \right\rbrace .
\end{aligned}
\end{equation}
An equivalent approach would be to consider $\mathcal{G}_1 \times SO(2)$ (and not $O(2)$),
\begin{equation}
\begin{aligned}
V ={}& \lambda_{1111} (h_{11}^2 + h_{22}^2) + \lambda_{1122} h_{11} h_{22} + \lambda_{1221} h_{12} h_{21} + \Lambda (h_{12}^2 + h_{21}^2)\\
& + \lambda_{3333} h_{33}^2 + i \lambda_{1112} ( h_{11} + h_{22})(h_{12} - h_{21})\\
& + \left\lbrace  \lambda_{1313}( h_{13}^2 +h_{23}^2)  + \mathrm{h.c.} \right\rbrace .
\end{aligned}
\end{equation}
By utilising the basis transformation of $\mathcal{R}_{(\frac{\pi}{4},\, -\frac{\pi}{4},\, \frac{\pi}{4},\, 0)}$ we arrive at the identical scalar potential.

$\circ$  For $\mathcal{G}_1 \ast  [U(1) \rtimes \mathbb{Z}_2]$ we get
\begin{equation}
\begin{aligned}
V ={}& \lambda_{1111} (h_{11}^2 + h_{22}^2) + \lambda_{3333} h_{33}^2 + \lambda_{1122} h_{11} h_{22} + \lambda_{1221} h_{12} h_{21}\\
&  + \left\lbrace  \lambda_{1323}h_{13}h_{23}  + \mathrm{h.c.} \right\rbrace .
\end{aligned}
\end{equation}

$\circ$ For $\mathcal{G}_1 \ast  [SO(2) \rtimes \mathbb{Z}_2]_{(h_2{-}h_3)}$ we get
\begin{equation}
\begin{aligned}
V ={}& \lambda_{1111} h_{11}^2 + \lambda_{2222} (h_{22}^2 +h_{33}^2 + h_{23}^2 + h_{32}^2)\\&  + \left\lbrace  \lambda_{1212} ( h_{12}^2 + h_{13}^2) + \mathrm{h.c.} \right\rbrace .
\end{aligned}
\end{equation}

$\circ$ For $\mathcal{G}_3 \ast SO(2)$ we get
\begin{equation}
\begin{aligned}
V ={}& \lambda_{1111} (h_{11}^2 + h_{22}^2) + \lambda_{1122} h_{11} h_{22} + \lambda_{1221} h_{12} h_{21} + \Lambda (h_{12}^2 + h_{21}^2)\\& + \lambda_{3333} h_{33}^2  + i \lambda_{1233} (h_{12} - h_{21}) h_{33}\\& + i \lambda_{1332} (h_{23}h_{31} - h_{13}h_{32})   + \left\lbrace  \lambda_{1313}( h_{13}^2 +h_{23}^2)  + \mathrm{h.c.} \right\rbrace .
\end{aligned}
\end{equation}
By utilising the basis transformation of $\mathcal{R}_{(\frac{\pi}{4},\, -\frac{\pi}{4},\, \frac{\pi}{4},\, 0)}$ we arrive at
\begin{equation}
\begin{aligned}
V ={}& \lambda_{1111} (h_{11}^2 + h_{22}^2) + \lambda_{1122} h_{11} h_{22} + \lambda_{1221} h_{12} h_{21} + \lambda_{3333} h_{33}^2\\
& + \lambda_{1133} (h_{22} - h_{11}) h_{33} + \lambda_{1331} (h_{13}h_{31} - h_{23}h_{32}) \\& + \left\lbrace  \lambda_{1323} h_{13}h_{23}  + \mathrm{h.c.} \right\rbrace .
\end{aligned}
\end{equation}

$\circ$ For $\mathcal{G}_3 \ast [SO(2) \times U(1)]$ we get
\begin{equation}
\begin{aligned}
V ={}& \lambda_{1111} (h_{11}^2 + h_{22}^2) + \lambda_{1122} h_{11} h_{22} + \lambda_{1221} h_{12} h_{21} + \Lambda (h_{12}^2 + h_{21}^2)\\& + \lambda_{3333} h_{33}^2  + i \lambda_{1233} (h_{12} - h_{21}) h_{33}\\& + i \lambda_{1332} (h_{23}h_{31} - h_{13}h_{32}).
\end{aligned}
\end{equation}
By utilising the basis transformation of  $\mathcal{R}_{(\frac{\pi}{4},\, -\frac{\pi}{4},\, \frac{\pi}{4},\, 0)}$ we arrive at
\begin{equation}
\begin{aligned}
V ={}& \lambda_{1111} (h_{11}^2 + h_{22}^2) + \lambda_{1122} h_{11} h_{22} + \lambda_{1221} h_{12} h_{21} + \lambda_{3333} h_{33}^2\\
& + \lambda_{1133} (h_{22} - h_{11}) h_{33} + \lambda_{1331} (h_{13}h_{31} - h_{23}h_{32}).
\end{aligned}
\end{equation}

$\circ$ For $\mathcal{G}_4 \ast U(1)_1$ we get
\begin{equation}
\begin{aligned}
V ={}& \lambda_{1111} (h_{11}^2 - h_{22}^2) + \lambda_{1122} h_{11} h_{22} + \lambda_{1221} h_{12} h_{21} + \lambda_{3333} h_{33}^2\\
& + \left\lbrace  \lambda_{1323} h_{13}h_{23}  + \mathrm{h.c.} \right\rbrace .
\end{aligned}
\end{equation}

$\circ$ For $\mathcal{G}_4 \ast O(2)$ (as $U(1) \rtimes \mathbb{Z}_2$) we get
\begin{equation}
\begin{aligned}
V ={}& \lambda_{1122} h_{11} h_{22} + \lambda_{1221} h_{12} h_{21} + \lambda_{3333} h_{33}^2 + \left\lbrace  \lambda_{1323} h_{13}h_{23}  + \mathrm{h.c.} \right\rbrace.
\end{aligned}
\end{equation}

\end{itemize}

\item $\mathbf{2_2 4_1}$

\begin{itemize}

\item $\mathbf{|1_4 1_2 2_1|}$

$\circ$ For $\mathcal{G}_1 \ast [D_4]_{(h_2{-}h_3)}$ we get
\begin{equation}
\begin{aligned}
V ={}& \lambda_{1111} h_{11}^2 + \lambda_{2222} (h_{22}^2 + h_{33}^2) + \lambda_{2323} (h_{23}^2+ h_{32}^2)\\
& + \left\lbrace \lambda_{1313} (h_{12}^2 + h_{13}^2) + \mathrm{h.c.} \right\rbrace.
\end{aligned}
\end{equation}

\item $\mathbf{|1_4 4_1|}$

$\circ$ For $\mathcal{G}_1 \ast \mathbb{Z}_4$ we get
\begin{equation}
\begin{aligned}
V ={}& \lambda_{1111} h_{11}^2 + \lambda_{2222} h_{22}^2 + \lambda_{3333} h_{33}^2 + \lambda_{1221} h_{12} h_{21} + \lambda_{1122}h_{11}h_{22} \\
& + \left\lbrace \lambda_{1323} h_{13}h_{23} + \lambda_{1212} h_{12}^2 + \mathrm{h.c.} \right\rbrace.
\end{aligned}
\end{equation}

$\circ$ For $\mathcal{G}_1 \ast D_4$ we get
\begin{equation}
\begin{aligned}
V ={}& \lambda_{1111} (h_{11}^2 + h_{22}^2) + \lambda_{3333} h_{33}^2 + \lambda_{1221} h_{12} h_{21} + \lambda_{1122}h_{11}h_{22} \\
& + \lambda_{1212} (h_{12}^2+ h_{21}^2) + \left\lbrace \lambda_{1323} h_{13}h_{23} + \mathrm{h.c.} \right\rbrace.
\end{aligned}
\end{equation}

$\circ$ For $\mathcal{G}_3 \ast U(1)_{h_3}$ we get
\begin{equation}
\begin{aligned}
V ={}& \lambda_{1111} h_{11}^2 + \lambda_{2222} h_{22}^2 + \lambda_{3333} h_{33}^2 + \lambda_{1221} h_{12} h_{21} + \lambda_{1122}h_{11}h_{22}\\
& + \left\lbrace \lambda_{1332} h_{13}h_{32} + \lambda_{1233} h_{12}h_{33} + \lambda_{1212} h_{12}^2 + \mathrm{h.c.} \right\rbrace.
\end{aligned}
\end{equation}

$\circ$ For $\mathcal{G}_3 \ast [U(1) \times \mathbb{Z}_2]$ we get
\begin{equation}
\begin{aligned}
V ={}& \lambda_{1111} (h_{11}^2 + h_{22}^2) + \lambda_{3333} h_{33}^2 + \lambda_{1221} h_{12} h_{21} + \lambda_{1122}h_{11}h_{22}\\
& + \lambda_{1332} (h_{13}h_{32} + h_{23}h_{31}) + \lambda_{1233} (h_{12} + h_{21})h_{33} + \lambda_{1212} (h_{12}^2+ h_{21}^2).
\end{aligned}
\end{equation}

$\circ$ For $\mathcal{G}_4 \ast U(1)_{h_3}$ we get
\begin{equation}
\begin{aligned}
V ={}& \lambda_{1111}(h_{11}^2 - h_{22}^2 ) + \lambda_{3333} h_{33}^2 + \lambda_{1221} h_{12} h_{21} + \lambda_{1122}h_{11}h_{22} \\
& + \lambda_{1212} (h_{12}^2 + h_{21}^2 ) + \lambda_{1332} (h_{13}h_{32} + h_{23}h_{31}) + \lambda_{1233} (h_{12} + h_{12}) h_{33}.
\end{aligned}
\end{equation}

$\circ$ For $\mathcal{G}_4 \ast \mathbb{Z}_4$ we get
\begin{equation}
\begin{aligned}
V ={}& \lambda_{1111}(h_{11}^2 - h_{22}^2 ) + \lambda_{3333} h_{33}^2 + \lambda_{1221} h_{12} h_{21} + \lambda_{1122}h_{11}h_{22}  \\
& + \lambda_{1212} (h_{12}^2 + h_{21}^2) + \left\lbrace \lambda_{1323} h_{13}h_{23} + \mathrm{h.c.} \right\rbrace.
\end{aligned}
\end{equation}

$\circ$ For $\mathcal{G}_4 \ast [U(1) \times \mathbb{Z}_2]$ (as $S_2$) we get
\begin{equation}
\begin{aligned}
V ={}& \lambda_{3333} h_{33}^2 + \lambda_{1221} h_{12} h_{21} + \lambda_{1122}h_{11}h_{22} + \lambda_{1212} (h_{12}^2 + h_{21}^2) \\
&+ \lambda_{1332} (h_{13}h_{32} + h_{23}h_{31}) + \lambda_{1233} (h_{12} + h_{21}) h_{33}.
\end{aligned}
\end{equation}

$\circ$ For $\mathcal{G}_4 \ast D_4$ we get
\begin{equation}
\begin{aligned}
V ={}& \lambda_{3333} h_{33}^2 + \lambda_{1221} h_{12} h_{21} + \lambda_{1122}h_{11}h_{22} + \lambda_{1212} (h_{12}^2+ h_{21}^2)\\
& + \left\lbrace \lambda_{1323} h_{13}h_{23} + \mathrm{h.c.} \right\rbrace.
\end{aligned}
\end{equation}

\item $\mathbf{|2_3 2_1|}$

$\circ$ For $\mathcal{G}_6$ we get
\begin{equation}
\begin{aligned}
V ={}& \lambda_{1111} (h_{11}^2 + h_{22}^2 + h_{33}^2) + \lambda_{1122} ( - h_{11} h_{22} + h_{11} h_{33} + h_{22} h_{33}) \\
& + \lambda_{1221}(-h_{12}h_{21} + h_{13}h_{31} +h_{23}h_{32})\\& + \left\lbrace \lambda_{1212}(-h_{12}^2 +h_{13}^2 -h_{23}^2 ) + \mathrm{h.c.} \right\rbrace.
\end{aligned}
\end{equation}

$\circ$ For $\mathcal{G}_6 \ast D_4$ we get
\begin{equation}
\begin{aligned}
V ={}& \lambda_{1111} (h_{11}^2 + h_{22}^2 + h_{33}^2) + \lambda_{1122} ( - h_{11} h_{22} + h_{11} h_{33} + h_{22} h_{33}) \\
& + \lambda_{1221}(-h_{12}h_{21} + h_{13}h_{31} +h_{23}h_{32})\\& +  \lambda_{1212}(-h_{12}^2 +h_{13}^2 -h_{23}^2 + \mathrm{h.c.}) .
\end{aligned}
\end{equation}

\item $\mathbf{|3_2 2_1|}$

$\circ$ For $\mathcal{G}_1 \ast [U(1) \times \mathbb{Z}_2]_{(h_2{-}h_3)}$ we get
\begin{equation}
\begin{aligned}
V ={}& \lambda_{1111} h_{11}^2 + \lambda_{2222} h_{22}^2 + \lambda_{3333} h_{33}^2 + \lambda_{1221} h_{12} h_{21} + \lambda_{1122}h_{11}h_{22} \\
& + \left\lbrace\lambda_{2323} h_{23}^2 + \mathrm{h.c.} \right\rbrace.
\end{aligned}
\end{equation}

\end{itemize}

\item $\mathbf{8_1},~\mathbf{|2_2 4_1|}$

$\circ$ For $\mathcal{G}_1$ we get
\begin{equation}
\begin{aligned}
V ={}& \lambda_{1111} h_{11}^2 + \lambda_{2222} h_{22}^2 + \lambda_{3333} h_{33}^2 + \lambda_{1221} h_{12} h_{21} + \lambda_{1122}h_{11}h_{22} \\
&+ \bigg\{ \lambda_{1112} h_{11}h_{12} + \lambda_{1222} h_{12}h_{22} + \lambda_{1323} h_{13}h_{23} \\
&\qquad+ \lambda_{1212} h_{12}^2 + \lambda_{1313} h_{13}^2 + \lambda_{2323} h_{23}^2 + \mathrm{h.c.} \bigg\}.
\end{aligned}
\end{equation}

$\circ$ For $\mathcal{G}_1 \ast [\mathbb{Z}_2 \times \mathbb{Z}_2]$ we get
\begin{equation}
\begin{aligned}
V ={}& \lambda_{1111} h_{11}^2 + \lambda_{2222} h_{22}^2 + \lambda_{3333} h_{33}^2 + \lambda_{1221} h_{12} h_{21} + \lambda_{1122}h_{11}h_{22}\\
&+ \left\lbrace \lambda_{1212} h_{12}^2 + \lambda_{1313} h_{13}^2 + \lambda_{2323} h_{23}^2 + \mathrm{h.c.} \right\rbrace.
\end{aligned}
\end{equation}

$\circ$ For $\mathcal{G}_1 \times \mathcal{G}_1$ we get
\begin{equation}
\begin{aligned}
V ={}& \lambda_{1111} h_{11}^2 + \lambda_{2222} h_{22}^2 + \lambda_{3333} h_{33}^2 \\
&+ \left\lbrace \lambda_{1212} h_{12}^2 + \lambda_{1313} h_{13}^2 + \lambda_{2323} h_{23}^2 + \mathrm{h.c.} \right\rbrace.
\end{aligned}
\end{equation}

$\circ$ For $\mathcal{G}_3$ we get
\begin{equation}
\begin{aligned}
V ={}& \lambda_{1111} h_{11}^2 + \lambda_{2222} h_{22}^2 + \lambda_{3333} h_{33}^2 + \lambda_{1221} h_{12} h_{21} + \lambda_{1122}h_{11}h_{22}\\
& + \left\lbrace \lambda_{1332} h_{13}h_{32} + \lambda_{1233} h_{12}h_{33} + \lambda_{1212} h_{12}^2 + \lambda_{1313} h_{13}^2 + \lambda_{2323} h_{23}^2 + \mathrm{h.c.} \right\rbrace.
\end{aligned}
\end{equation}

$\circ$ For $\mathcal{G}_3 \ast [\mathbb{Z}_2 \times \mathbb{Z}_2]$ (applied as $(S_2)_{(h_1,\,h_2)}$ and $(\mathbb{Z}_2)_{h_3}$) we get
\begin{equation}
\begin{aligned}
V ={}& \lambda_{1111} (h_{11}^2 + h_{22}^2) + \lambda_{3333} h_{33}^2 + \lambda_{1221} h_{12} h_{21} + \lambda_{1122}h_{11}h_{22} \\
&+ \lambda_{1332} (h_{13}h_{32} + h_{23}h_{31}) + \lambda_{1233} (h_{12} + h_{21})h_{33} + \lambda_{1212} (h_{12}^2+ h_{21}^2) \\
&+ \left\lbrace \lambda_{1313} (h_{13}^2 + h_{23}^2) + \mathrm{h.c.} \right\rbrace.
\end{aligned}
\end{equation}

$\circ$ For $\mathcal{G}_4$ we get
\begin{equation}
\begin{aligned}
V ={}& \lambda_{1111} (h_{11}^2 - h_{22}^2) + \lambda_{3333} h_{33}^2 + \lambda_{1221} h_{12} h_{21} + \lambda_{1122}h_{11}h_{22} \\
&+ \lambda_{1332} (h_{13}h_{32} + h_{23}h_{31}) + \lambda_{1233} (h_{12} + h_{21})h_{33} + \lambda_{1212} (h_{12}^2+ h_{21}^2) \\
&+ \left\lbrace \lambda_{1323} h_{13}h_{23} + \mathrm{h.c.} \right\rbrace.
\end{aligned}
\end{equation}

$\circ$ For $\mathcal{G}_4 \ast [\mathbb{Z}_2 \times \mathbb{Z}_2]$ (applied as $(S_2)_{(h_1,\,h_2)}$ and $(\mathbb{Z}_2)_{h_3}$) we get
\begin{equation}
\begin{aligned}
V ={}& \lambda_{3333} h_{33}^2 + \lambda_{1221} h_{12} h_{21} + \lambda_{1122}h_{11}h_{22} \\
& + \lambda_{1212} (h_{12}^2+ h_{21}^2) + \lambda_{1332} (h_{13}h_{32} + h_{23}h_{31}) + \lambda_{1233} (h_{12} + h_{21})h_{33}\\
&+ \left\lbrace \lambda_{1323} h_{13}h_{23} + \mathrm{h.c.} \right\rbrace.
\end{aligned}
\end{equation}

$\circ$ For $\mathcal{G}_5$ we get
\begin{equation}
\begin{aligned}
V ={}& \lambda_{1111} h_{11}^2 + \lambda_{2222} h_{22}^2 + \lambda_{3333} h_{33}^2 + \lambda_{1221} h_{12} h_{21} + \lambda_{1122}h_{11}h_{22}\\
&  + \left\lbrace \lambda_{1323} h_{13}h_{23} + \lambda_{1332} h_{13}h_{32} + \lambda_{1233} h_{12}h_{33} + \lambda_{1212} h_{12}^2 + \mathrm{h.c.} \right\rbrace.
\end{aligned}
\end{equation}

\end{itemize}

\subsection{GCP-like invariant potentials}\label{App:GCP_TGOOFy}

Categorised in terms of the eigenvalues patterns, different identified scalar potentials are:

\begin{itemize}

\item $\mathbf{1_4 4_1}$

$\circ$ For $\mathcal{G}_7^\ast  \ast \left[ ([U(1) \times D_4]/ \mathbb{Z}_2) \rtimes \mathbb{Z}_2^\ast \right]$ we get
\begin{equation}
\begin{aligned}
V ={}& \lambda_{1111} (h_{11}^2 + h_{22}^2) + \lambda_{3333} h_{33}^2 + \lambda_{1221} h_{12} h_{21} + \lambda_{1122}h_{11}h_{22} \\ & +  \left\lbrace \lambda_{1212} h_{12}^2 + \mathrm{h.c.} \right\rbrace.
\end{aligned}
\end{equation}

\newpage
\item $\mathbf{2_2 4_1}$

\begin{itemize}

\item $\mathbf{1_4 1_2 2_1}$

$\circ$ For $\mathcal{G}_{8}^\ast$ we get
\begin{equation}
\begin{aligned}
V ={}& \lambda_{1111} (h_{11}^2 + h_{22}^2 - h_{12}^2 - h_{21}^2) + \lambda_{3333} h_{33}^2 + \lambda_{1122} (h_{11}h_{22} + h_{12} h_{21}) \\
& + \lambda_{1112} (h_{22} - h_{11})(h_{12} + h_{21}) + \lambda_{1323} (h_{13}h_{23} + h_{31}h_{32})\\
& + \lambda_{1313} ( h_{13}^2 - h_{23}^2 + \mathrm{h.c.}).
\end{aligned}
\end{equation}

\item $\mathbf{1_4 4_1}$

$\circ$ For $\mathcal{G}_{1}^\ast  \ast U(1)_{h_3}$ we get
\begin{equation}
\begin{aligned}
V ={}& \lambda_{1111} h_{11}^2 + \lambda_{2222} h_{22}^2 + \lambda_{3333} h_{33}^2 + \lambda_{1221} h_{12} h_{21} + \lambda_{1122}h_{11}h_{22} \\ &  + \lambda_{1112} h_{11}(h_{12} + h_{21}) + \lambda_{1222} (h_{12} + h_{21})h_{22} + \lambda_{1212}(h_{12}^2 + h_{21}^2)\\
& + i \lambda_{1233} (h_{12} - h_{21})h_{33} + i \lambda_{1332} (h_{23}h_{31} - h_{13} h_{32}).
\end{aligned}
\end{equation}

$\circ$ For $\mathcal{G}_{1}^\ast  \ast \mathbb{Z}_4^\ast$ (CP4 as in eq.~\eqref{Eq:CP4_gen_pm_1}) we get
\begin{equation}
\begin{aligned}
V ={}& \lambda_{1111} (h_{11}^2 +  h_{22}^2) + \lambda_{3333} h_{33}^2 + \lambda_{1221} h_{12} h_{21} + \lambda_{1122}h_{11}h_{22} \\
& + \lambda_{1212} (h_{12}^2 + h_{21}^2) + \lambda_{1313} (h_{13}^2 + h_{23}^2 + \mathrm{h.c.})\\
& + \lambda_{1112} (h_{22} - h_{11}) (h_{12} + h_{21}).
\end{aligned}
\end{equation}

$\circ$ For $\mathcal{G}_7^\ast  \ast U(1)_{h_3}$ we get
\begin{equation}
\begin{aligned}
V ={}& \lambda_{1111} (h_{11}^2 + h_{22}^2) + \lambda_{3333} h_{33}^2 + \lambda_{1221} h_{12} h_{21} + \lambda_{1122}h_{11}h_{22} \\ &  + \lambda_{1133} (h_{22} - h_{11})h_{33} +  \lambda_{1331} (h_{13}h_{31} - h_{23}h_{32}) \\
& + \bigg\{ \lambda_{1212} h_{12}^2 + \lambda_{2331} h_{23} h_{31} + \lambda_{1112} (h_{11} - h_{22})h_{21} +  \lambda_{1233} h_{12}h_{33} + \mathrm{h.c.} \bigg\}.
\end{aligned}
\end{equation}

$\circ$ For $\mathcal{G}_7^\ast  \ast  \mathbb{Z}_4^\ast$ (CP4 as in eq.~\eqref{Eq:CP4_gen_pm_i}) we get
\begin{equation}
\begin{aligned}
V ={}& \lambda_{1111} (h_{11}^2 + h_{22}^2) + \lambda_{3333} h_{33}^2 + \lambda_{1221} h_{12} h_{21} + \lambda_{1122}h_{11}h_{22} \\ &  +  \lambda_{1323} (h_{13}h_{23} + h_{31}h_{32}) \\
& + \bigg\{ \lambda_{1112} (h_{22} - h_{11}) h_{12} + \lambda_{1212} h_{12}^2 +  \lambda_{1313} (h_{13}^2 + h_{23}^2) + \mathrm{h.c.} \bigg\}.
\end{aligned}
\end{equation}

$\circ$ For $\mathcal{G}_7^\ast  \ast \mathbb{Z}_4$ we get
\begin{equation}
\begin{aligned}
V ={}& \lambda_{1111} (h_{11}^2 + h_{22}^2) + \lambda_{3333} h_{33}^2 + \lambda_{1221} h_{12} h_{21} + \lambda_{1122}h_{11}h_{22} \\ &  + \lambda_{1133} (h_{22} - h_{11})h_{33} +  \lambda_{1331} (h_{13}h_{31} - h_{23}h_{32})\\
& + \lambda_{1323} (h_{13}h_{23} + h_{31}h_{32})  + \bigg\{ \lambda_{1212} h_{12}^2 + \mathrm{h.c.} \bigg\}.
\end{aligned}
\end{equation}

$\circ$ For $\mathcal{G}_7^\ast  \ast [V_4 \rtimes \mathbb{Z}_2^\ast ]$ (as in eq.~\eqref{Eq:GCP_CP4_Z2_Z2}) we get
\begin{equation}
\begin{aligned}
V ={}& \lambda_{1111} (h_{11}^2 + h_{22}^2) + \lambda_{3333} h_{33}^2 + \lambda_{1221} h_{12} h_{21} + \lambda_{1122}h_{11}h_{22} \\ & + \lambda_{1212} (h_{12}^2 + h_{21}^2) + \lambda_{1313} (h_{13}^2 - h_{23}^2 + \mathrm{h.c.})\\
& + \lambda_{1323} (h_{13}h_{23} + h_{31}h_{32}) + \lambda_{1112} (h_{22} - h_{11})(h_{12} + h_{21}).
\end{aligned}
\end{equation}

\end{itemize}

\item $\mathbf{8_1},~ \mathbf{|2_2 4_1|}$

$\circ$ For $\mathcal{G}_1^\ast $ we get
\begin{equation}
\begin{aligned}
V ={}& \lambda_{1111} h_{11}^2 +  \lambda_{2222} h_{22}^2 + \lambda_{3333} h_{33}^2 + \lambda_{1221} h_{12} h_{21} + \lambda_{1122}h_{11}h_{22} \\
& + \lambda_{1212} (h_{12}^2 + h_{21}^2) + \lambda_{1313} (h_{13}^2 + h_{31}^2) + \lambda_{2323} (h_{23}^2 + h_{32}^2)\\
&  + \lambda_{1323} (h_{13}h_{23} + h_{31}h_{32}) + \lambda_{1112} h_{11} (h_{12} + h_{21}) + \lambda_{1222} (h_{12} + h_{21}) h_{22}\\
& + i \lambda_{1233} (h_{12} - h_{21}) h_{33} + i \lambda_{1332} (h_{23}h_{31} - h_{13}h_{32}).
\end{aligned}
\end{equation}

$\circ$ For $\mathcal{G}_7^\ast$ we get
\begin{equation}
\begin{aligned}
V ={}& \lambda_{1111} (h_{11}^2 + h_{22}^2) + \lambda_{3333} h_{33}^2 + \lambda_{1221} h_{12} h_{21} + \lambda_{1122}h_{11}h_{22}\\
&  + \lambda_{1133} (h_{22} - h_{11})h_{33} +  \lambda_{1331} (h_{13}h_{31} - h_{23}h_{32}) + \lambda_{1323} (h_{13}h_{23} + h_{31}h_{32})\\
& + \bigg\{ \lambda_{1212} h_{12}^2 + \lambda_{2331} h_{23} h_{31} + \lambda_{1112} (h_{11} - h_{22})h_{21}\\
& \qquad+  \lambda_{1233} h_{12}h_{33} + \lambda_{1313} (h_{13}^2 - h_{32}^2) + \mathrm{h.c.} \bigg\}.
\end{aligned}
\end{equation}

$\circ$ For $\mathcal{G}_7^\ast \ast [\mathbb{Z}_2 \times \mathbb{Z}_2]$ we get
\begin{equation}
\begin{aligned}
V ={}& \lambda_{1111} (h_{11}^2 + h_{22}^2) + \lambda_{3333} h_{33}^2 + \lambda_{1221} h_{12} h_{21} + \lambda_{1122}h_{11}h_{22}\\
&  + \lambda_{1133} (h_{22} - h_{11})h_{33} +  \lambda_{1331} (h_{13}h_{31} - h_{23}h_{32})\\
& + \bigg\{ \lambda_{1212} h_{12}^2  + \lambda_{1313} (h_{13}^2 - h_{32}^2) + \mathrm{h.c.} \bigg\}.
\end{aligned}
\end{equation}
In another basis, utilising $\mathcal{R}_{(\alpha,\, 0,\, \theta,\, 0)}$ transformation, it can be written as
\begin{equation}
\begin{aligned}
V ={}& \lambda_{1111} (h_{11}^2 + h_{22}^2) + \lambda_{3333} h_{33}^2 + \lambda_{1221} h_{12} h_{21} + \lambda_{1122}h_{11}h_{22}\\
&  + \lambda_{1212} (h_{12}^2 + h_{21}^2) + \lambda_{1313} (h_{13}^2 + h_{23}^2 + \mathrm{h.c.})\\
& + \lambda_{1233} (h_{12} + h_{21})h_{33} + \lambda_{1332} (h_{13}h_{32} + h_{23}h_{31})\\& + i \lambda_{1323} (h_{13}h_{23} - h_{31}h_{32}).
\end{aligned}
\end{equation}

\end{itemize}

\bibliographystyle{JHEP}
\bibliography{ref}

\end{document}